\documentclass[twocolumn]{aastex701}

\usepackage{amsmath}
\usepackage{physics}
\usepackage[T1]{fontenc}
\usepackage[utf8]{inputenc}
\usepackage{tabularx}
\usepackage{ragged2e}   
\newcolumntype{Y}{>{\RaggedRight\arraybackslash}X}

\graphicspath{ {./} }
\usepackage{colortbl}

\begin{document}

\title{A New Search Pipeline for Short Gamma Ray Bursts in Fermi/GBM Data - A 50\% Increase in the Number of Detections}

\author[0009-0004-8898-248X]{Ariel Perera}
\affiliation{Department of Particle Physics \& Astrophysics, Weizmann Institute of Science, Rehovot 76100, Israel}
\email[show]{ariel.perera@weizmann.ac.il}

\author[0000-0001-5162-9501]{Barak Zackay}
\affiliation{Department of Particle Physics \& Astrophysics, Weizmann Institute of Science, Rehovot 76100, Israel}
\email{barak.zackay@weizmann.ac.il}

\author[0000-0002-1661-2138]{Tejaswi Venumadhav}
\affiliation{Department of Physics, University of California at Santa Barbara, Santa Barbara, CA 93106, USA}
\affiliation{International Centre for Theoretical Sciences, Tata Institute of Fundamental Research, Bangalore 560089, India}
\email{teja@ucsb.edu}

\begin{abstract}

In this paper, we present the development and the results of a new search pipeline for short gamma-ray bursts (sGRBs) in the publicly available data from the Gamma-Ray Burst Monitor (GBM) on board the Fermi satellite. This pipeline uses rigorous statistical methods that are designed to maximize the information extracted from the Fermi/GBM detectors. 
Our approach differs substantially from existing search efforts in several aspects:
The pipeline includes the construction of template banks, Poisson matched filtering, background estimation, background misestimation correction, automatic routines to filter contaminants, statistical estimation of the signal location and a quantitative estimator of the signal probability to be of a cosmological, terrestrial, or solar origin. Our analysis also includes operating the pipeline on "time-slided" copies of the data, which allows exact significance assessment and $p_{\text{astro}}$ computation, akin to the state-of-the-art gravitational waves (GW) data analysis pipelines. Depending on the spectral properties of the bursts, our pipeline achieves a signal-to-noise ratio (SNR) improvement by a factor of 2 to 15 over the onboard GBM triggering algorithm. This enhancement increases the detectable volume for sGRBs and results in an approximate 50\% increase in sGRB detections in the 2014 GBM dataset. As a further consequence of the sensitivity increase, we detect hundreds of soft gamma-ray flares of galactic origin. This improved sensitivity enhances the chances of detecting fainter, off-axis GRBs that would likely fall below the standard triggering thresholds. Applying this pipeline to the full GBM archive is expected to expand further the joint sGRB–GW detection volume.


\end{abstract}


\section{Introduction}\label{sec:intro}
Ever since the very first serendipitous detection of Gamma-ray Bursts (GRBs) in the 1970s (\cite{1973ApJ...182L..85K}), a few thousand events have been observed (\cite{10.1088/2514-3433/aae164}). These are brief, energetic bursts of electromagnetic (EM) radiation with energies in the gamma-ray band - typically from tens of KeV to one MeV - and durations lasting from a fraction of a second to a few hundred seconds (\cite{kouveliotou1993identification, paciesas1999fourth, von2020fourth, lien2016third}). The growing number of GRB detections, and the development of instruments with better spectral resolution, have allowed for the identification of two distinct classes based on duration and spectral hardness (\cite{kouveliotou1993identification}). These classes have been further analyzed and refined with enhanced spectral data from detectors and improved statistics from larger samples (\cite{von2020fourth}). In addition, the increased number of GRBs, including their localizations, allows for better characterization of their rates, hosts, and progenitors (\cite{berger2014short, blanchard2016offset, fong2009hubble, fong2013locations}). 
It is now believed that long/soft GRBs arise from the collapse of a massive star (\cite{woosley1993gamma, galama1998unusual, kulkarni1998radio, lunnan2014hydrogen}), which is based on the GRB-supernova association of tens of events (\cite{cano2017observer}). For short/hard bursts, the progenitor was predicted to be a binary neutron star merger (\cite{paczynski1986gamma, goodman1986gamma, eichler1989nucleosynthesis}) which was also expected to be the site of heavy element production - a so-called kilonova (\cite{li1998transient, kulkarni2005modeling}, also see \cite{metzger2020kilonovae} for a review). 
This hypothesis was confirmed observationally by 
the joint observation of the gravitational wave event GW170817 (\cite{abbott2017gw170817}), an associated short GRB (\cite{goldstein2017ordinary, savchenko2017integral}), and a kilonova from the same source (\cite{smartt2017kilonova, tanvir2017emergence}). 
This is the only GW signal event to date with a confirmed EM counterpart -- the joint event was remarkably successful in constraining the central engine of GRBs in ways that would be impossible using EM signals alone (\cite{abbott2017gravitational}), but it also raised new questions that remain unanswered. For example, the luminosity of GRB170817A was several orders of magnitude lower than that of any previously observed GRB (\cite{abbott2017gravitational}). Additionally, the EM signal was detected approximately 1.7 seconds after the GW signal ended, a timing offset that remains a topic of debate (\cite{zhang2019delay, salafia2018interpreting, gottlieb2018cocoon}). 
These scientific questions underscore the importance of multimessenger astronomy, and specifically motivate investigations into maximizing our capability to detect GRBs both alone and in coincidence/follow-up with GWs.

Today, the most sensitive all-sky GRB detector is the Gamma-Ray Burst Monitor (GBM; \cite{meegan2009fermi}) on board the Fermi satellite. 
GBM consists of 14 scintillation detectors, 12 of which are Sodium Iodide (NaI) crystals covering an energy range of $8-1000$ KeV, and two Bismuth Germanate (BGO) crystals covering the higher energy range of $200$ KeV - $40$ MeV. At any given time, this set of detectors is sensitive to the entire sky unocculted by the Earth, corresponding to a solid angle of approximately 8 steradians.
Given that gravitational-wave detectors also operate with all-sky coverage, GBM is well-suited for finding coincident GRB–GW detections. 

The yearly GRB detection rate of search pipelines based on GBM's on-board triggering algorithms is $\sim 240$ (\cite{von2020fourth}), out of which $\sim1/6$ are short-duration GRBs. 
The flight software on board the spacecraft monitors the detected rates of gamma-ray photons and triggers if the count rates in two detectors or more exceed a fixed threshold above the background for different timescales (\cite{von2020fourth}). 
There are 119 combinations of timescale, energy range, and threshold used for triggering, with some of them inactive during most of the mission. 
Due to the triggering logic adopted, the significance of a GRB is determined only by the second loudest detector, resulting in a reduction of the signal-to-noise ratio (SNR) by approximately a factor of $2$ compared to a coherent analysis of all the detectors (\cite{blackburn2015high}). 
Additional efforts to detect GRBs of high interest are made by targeted searches (\cite{kocevski2018analysis, burns2019fermi}). These are offline (i.e. performed on the ground after the relevant data has been transmitted to Earth) searches around GW triggers from the Laser Interferometer Gravitational-wave Observatory (LIGO). 
The statistical procedure is described in \cite{blackburn2015high}, and updates to different LIGO observing runs are being made (\cite{goldstein2016updates, goldstein2019updates}).

GRBs can have very short durations ($\lesssim 0.1$ sec) and a wide range of spectral shapes. 
Hence, additional methods that are optimized for low photon number counts and targeted to the diverse spectra can, in-principle, further enhance detection capabilities.
This motivates the development of a scheme that optimizes the detection statistic SNR across all photon count regimes and searches all the available data. 
Such a scheme can both increase the number of GRB detections in regions of parameter space that are currently searched in, and allow us to probe the extreme lower tail of the GRB duration distribution, which was inaccessible by currently used detection methods.

In this paper, we present the development of a fully automated GRB detection pipeline that coherently analyzes the Time Tagged Event (TTE) data provided by GBM and searches for GRBs with durations ranging from 3 milliseconds up to 3.606 seconds. Our pipeline includes a rigorous detection statistic, which is nearly optimal (in the statistical sense) in the low counts Poisson regime, a template bank for a wide range of GRB spectra, and a statistical estimation of the sky position and spectrum of the candidate GRBs, which facilitates a follow-up gravitational wave searches around the trigger time. Although such a scheme can become computationally intensive, our methods and approximations allow for a comprehensive search of the GBM archival data in a reasonable computing time.

The paper is organized as follows: Section \ref{sec:pipe outline} describes the pipeline stages, \ref{sec:pipe decs} gives a more in-depth view of each stage, and in section \ref{sec:res}, we show the results of operating the pipeline on a year of GBM data. We conclude in section \ref{sec:conc}.

\section{Pipeline Stages} \label{sec:pipe outline}
We will begin by providing a birds-eye view of the pipeline and its various stages. 
The central idea is to use the Neyman–Pearson framework (\cite{neyman1933ix}) for statistical hypothesis testing to make a decision between competing hypotheses in a principled and optimal manner. 
This involves constructing a test statistic in the form of a likelihood ratio, marginalized over nuisance parameters. 
We apply this approach throughout the pipeline, both to identify genuine transients and to construct tests that reject contaminating signals.

\begin{figure*}
    \centering
    \includegraphics[scale=0.7]{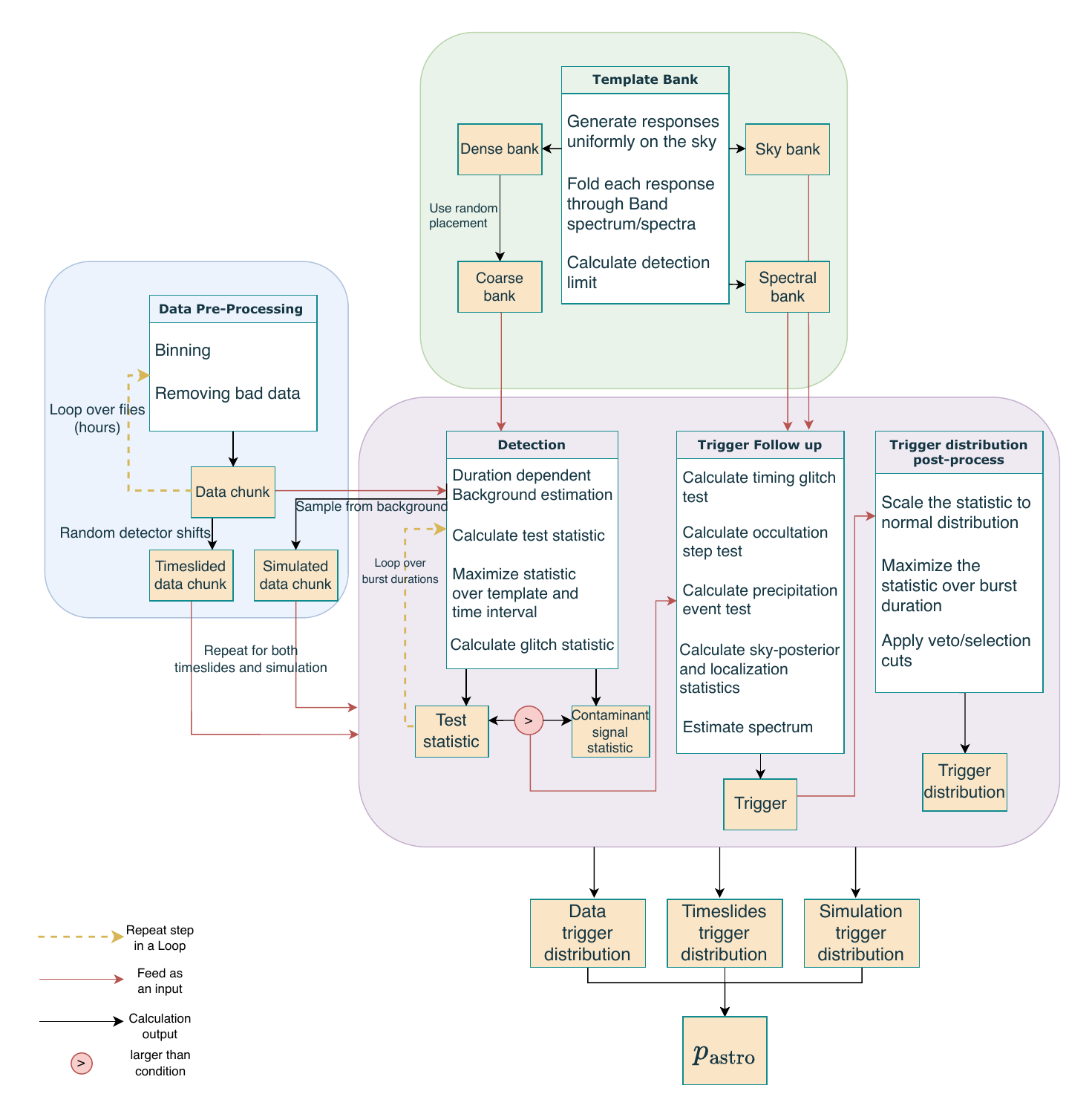}
    \caption{\textbf{A flowchart describing the pipeline structure}.}
    \label{fig: pipeflow}
\end{figure*}
The pipeline consists of the following stages, which are also summarized in Figure \ref{fig: pipeflow}.
\begin{enumerate}
    \item \textit{Construction of template banks}: 
    We create a representative collection of GRB spectra for a grid of points uniformly distributed in the sky and fold it through the detector response. We also model common non-GRB signals appearing in the data, corresponding to effects such as cosmic ray events, and make a "contaminant signal" (CS) bank, which is used to veto triggers. Two more banks were created for sky localization and estimation of spectral parameters.
    \item \textit{Data selection and pre-processing}: We apply cuts to the TTE data to ensure good data quality. This includes working only within the combined good time intervals (GTI) per detector provided by the GBM team, excluding times where more than 2 detectors were off for more than a few milliseconds, and times of detector throttling due to very high count rates. We then bin the data to either 10 ms or 1 ms, depending on the burst duration we search.
    \item \textit{Matched Filtering:}
        Our detection statistic is a Poisson match filter that optimally combines all the information from all the detectors in all the energy channels; we describe the statistic in Sec. \ref{sec: statistic} and provide a detailed derivation in Appendix \ref{appendix: test statistic}. We use a similar method to filter non-GRB events present in the data, where triggers with a CS statistic value exceeding the detection statistic are filtered out at this stage. We follow up on each trigger and estimate its position and spectrum by maximizing the marginalized posterior distributions.
    \item \textit{Background Counts Estimation:}
        To apply the detection statistic, a knowledge of the background count level is needed. We introduce a novel approach to local background count estimation when searching for a GRB, which will be detailed in Section \ref{sec: bkg est}.
    \item \textit{Background Spectral Drift Correction:} 
        A misestimation of the background leads to a loss in the sensitivity (which is defined as the SNR of the detection statistic, see appendix \ref{appendix: match}, Eq. \ref{eq: information content}) of the search. To correct for this, we re-normalize the matched-filter time series using its locally estimated mean and variance. The extent of the loss in sensitivity depends on the spectral shape of both the template and the background. We discuss this correction in section \ref{sec: drift correction}.
    
    \item \textit{Vetoing triggers}: To remove common contaminations, we devised more statistical tests to eliminate triggers. These are:
    \begin{enumerate}
        \item Single detector test: It is assumed that a GRB must appear in more than one detector. A statistical test rejects triggers coming from a single detector versus more than one detector.
        \item Timing glitches - GBM has timing glitches in their field programmable gate array (FPGA) logic\footnote{\url{https://fermi.gsfc.nasa.gov/ssc/data/analysis/GBM_caveats.html}} (\cite{von2014second}), resulting in peaks and dips due to data buffering, which may be separated by milliseconds to minutes. This implies that these triggers must have the background spectrum. This is also being statistically tested.
        \item Particle precipitation events: These occur when charged particles interact with the spacecraft materials or the Earth's atmosphere, producing bremsstrahlung radiation that can mimic astrophysical transients and lead to false triggers. We perform a statistical analysis to identify and remove such events.
        \item Occultation step: When a persistent gamma-ray source sets or rises over the Earth’s limb, it produces a sharp decrease or increase in the detector count rate. These changes can mimic transient signals and result in false triggers. We identify and remove such events through statistical tests.
        \item Trigger direction: Gamma rays from the Earth or Sun can also hit the detectors. These are the results of terrestrial gamma-ray flashes (TGFs) and solar flares (SFs), respectively. We apply a statistical test to assign a quantitative measure of the probability of a trigger being a result of cosmological, terrestrial, or solar gamma rays. This method also gives a localization map of the trigger as a product. 
    \end{enumerate}
    \item \textit{Background trigger distribution estimation}: To account for systematics, we empirically estimate the trigger distribution of our pipeline given no signal in the data by performing a full re-run of the pipeline on a `clean' dataset. 
    To achieve such a data realization that is devoid of GRBs, we use the method of \textit{timeslides} (\cite{2010CQGra..27a5005W}). 
    This uses the fact that astrophysical sources populate photons in the 14 detectors onboard GBM in a manner that is coincident in time; we temporally shift the different detectors with respect to one another to create a synthetic data series without these coincident bursts. We also validate this method by generating simulated data with the estimated background levels, running our pipeline on it, and checking that the estimated background is consistent with that from timeslides \footnote{Note that when a real signal is present, time-slides can produce up to 14 single-detector events at different times. Our single-detector veto ensures that these signal-related artifacts are filtered out and do not contaminate the time-slid background distribution. See sec. \ref{sec: single detector veto}.}.
    \item \textit{Assigning probability of being of astrophysical origin}:
        $p_{\text{astro}}$ quantifies how likely it is for a trigger to be a real transient (of astrophysical origin) vs. to be a result of a statistical fluctuation (\cite{kapadia2020self, venumadhav2019new}). We estimate it empirically from the data by the formula
        \begin{equation}\label{eq: pastro}
            p_{\text {astro }}(\text {event})=\frac{P\left(\text {event} | \mathcal{H}_1\right)}{P\left(\text {event} | \mathcal{ H}_0\right)+P\left(\text {event} | \mathcal{H}_1\right)},
        \end{equation}
        where $\mathcal{H}_1$ is the signal hypothesis and $\mathcal{ H}_0$ is the background triggers hypothesis. One of the advantages of this formulation is that it allows for a probabilistic interpretation of the reliability of a trigger even in high background regimes, where signals are frequent, overlapping, or poorly separated from noise. In such scenarios, a rough inverse false alarm rate or ranking statistic cut may be inefficient. 
    
\end{enumerate}

\section{Description of the Pipeline Stages} \label{sec:pipe decs}

    \subsection{Template Banks} \label{sec: template banks}
\begin{figure}
            \centering
            \includegraphics[width=1\linewidth]{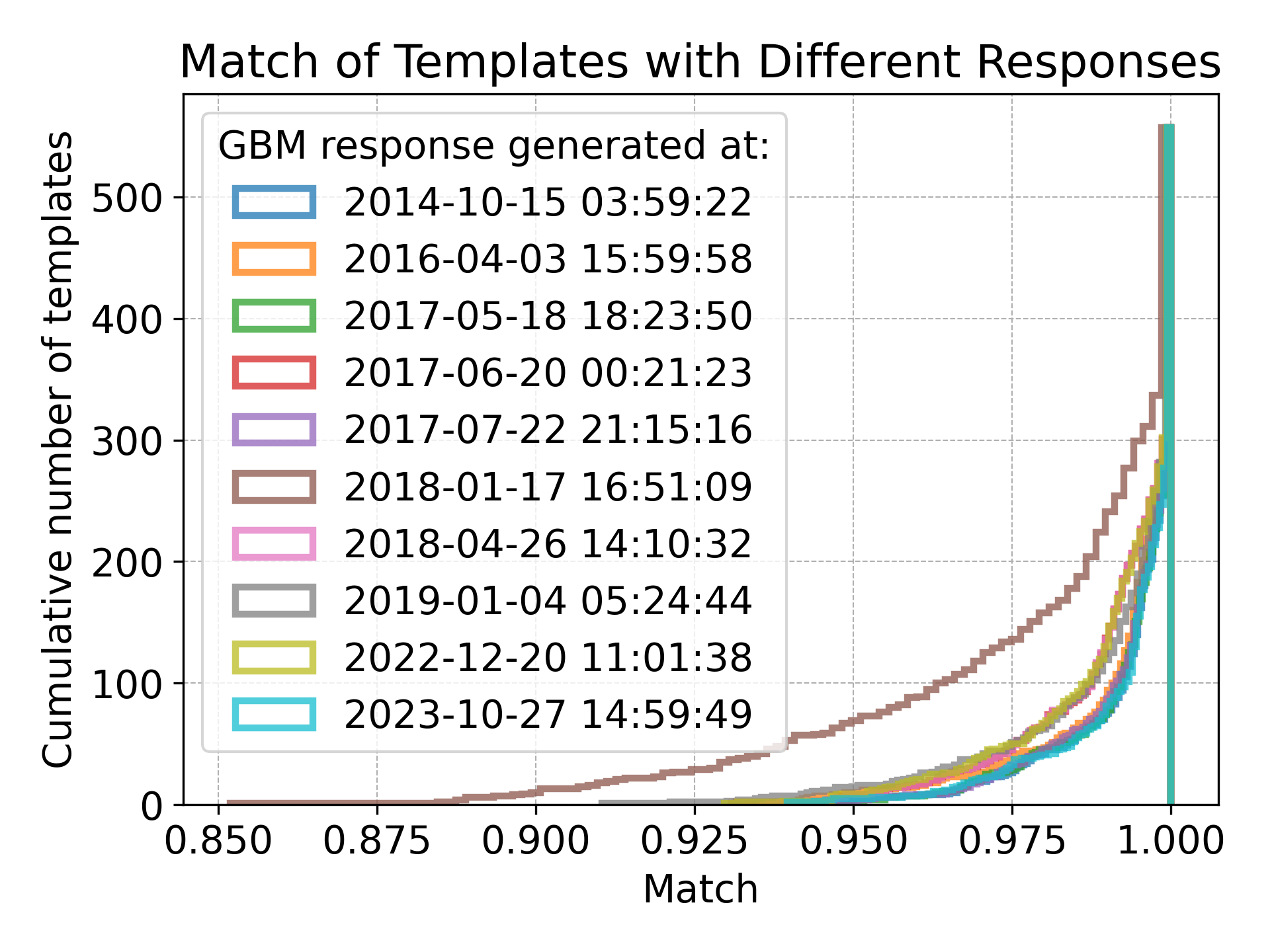}
            \caption{\textbf{Comparison of templates generated with different software at different times.} We matched (as in Eq. \ref{eq: match}) every template in the coarse bank, generated using the response generator \textit{gbm\_drm\_gen} at met=654393605.0 sec (2021-09-27 00:00:00), with templates having the same spectral parameters and sky positions but generated using GBM's response generator at the specified dates, which were randomly chosen. One can see that the maximal mismatch is only 15\%, and that for the most part, the time in which the response was generated does not affect the template significantly. On 2018-01-17 at 16:51:09 (brown line), Fermi was in an offset-pointing position due to a Target of Opportunity (ToO) observation (Obs ID 100601-1-1), which resulted in a slightly higher mismatch. A full table describing the observing modes during these times is presented in Appendix \ref{appendix: tables}.}
            \label{fig: gbm vs balrog}
        \end{figure}

        GBM provides arrival times of photons in each of their energy channels. 
        If we have knowledge of the spectral shape of the putative signal, we can use these photon counts in an optimal manner to obtain more information about the signal. To this end, one needs to know both the spectral shape and how this shape presents itself when incident on the detector. The transformation between the physical space and count space (i.e. the photon counts) is commonly known as the \textit{detector response matrix} (DRM) and it depends on both the energy of the incident photons and their direction of arrival. This means that our templates will be a function of the 2 sky positions as well as the 3 spectral parameters described below. Our template is thus $T = T(\alpha, \beta, E_{\text{peak}}, \text{RA}, \text{DEC})$. In general, the DRM is also influenced by the atmospheric scattering of photons, which makes the template time-dependent. However, as shown in Fig. \ref{fig: gbm vs balrog}, we find that this time dependence leads to only a modest loss in SNR, which is subdominant to other losses in the pipeline and hence can be safely neglected at the level of detection. For ease of notation, we may refer to the set of parameters aforementioned simply as $\theta$.
        A template bank is a set of templates with parameters covering, ideally, as much parameter space as possible. Since a high completeness bank may require a large number of templates, resulting in a high computational complexity, we also wish to limit the number of templates from the high completeness bank we use when performing a search to the lowest amount possible with the least amount of SNR loss. This is achievable by selecting a smaller subset of templates representative of the entire set. Below, we describe the process of generating a large bank of templates and selecting the representative subset we ultimately use to detect GRBs. \\
        
        \textit{Response generator}: We use the \textit{gbm\_drm\_gen} python package provided by \cite{burgess2018awakening} to generate a response function for each of the GBM detectors for a uniform grid of points on the sky. That is, to account for GRBs coming from each direction to the detectors. To this end, we use the Fibonacci sphere algorithm (\cite{gonzalez2010measurement}) to uniformly place points on the sphere. The approach of generating the templates in advance and storing them on disk rather than generating them for each data segment significantly improves run times as their generation is computationally expensive. This approach comes with a loss in detection sensitivity due to changes in the detector response in time and due to atmospheric scattering. Nevertheless, as shown in Fig. \ref{fig: gbm vs balrog}, the mismatch between our templates, generated at a fixed time, and the same templates generated at different times are small most of the time (less than 5\%). In addition, this shows that the responses generated by GBM's response generator and by \textit{gbm\_drm\_gen} are very similar. \\
        
        \textit{GRB spectra}: GRBs exhibit nonthermal spectra. In the Burst And Transient Source Experiment (BATSE) era, it was found that most GRBs are fit to the phenomenological Band function (\cite{1993ApJ...413..281B}), consisting of two smoothly joint power laws with exponential cutoff given by the functional form:
        \begin{equation}\label{eq: Band function}
        \begin{aligned}
            F_E = A\begin{cases}
            \left(\frac{E}{100 \mathrm{KeV}}\right)^\alpha \exp \left(-\frac{E}{E_0}\right), E<(\alpha-\beta) E_0, 
            \\
            \left[\frac{(\alpha-\beta) E_0}{100 \mathrm{KeV}}\right]^{\alpha-\beta} \exp (\beta-\alpha)\left(\frac{E}{100 \mathrm{KeV}}\right)^\beta,  \\\qquad\qquad\qquad\qquad\qquad E \geq(\alpha-\beta) E_0,
            \end{cases}
        \end{aligned}
        \end{equation}
        Here $\alpha$ and $\beta$ are the low-energy and high-energy spectral slopes, respectively, $A$ is the amplitude, and $E_0$ is the break energy. 
        The break energy is related to the peak of the $\nu F_{\nu}=EF_E=E^2N_E$ (with $N_E$ the number of photons) spectrum by $E_p = (2+\alpha)E_0$. Since this function describes most GRBs, we use it as our base model for the signal detection. This model has three parameters: $\alpha$, $\beta$ and $E_{\text{peak}}$, and we fix the amplitude to the detection limit that we calculate for each template. To create our bank, for each sky position, we randomize $\alpha \in [-1.5, 3]$, $\beta \in [-5, -1.6]$, and $E_{\text{peak}} \in [10, 3000]$ KeV to generate a large sample of spectra. The choice in this parameter range was motivated by \cite{preece2000batse} and the spectral fits accompanied to \cite{von2020fourth}. We further increased the parameter ranges to allow the detection of outlier events and more types of transients (such as TGFs that have a power law behavior in the GBM range). Each spectrum is then folded through the DRM using the 3ML python package (\cite{vianello2015multi}). \\
        
        \textit{Temporal shape}: To account for different burst durations, the templates also carry a temporal shape. We assume that the spectral and temporal components of the signal factorize over the short duration we search for, allowing us to model them separately. As GRB lightcurves are diverse and may vary significantly from one another, we use a box-shaped template in time. This will result in an SNR loss that depends on the shape mismatch between the box and the burst. To account for different burst durations, we use multiple box lengths in our search. The optimal box length choice is logarithmically spaced (see Appendix \ref{appendix: match}). We choose the multiplication factor to be 1.35 (that is, each box width is 1.35 times longer than the previous one) so that the maximal SNR loss between different widths is $\sim 7\%$. This choice is a compromise between the number of templates and our loss tolerance. \\

\begin{figure}
            \centering
            \includegraphics[width=1\linewidth]{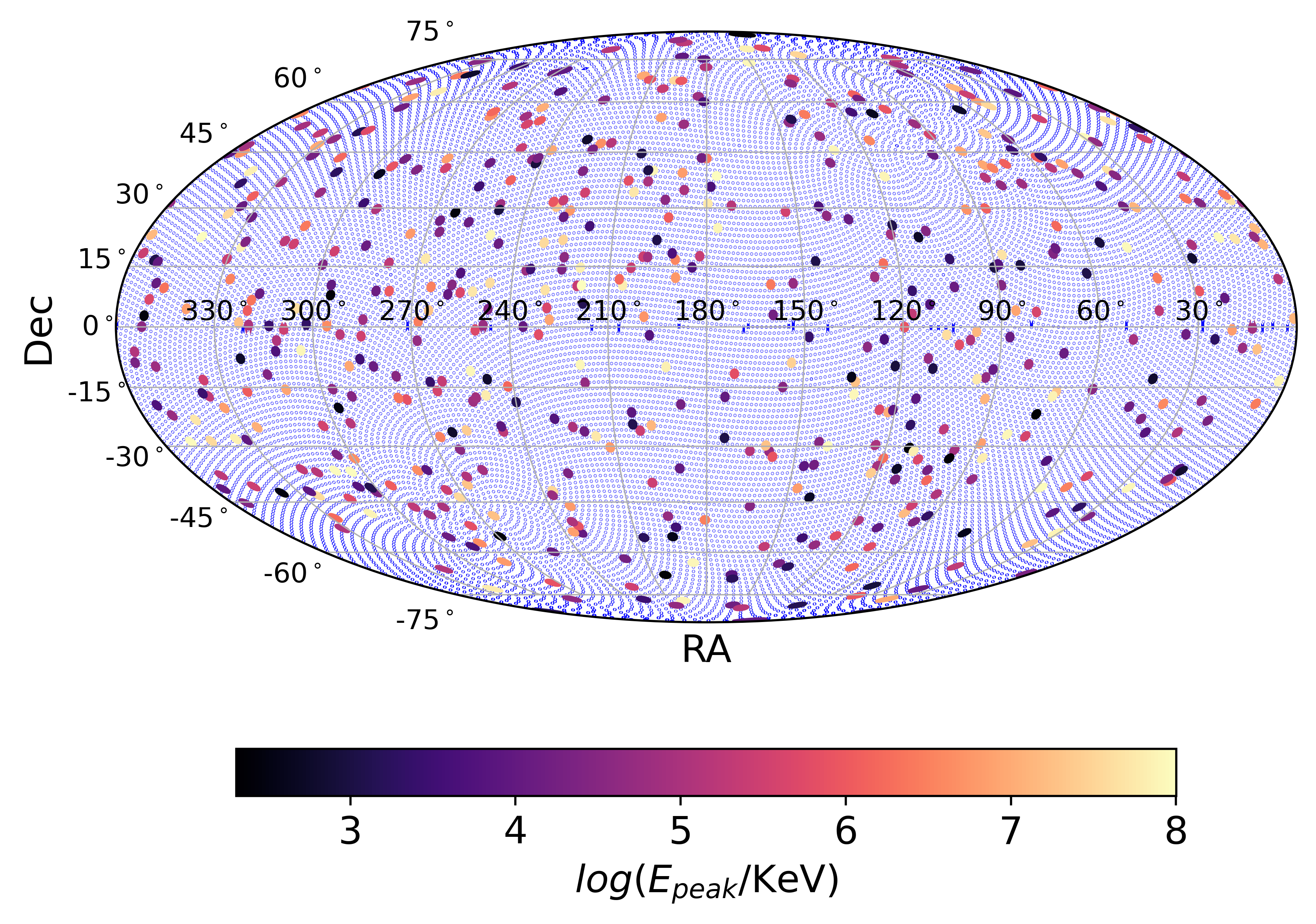}
            \caption{\textbf{Visualization of the constructed template bank.} The small blue dots represent the dense bank template distribution over the sky and the larger points represent the coarse bank created by applying the random placement algorithm. Each point in this plot represents a single point in the five-dimensional parameter space $(\alpha, \beta, E_{\text{peak}}, \text{RA}, \text{DEC})$. The coarse bank points are colored by one of the spectral parameters - $E_{\text{peak}}$.}
            \label{fig:tbank}
        \end{figure}

\begin{figure}
            \centering
            \includegraphics[scale=0.5]{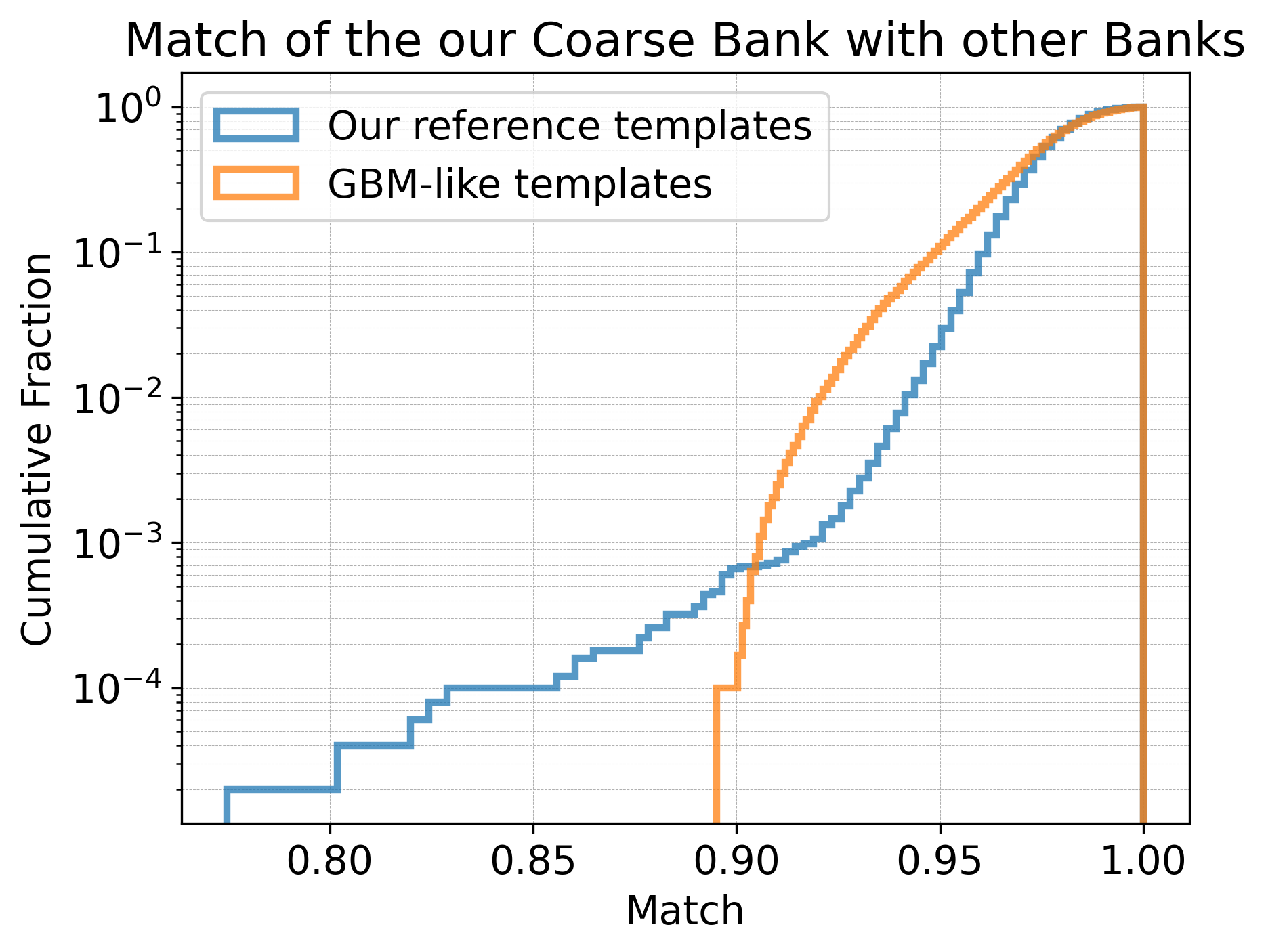}
            \caption{\textbf{Template space coverage checks.} 
            Match of each template in the reference template bank and each template in the GBM-like bank with all the templates in the coarse bank, keeping the maximal value. The y-axis is the cumulative fraction of templates with a maximal match larger than the number specified by the x-axis. The reference bank is a larger bank, containing $5\times10^4$ templates. We see that only a small fraction, $\sim 0.02$, has a match less than $0.95$. Compared with the GBM-like templates, our $\sim 500$ templates cover more than $90\%$ of the parameter space covered by the GBM search bank containing $3$ spectral shapes for each of the $10^4$ sky positions required for $\sim 4$ deg${}^2$ resolution.}
            \label{fig:tbank match}
        \end{figure}

        \begin{deluxetable}{ccccc}
        \setlength{\tabcolsep}{3pt}
        \tablecaption{Template Banks\label{table: banks}}
        \tablehead{
        \colhead{Bank} & \colhead{Structure}& \colhead{\#Templates} & \colhead{Resolution} & \colhead{\#Spectra}
        }
            \startdata
            Detection & Random & 557 & - & -
            \\
            Sky & Grid & $4.5\times10^5$ & $\sim16.5$ deg${}^{2}$ & 180
            \\
            Spectrum & Grid & $5\times10^5$ & $\sim40$ deg${}^{2}$ & 500
            \\
            \enddata
        \tablecomments{For the detection bank we have a single, random parameter for each sky position, and after the reduction algorithm we remain with a subset of 557 templates instead of the full $2\times10^4$ templates. The sky bank and spectral bank are used for parameter estimation.}
        \end{deluxetable}
        \textit{Bank reduction algorithm}: After obtaining the dense template bank by executing the above two steps, we need to reduce the number of templates. We use the random placement algorithm (\cite{harry2009stochastic}) with modifications to account for the relevant statistics for our data. We use the ratio of the SNR of the test statistic when using different templates and define it as $\mathcal{M}$, the "match" between the templates. For two templates $T$ and $T'$, it is given by (see Appendix {\ref{appendix: match}})
        \begin{align}\label{eq: match}
            &\mathcal{M}(T, T') = \frac{\sum_n T_n \log(1+ T'_n/b_n)}{\sum_n T_n \log(1+ T_n/b_n)}
            \cdot \nonumber\\
            &\qquad\qquad\qquad\qquad \cdot\sqrt{\frac{\sum_n b_n\log^2(1 + T_n/b_n)}{\sum_n b_n\log^2(1 + T'_n/b_n)}},
        \end{align}
        where the subscript $n \in [0, 14\times128]$ represents the combined detector-energy channel dimension and $b_n$ is the background. To calculate this quantity, we take a well-behaved data segment without any event (according to the GBM catalog) to estimate the background by taking the mean count rate for each $n$ over the interval. The algorithm is as follows:
        \begin{enumerate}
            \item At step 0, select a random template and place it in the coarse bank.
            \item For all the templates in the coarse bank, calculate their "match" with a template in the dense bank.
            \item If there is a template from the coarse bank without a match higher than $95\%$ with the template from the dense bank, add it to the coarse bank.
            \item Repeat from step 2 with a different template from the dense bank.
        \end{enumerate}
        We started with $2\times10^4$ uniform points on the sphere, such that each template covers 2 square degrees, and carries a random Band spectrum.
        Applying the random placement algorithm, we ended up with 557 templates that cover the same parameter space up to a $5\%$ mismatch (see Fig. \ref{fig:tbank}). To verify that we get the required coverage of the parameter space, we generated another set of templates, using a different reference date for the response and a different background (estimated in another time interval) and calculated their match (Fig. \ref{fig:tbank match} left panel). To further assess the coverage of our template bank, we compared it to the template bank used by the GBM subthreshold search (\cite{goldstein2016updates}) by matching it with our template bank. Their template bank consists of a $\sim 5^{\circ}$ sky resolution of 3 spectra - soft, normal, and hard, where the soft is a Band function with $\alpha=-1.9$, $\beta=-3.7$, $E_{\text{peak}}=70$ KeV, the normal is a Band function with $\alpha=-1$, $\beta=-2.3$, $E_{\text{peak}}=230$ KeV, and the hard is a cutoff power-law with $\alpha=-1.5$, and $E_{\text{peak}}=1500$ KeV. The right panel of Fig. \ref{fig:tbank match} shows that our 557 templates cover the same parameter space as the $\sim 3\times 10^{4}$ templates used in the GBM subthreshold search to a $10\%$ mismatch.
        In addition to the bank constructed for the detection of bursts, we constructed two more banks that we use for parameter estimation (PE). These two banks represent a grid of sky positions and Band spectra, where one is denser in the sky position axis, and one is denser in the Band spectrum axis. A summary of the bank choices is presented in table \ref{table: banks}.
        Furthermore, we construct a single-detector bank to filter out non-GRB loud triggers. There are occasionally single-detector events, probably caused by charged particles exciting the low energy channels of the NAI detectors and causing phosphorescent spikes (\cite{goldstein2019updates}). We model these spikes as a Gaussian in the first three channels of the NAI detectors. For the BGO detectors, we observed that there are transient signals appearing in only one BGO detector with no counterpart at the NAI detectors. We collected a large sample of these triggers and modeled their spectral shapes using Singular Value Decomposition (SVD). These templates are shown in Fig. \ref{fig: bgo glitches} in Appendix \ref{appendix: glitch bank}.

    \subsection{Data pre-processing}
        The data format we use in our analysis is TTE, provided by the GBM team, and is publicly available on the HEASARC FTP server. This data format provides the highest energy resolution, as well as temporal resolution. We also use some of the utilities of \textit{GBM data tools} python package (\cite{GbmDataTools}) provided by the GBM team, such as binning the TTE data, plotting sky maps and converting angles to and from the spacecraft coordinate system.
        Our basic analysis unit is one TTE file, consisting of one hour of photon counts per detector. We bin the data to either 0.01 seconds or 0.001 seconds, depending on the GRB duration we search for. The TTE file also holds information about the detector's Good Time Intervals (GTI). Since we analyze all the detectors coherently, we use the intersection of all the detectors' GTIs as the GTI for the data segment. To avoid edge effects, we exclude the first and last 15 seconds around any GTI. We also exclude data segments in which at least one detector is off for more than several tens of milliseconds. Another bad data segment we filter out relates to extremely high count rates that cause pulse pile-up and, therefore, non-reliable detector behavior (similar to what happened with GRB221009A. See e.g. \cite{lesage2023fermi}). This effect occasionally occurs due to strong SFs and is not always included in the GTI reported. Lastly, in the time-slided data and simulated data, we exclude all catalog events (e.g., SFs and TGFs), whereas in the original data, we only exclude catalog SFs. 

    \subsection{Detection, Filtration \& Background Estimation}
        \subsubsection{Match Filtering}\label{sec: statistic}
            Our detection statistic is derived within the Neyman-Pearson framework (\cite{neyman1933ix}), which yields the optimal test in terms of true detection rate at a fixed false alarm rate for discriminating between a null hypothesis $\mathcal{H}_0$ and an alternative hypothesis $\mathcal{H}_1$. While the likelihood ratio test is strictly optimal only for simple hypotheses (i.e., with no free parameters), our alternative hypothesis is composite, involving unknown model parameters. To address this, we marginalize the likelihoods over the relevant parameter space. At the short timescales considered here, the number of photon counts per detector and per energy channel is typically small, so the underlying statistics follow a Poisson distribution. The hypotheses we evaluate are therefore defined in terms of Poisson likelihoods
            \begin{equation} \label{eq: detection hypotheses}
                \begin{aligned}
                   &\mathcal{H}_0: \; d_{t,n} \sim \text{Poiss}(b_{t,n}); \\
                   &\mathcal{H}_1: \; d_{t,n} \sim \text{Poiss}(b_{t,n} + AT_{t,n}(\theta)),
                \end{aligned}
            \end{equation}
            where $d_{t, n}$ is the number photon counts in time bin $t$ in detector - energy channel index $n$, $b$ is the background and $T(\theta)$ are the templates, multiplied by the signal amplitude $A$.. Instead of a full marginalization, we can approximate the statistic to be
            \begin{equation}\label{eq: statistic}
                \mathcal{S}_t(\theta) = \frac{\sum_{n} (d_{t,n} - b_{t,n}) \circledast_t \log\left(1 + \frac{AT_{t,n}(\theta)}{b_n}\right)}
                {\sqrt{\sum_{n, t} b_{t,n} \circledast_t \log^2\left(1 + \frac{AT_{t,n}(\theta)}{b_n}\right)}},
            \end{equation}
            which is normalized to have zero mean and unit variance. The $\circledast_t$ represents a convolution over the time axis, and the amplitude $A$ is set to the detection threshold. A full derivation of the detection statistic is described in Appendix \ref{appendix: test statistic}, along with a description of how it is used to estimate the signal's amplitude. 
            
            
        \subsubsection{Local Background Count Estimation} \label{sec: bkg est}
\begin{figure}
                \centering
                \includegraphics[width=1\linewidth]{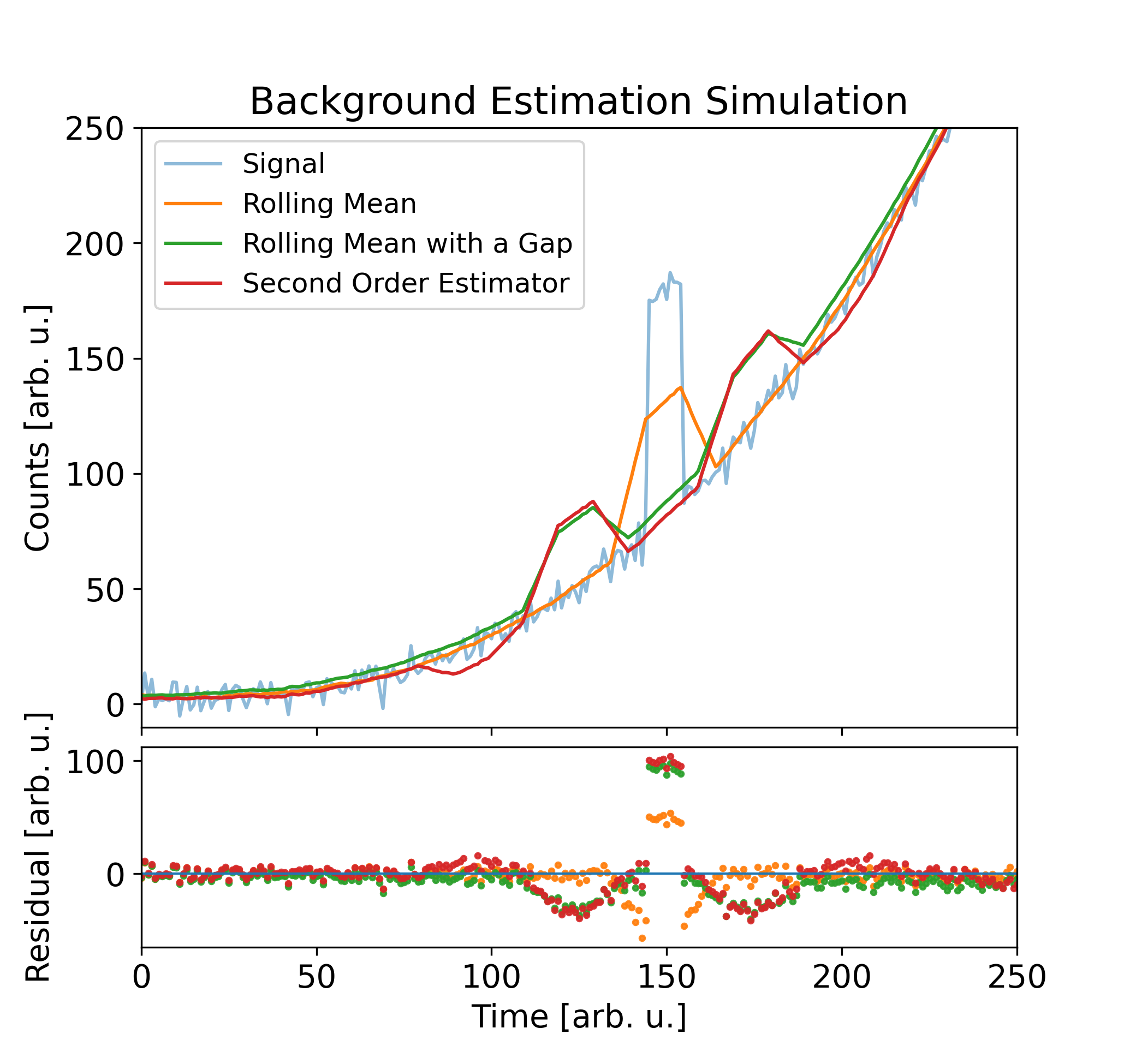}
                \caption{\textbf{Local Background Estimation Example.} The figure shows a simulated signal with noise and the local background estimation using a standard rolling mean, a rolling mean with a gap, and a second-order filter. It is clear that when a burst occurs, the gapped version captures the real background more accurately at the time of the burst. Furthermore, by applying the second-order filter (see main text and Appendix \ref{appendix: bkg}), we improve the estimation, especially where the background varies in a non-linear fashion.}
                \label{fig: local background}
            \end{figure}

        The test statistic \ref{eq: statistic} requires an estimation of the local background counts. To estimate it, we use a rolling mean with a gap $\Delta$ in the middle of the window (see Fig. \ref{fig: local background})
            \begin{equation}\label{eq: bkg estimator}
                \hat{b}_{t, n}(\Delta) = \frac{1}{N} \left[\sum_{\tau = t- (N+\Delta)/2}^{t-\Delta/2} d_{\tau, n} + \sum_{\tau =t+\Delta/2}^{ t + (N+\Delta)/2} d_{\tau, n}\right].
            \end{equation}
            This method allows for estimating the background from the data without explicit fitting. The gap in the middle makes the background estimation free of any influence by the burst itself and is chosen according to the burst duration. To improve the local estimation further, we use a linear combination of estimators with different gaps to construct a second-order filter that better accounts for non-linear temporal variations in the background. We can think of the gapped rolling mean discussed above as taking the mean of the estimators in the regions around the gap, which is similar to a linear fit. To get a quadratic fit, we use the following combination (see Appendix \ref{appendix: bkg})
            \begin{equation}\label{eq: bkg quad}
                \tilde {\hat{b}}_{t, n} = \frac{4\hat{b}_{t, n}(\Delta) - \hat{b}_{t, n}(3\Delta)}{3},
            \end{equation}
            where $\Delta$ is the gap, as defined in Eq. \ref{eq: bkg estimator}.
            
\begin{figure}
                \centering
                \includegraphics[width=1\linewidth]{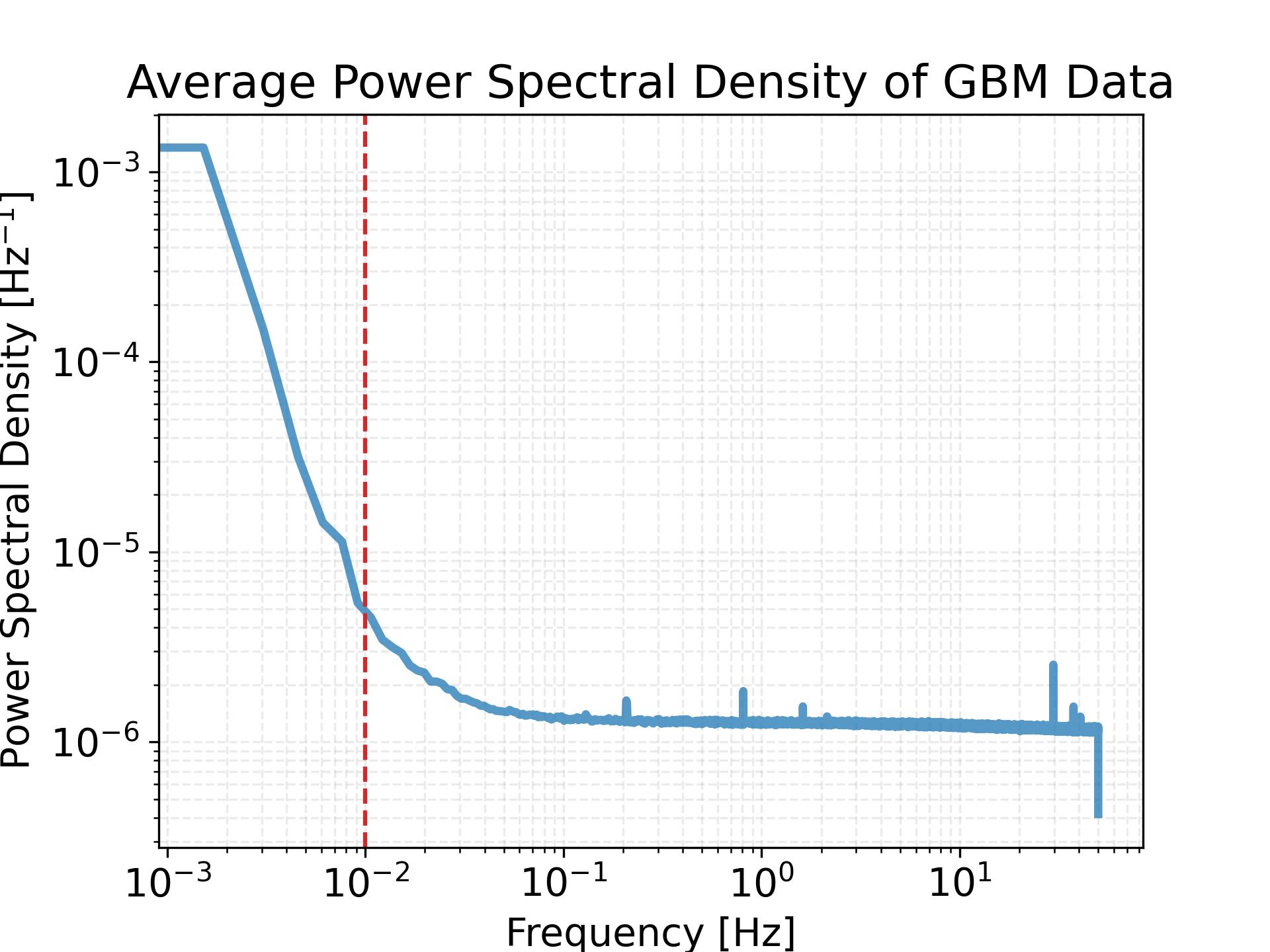}
                \caption{\textbf{Power spectral density (PSD) of the count rate summed over the energy channels}. It was computed using Welch’s method with a segment length of $\sim 650$ seconds and median averaging. The PSD is averaged over 1000 data segments with at least 60 minutes of continuous, trigger-free data. The spectrum shows a steep increase at low frequencies, indicative of red noise fluctuations in the background, which limits our ability to increase the averaging window. The red line marks the $10^{-2}$ Hz scale, showing that significant background fluctuations are already dominating the Poisson-noise in timescales longer than 100 sec. As a result, the validity of our background estimation is limited to shorter timescales, further shortening the validity of the matched filtering results to the order of a few seconds.}
                \label{fig: psd}
            \end{figure}

\begin{figure*}
                \centering
                \includegraphics[scale=0.4]{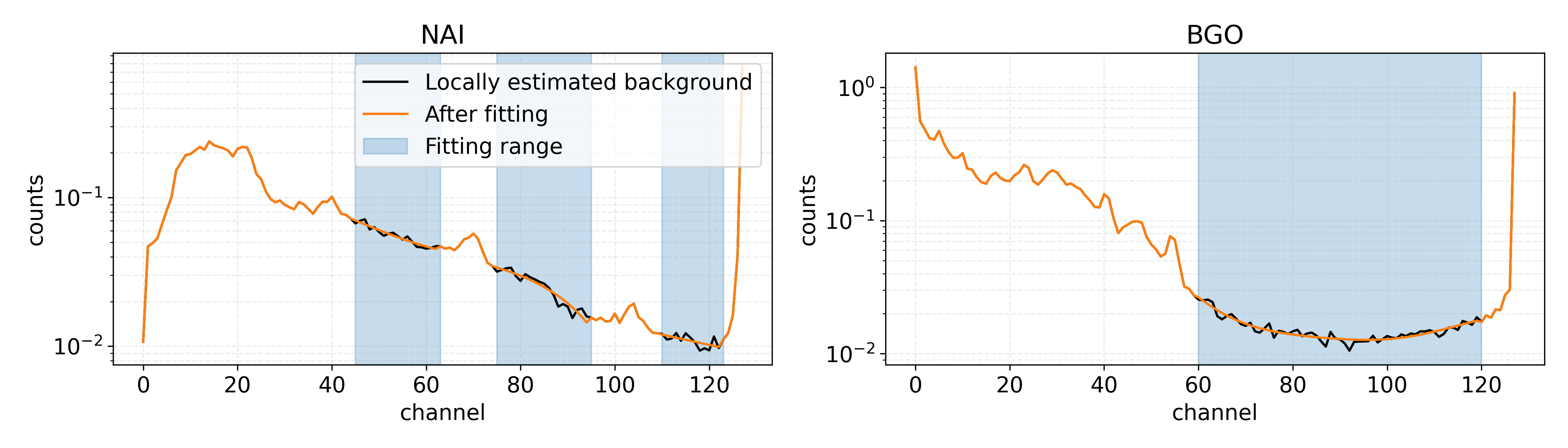}
                \caption{\textbf{Background Fitting}. After obtaining the background estimator (eq. \ref{eq: bkg quad}), we smooth out background fluctuations by fitting a polynomial to the selected energy channels. The fitting ranges where chosen in featureless segments of the spectrum.}
                \label{fig: bkg fit}
            \end{figure*}

            The background estimation accuracy is determined by the number of samples used around the gap in the rolling mean part and, naively, becomes more exact as we increase the number of samples used. However, the background changes on timescales of $\sim 100$ seconds, as can be seen in the noise's power spectral density (PSD) (Fig. \ref{fig: psd}).
            This limits the amount of time available for averaging in the background estimation. To increase the accuracy of the estimation without using more time-samples, we need more counts in the channels with the lowest background levels. 
            We fit a polynomial dependence in terms of energy to parts of the spectrum with the least observed count rate and a relatively featureless energy dependence (see Fig. \ref{fig: bkg fit}) -- this way, we effectively trade off freedom to represent an arbitrary energy dependence for improved precision of the background estimates.
            For the constant background term in eq. \ref{eq: statistic}, we take the average of the background estimator \ref{eq: bkg quad} in short windows (generally $2-4$ min), effectively splitting the analysis into independent segments $\hat{b}_n = 1/T\sum_t \hat{b}_{t, n}$. The estimation error in the log term of the statistic is less sensitive to estimation errors.

            \subsubsection{Background Spectral Drift Correction}\label{sec: drift correction}
\begin{figure*}
                    \centering
                    \includegraphics[scale=0.4]{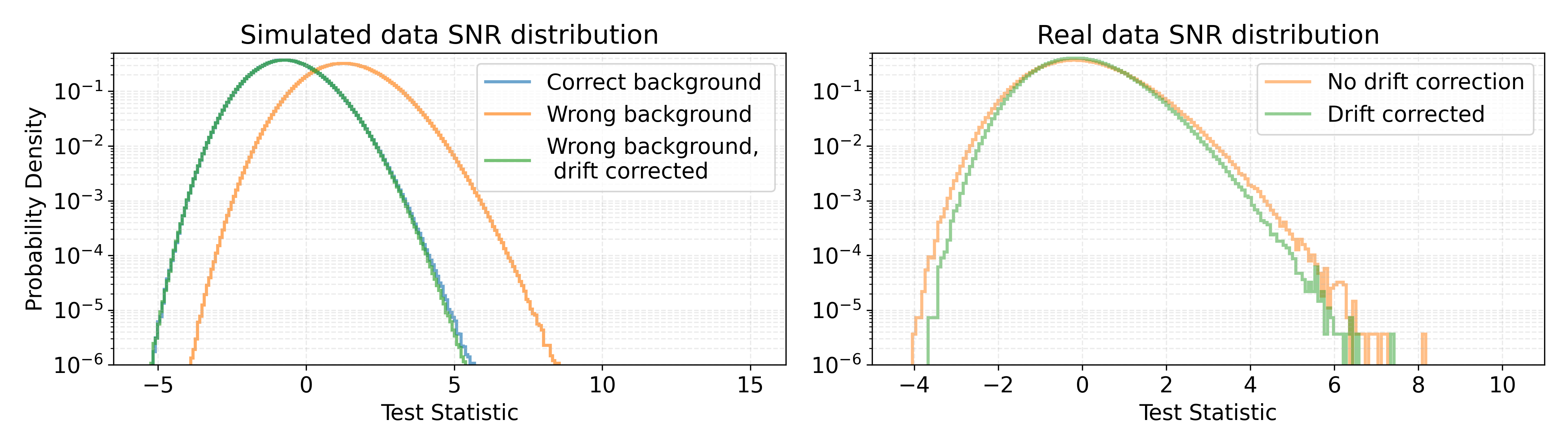}
                    \caption{\textbf{Wrong Background Effect on the Statistic Distribution}. Left: Simulation of the background misestimation effect. We sampled data from a background estimation of real data and applied the test statistic. The blue curve represents the result when the true background was used in the statistic expression (Eq. \ref{eq: statistic}), while the orange curve corresponds to a case where the background was misestimated with a 25\% negative error. The use of an incorrect background introduces a tail in the trigger distribution, leading to a loss in sensitivity. This effect is significantly reduced when the drift correction is applied. The green curve shows the result after applying the drift correction to the misestimated background statistic. Right: The statistic was calculated on actual data, with and without applying the correction. It can be seen that the behavior is similar to the simulation but with a much smaller drift.} 
                    \label{fig: drift correction effect}
                \end{figure*}

                As noted in the previous section, the background may vary on various timescales that can be comparable to our averaging window. Misestimation of the background leads to a tail in the trigger distribution and, therefore, lowers the search sensitivity. To account for that, we take a similar approach to \cite{zackay2021detecting} and apply the "drift correction" to our match-filter time series. By the application of this correction, the misestimation error from the unknown background reduces from linear to quadratic in the case where the noise is Gaussian. For example, in this case, a $10\%$ error in the background misestimation results in only $\sim 1\%$ reduction in SNR. In the Poisson regime, however, a cancellation of the linear error term is not achievable (see Appendix \ref{appendix: drift correction}). This is because the optimal test statistic, derived from the likelihood ratio, maximizes the power of the test\footnote{but rather to minimize the probability of erroneously rejecting the alternative hypothesis when it should not be rejected, at a given false alarm rate.}, which does not coincide with maximizing the SNR in the low-count Poisson regime.
                Nevertheless, this procedure still significantly reduces the error even in the Poisson regime, as we show by both simulating the effect using real data, and by a direct comparison of the statistic calculated with and without the correction, using arbitrary times (Fig. \ref{fig: drift correction effect}).
                
                The correction we apply is a re-normalization of the test statistic to have zero mean and unit variance, with the mean and variance estimated from the match-filter time series itself
                \begin{equation}
                    \label{Eq.Drift_correction}
                    \tilde{\mathcal{S}}_t(\theta) = \frac{\mathcal{S}_t(\theta) - \left\langle \mathcal{S}_t(\theta) \right\rangle_t}{\sqrt{\left\langle \mathcal{S}_t^2(\theta) \right\rangle_t - \left\langle \mathcal{S}_t(\theta) \right\rangle_t^2}}.
                \end{equation}
                This process is repeated for each of the template banks wherever it is used (for example, for the dense sky bank, when it is being used to calculate the posterior distribution for sky-position). 
                A technical point is that the process is computationally expensive in general, so we use SVD to reduce the computational complexity while retaining the precision of the estimate to $\sim 1\%$ error (see Appendix \ref{appendix: drift SVD}).
                
                This correction also introduces an error term, which is the estimation error in the denominator of Eq.~\ref{Eq.Drift_correction}. This error term becomes too large when we consider burst times above a few seconds, leading to a dramatic tail in the final background distribution for long-duration integrations. This fact limited the validity of the pipeline to 3.6 seconds. To detect longer bursts, one would have to build a more suitable background estimator that accounts for the long timescale variations. Also, one would have to consider the variation of the DRM over time, which is outside the scope of this paper.  

    \subsection{Vetoing triggers}\label{sec: vetoes}
        After obtaining the raw triggers from the detection step, non-GRB events still contaminate our sample. For this, we developed several specialized tests that reject these events. In this section, we will provide a description of these tests.
        
        \subsubsection{Single Detector Test}\label{sec: single detector veto}
            A common type of false trigger occurs when only a single detector registers a sufficiently strong signal. For example, due to a charged particle inducing a phosphorescent spike in the detector (\cite{goldstein2019updates}). Given the presence of background fluctuations at all times, it may not be obvious whether such events are genuine GRBs or single-detector events.
            Another common occurrence of single detector events is in the time-slided version of the data. When a real event is present, shifting the detectors leaves the signal in only one detector as long as the shift is longer than the burst. We wish to filter those out. This problem can be phrased as another hypothesis-testing problem. We test for a signal from one detector, denoted by $n_j$, against a signal coming from all detectors
            \begin{equation}
                \begin{aligned}
                   &\mathcal{H}_0: \; d_{t^*,n_j} \sim \text{Poiss}(b_{t^*,n_j} + AT_{t^*,n_j}(\theta^*)); \\
                   &\mathcal{H}_1: \; d_{t^*,n} \sim \text{Poiss}(b_{t^*,n} + AT_{t^*,n}(\theta^*)),
                \end{aligned}
                \end{equation}
            where the star refers to the best-fit parameters found when triggering, and the test is repeated for each of the detectors. The likelihood ratio provides the optimal test statistic for deciding if the trigger should be filtered. The hypotheses above, results in a simple test; it is simply the SNR of all the detectors except for the the single detector being tested
            \begin{equation}\label{eq: single detector statistic}
                \mathcal{S}^j_{t^*}(\theta^*) = \frac{\sum_{n\neq n_j} (d_{t^*,n} - b_{t^*,n}) \log\left(1 + \frac{AT_{t^*,n}(\theta)}{b_n}\right)}
                {\sqrt{\sum_{n\neq n_j, t} b_{t^*,n} \log^2\left(1 + \frac{AT_{t^*,n}(\theta)}{b_n}\right)}}.
            \end{equation}
            The threshold for this test was conservatively set at $4\sigma$. A histogram illustrating its performance is provided in Appendix \ref{app: vetoes}.
            
        \subsubsection{Timing Glitches}
\begin{figure}
                \centering
                \includegraphics[width=1\linewidth]{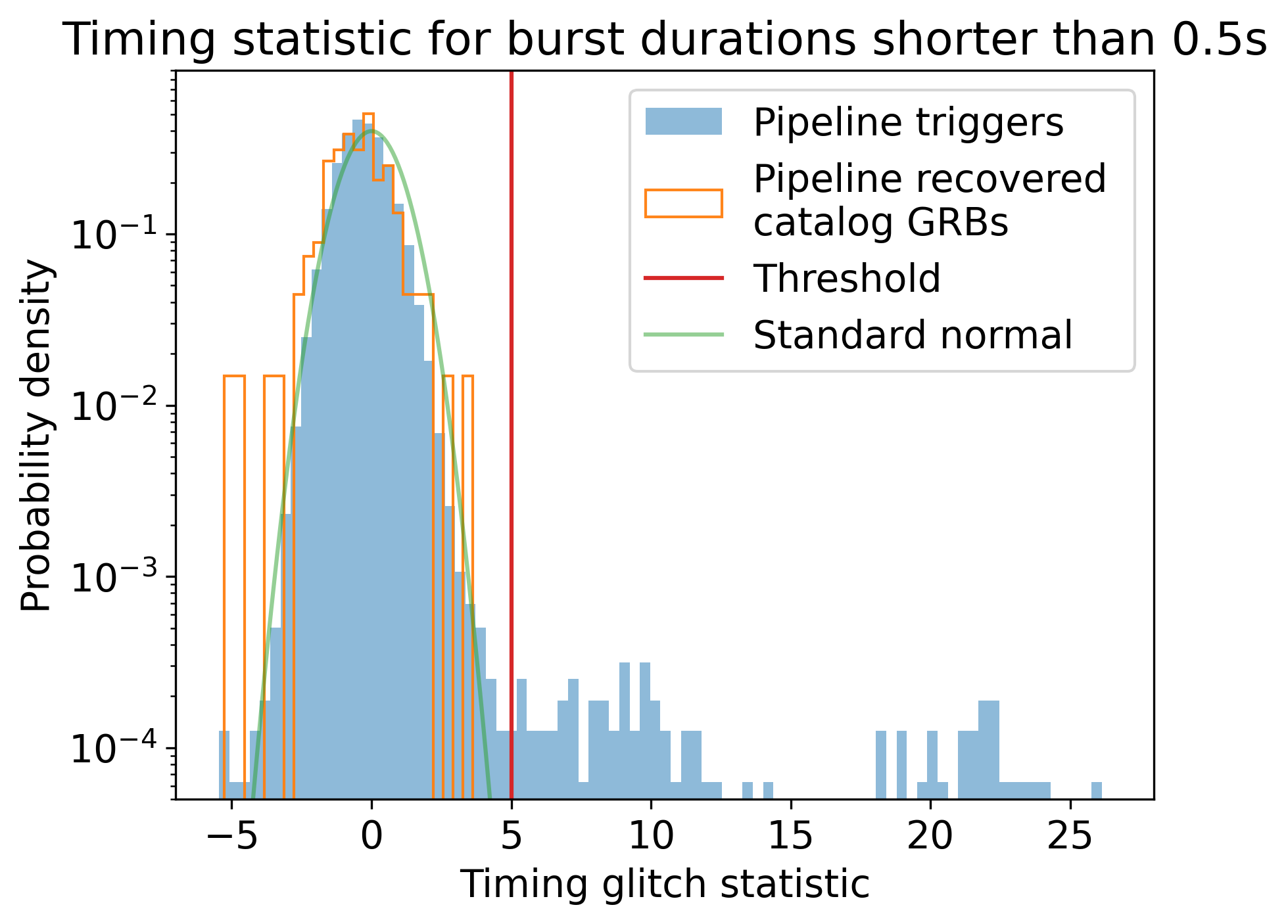}
                \caption{\textbf{Timing Glitch Test.} The figure shows a histogram of the timing glitch test statistic for 2014 triggers with a duration of less than 0.5 sec. Overlaid are the triggers that were classified by the GBM catalog as GRBs. The distribution seems to fit a normal distribution (with a small SNR-dependent bias), justifying the $5\sigma$ selection of the threshold.}
                \label{fig:timing glitches}
            \end{figure}

            According to \cite{von2014second}, GBM TTE data suffer from timing glitches on timescales of $\sim0.10$ sec. These manifest themselves when binning the data, where it can be seen as pairs of adjacent spikes and holes. A hole is a deficit of counts, while a spike is an excess of counts, which makes it appear similar to a GRB and thus can contaminate the trigger distribution. Given that the spike-hole pairs are indeed adjacent, they can be easily filtered on the pre-processing level. Nonetheless, when analyzing TTE data, we found that some spike-hole pairs may be separated by much longer timescales - as long as thousands of seconds. Since there is no typical timescale for the separation, we devised another test to discriminate these events from the rest of the triggers. The idea is that the excess counts will have the same energy spectrum as the background if there is no real event at that time. We can therefore phrase our hypotheses as a GRB signal (i.e. using a template from the template bank) versus a "background template"
            \begin{equation}
                \begin{aligned}
               &\mathcal{H}_0: \; d_{t^*,n} \sim \text{Poiss}(b_{t^*,n} + AT_{t^*,n}(\theta^*)); \\
               &\mathcal{H}_1: \; d_{t^*,n} \sim \text{Poiss}(b_{t^*,n} + b_{t^*,n}),
               \end{aligned}
            \end{equation}
            where, as before, we use the best-fit parameters at the time of the trigger. The resulting test statistic is
            \begin{equation}
                \mathcal{S} = \frac{\sum_n (d_n - AT_n - b_n) \log \left( \frac{2}{1+ AT_n/b_n} \right)}{\sqrt{\sum_n (AT_n + b_n) \log^2 \left( \frac{2}{1+ AT_n/b_n} \right)}}.
            \end{equation}
            Since the SNR may be very high for a real burst, we may erroneously reject it as the high glitch statistic value may be due to a mismatch between the real GRB signal and our template. To account for that, we downscale the statistic value by $5\%$ of the trigger SNR. The detection threshold is then set to be $5\sigma$. As can be seen from the empirical distribution of the test on all of the triggers from the 2014 data, the distribution closely follows a standard normal distribution, justifying the $5\sigma$ threshold (see Fig. \ref{fig:timing glitches}). Since this effect is expected to be present for timescales $\sim 0.10$ sec, we use this test to filter triggers with durations up to 0.5 sec.

        \subsubsection{Particle Precipitation Events}
            Charged particles interacting with the spacecraft materials or the Earth's atmosphere can lose energy through bremsstrahlung radiation. Such events constituted a significant source of false triggers in the BATSE data (\cite{horack1991batse}) and are similarly expected in GBM observations\footnote{\url{https://fermi.gsfc.nasa.gov/science/resources/swg/bwgreview/Classification_Report.pdf}}. Because these interactions occur near or within the spacecraft, the resulting signals are typically observed uniformly across all detectors, distinguishing them from true astrophysical transients. To classify such events, we developed a statistical test based on the following logic: for each trigger, we compute our detection test statistic (eq . \ref{eq: statistic}) individually for each detector. We then construct a new template by replicating the template of the detector that exhibited the highest SNR at the trigger time. Finally, we perform hypothesis testing to compare the original triggered template with this constructed template, allowing us to identify events consistent with a spacecraft or local origin. A histogram illustrating its performance is provided in Appendix \ref{app: vetoes}.

        \subsubsection{Occultation step}
            Gamma-ray detectors observe step-like occultation features when a gamma-ray point source passes behind the Earth's limb, resulting from variations in atmospheric absorption (\cite{harmon2002burst}). This phenomenon has been utilized in instruments such as BATSE and GBM for the monitoring, characterization, and imaging of gamma-ray sources (\cite{harmon2002burst, rodi2014earth}). However, within the context of our detection pipeline, these features are regarded as background artifacts, as they are not associated with gamma-ray transients. Due to the sharp increases or decreases they induce in count rates, such events may nonetheless trigger false detections. To address this, we developed a statistical test designed to identify and exclude these spurious signals. We exploit the asymmetry in the signal: a persistent increase is present on one side of the trigger, while the other side contains only the background. This contrasts with genuine gamma-ray transients, which are expected to be temporally localized, with background-dominated data on both sides of the trigger. For each trigger identified by the pipeline, we test whether the data preceding the trigger is better described by the background estimated after the trigger plus the best-fitting template, or by the background estimated from the same side. Specifically, we consider the following hypotheses:
            \begin{equation}
                \begin{aligned}
                   &\mathcal{H}_0: \; d_{t^*-\tau,n} \sim \text{Poiss}(b_{t^*-\tau,n}); \\
                   &\mathcal{H}_1: \; d_{t^*-\tau,n} \sim \text{Poiss}(b_{t^*+\tau,n} + A^*T_{n}(\theta)),
                \end{aligned}
            \end{equation}
            where $\tau$ is the offset from the trigger time $t^{*}$. The test statistic is evaluated across several time offsets, large enough so that no data from the triggered interval is used and short enough so that all background estimates are drawn from the original estimation window. To avoid the impact of outliers, we take the median value of the test statistic across all offsets. Again, the threshold for this test was set at $4\sigma$. A histogram illustrating the test's performance is provided in Appendix \ref{app: vetoes}. 

        \subsubsection{Trigger Direction}\label{sec:locstat}
            The broad spectral range of GBM allows it to detect more kinds of gamma-ray transients arising from different sources. Two sources of such gamma-ray emissions are the Sun and the Earth. Solar flares (SFs) are electromagnetic emissions from the Sun that can be as short as milliseconds and as long as tens of minutes (\cite{fletcher2011observational}) and cover the whole electromagnetic spectrum, from radio to gamma-rays. Terrestrial gamma ray flashes (TGFs) are coming from lightning flashes on Earth (\cite{dwyer2012high, briggs2013terrestrial}) and are usually very short (sub-millisecond) and display a power-law spectrum with energies of tens of MeV (\cite{lindanger2021spectral}). Even though the spectrum of these two phenomena is generally different from a typical GRB spectrum, the best-fitting template to a real GRB may have spectral parameters that are similar to SFs or TGFs due to different factors such as inaccurate response and inaccurate spectral model. To quantify whether a trigger is of terrestrial, solar, or cosmological origin, we phrase this question as a Bayesian model selection problem. This amounts to calculating the Bayesian evidence ratio, weighted by the prior probabilities of each model (or the event rate ratio). In our case, the models are similar to previously stated tests in this paper, with the modification that now the allowed parameters for a given model are constrained by the direction the event is coming from. For example, the Earth model only integrates templates with responses that are consistent with the Earth's direction. Phrased another way, we can divide the sky into disjoint regions and integrate the posterior probability of the sky position only in those regions we want to test and take their ratio. A complete derivation of the test is shown in Appendix \ref{appendix: bayesian model selection}. The procedure described above amounts to calculate the Bayes factor,  which we use as our quantitative measure for trigger direction
            \begin{equation}
                \mathcal{B}_{\oplus} = \frac {\int d\Omega \; p(\Omega) \; I(\Omega \in \Omega_{\oplus}) p(\Omega | d, \mathcal{H}_1)}{\int d\Omega \; p(\Omega) \; I(\Omega \notin \Omega_{\odot, \oplus}) p(\Omega | d, \mathcal{H}_1) } ,
            \end{equation}
            where all the quantities are evaluated at the time of the trigger, $\xi$ are the spectral parameters, $\Omega$ is the sky position, $I(\Omega \in \Omega_{\oplus})$ is an indicator for positions that are occulted by the Earth (and similarly for the Sun, $\Omega_{\odot}$) and $p(\Omega | d, \mathcal{H}_1)$ is the posterior probability of the sky position given the data and the signal hypothesis. Similarly, $\mathcal{B}_{\odot}$ is the Bayes factor of a Sun-originating trigger. To perform the integration required for this test, we created a template grid with $4.5\times 10^5$ templates in total. There are $2500$ different sky positions, uniformly placed on the sky (using the same Fibonacci algorithm described in sec. \ref{sec: template banks}), giving a resolution of $\sim 16.5$ square degrees per template. Each of these templates has the same 180 spectral parameters which were sampled in the range described in sec. \ref{sec: template banks}. After calculating the posterior using all the templates in the grid, we interpolate it to create a HealPix (\cite{gorski2005healpix}) map with a resolution of 0.2 square degrees.
            
            It is worth mentioning that this approach gives us localization maps of our new candidate triggers from which we can extract confidence intervals and maximum likelihood estimation of the location of the trigger. This approach is similar to the RoboBA algorithm (\cite{connaughton2015localization}), but incorporates several differences: it utilizes the all the available energy channels, employs a broader set of spectral parameters in the marginalization, and adopts a different background estimation technique. Extensive testing of these localization maps are required in order to asses their reliability for follow-up observations. These are out of the scope of this paper.
    
    \subsection{Background Trigger Estimation}
        After applying all the above steps to the data and collecting the triggers, we are still at risk of misinterpreting the values of the test statistic as significant due to systematic errors and uncertainties. For example, one source of uncertainty is the detector response, which contains systematic errors. Indeed, due to our choice of a fixed time to generate responses for the template bank, this problem is further exaggerated as the relative position of detectors with respect to the Earth may change from the time of response generation to the trigger at hand. Moreover, maximizing over many templates and temporal bins also introduces a broadening of the underlying distribution, effectively lowering the trigger significance. Therefore, it would be desired to apply the exact same pipeline to a "clean" data realization that would serve as a benchmark of our detection pipeline's performance. By clean, we mean devoid of signals that are usually present in the data (GRBs, TGFs, SFs, etc.). To achieve this, we use two methods. The first is to generate synthetic data from our background estimation and then run the pipeline on this data realization (see Fig. \ref{fig: pipeflow})
        \begin{equation}
            \tilde{d}_{t, n} \sim \text{Poisson}(\hat{b}_n).
        \end{equation}
        We denote this data realization "simulation".
        The second approach is to apply "timeslides", which means to temporally de-correlate the data by applying nonphysical shifts between different detectors. Under the assumption that a real signal must be present in more than one detector, timeslides ensure that real signals will be filtered out (given the time shifts are larger than the GRB duration and smaller than the background variation timescale). As in the data simulation, we apply our pipeline on a time-slided data realization and repeat the same filtration steps described above. Timeslides provide a means to empirically measure the extreme tail of the background distribution, thereby allowing us to measure the false alarm rate and $p_{\text{astro}}$ of the triggered events.

\section{Results} \label{sec:res}
    In this section, we present the results of applying our pipeline to one year of Fermi/GBM data. Specifically, we focus on the year 2014 to demonstrate the pipeline's performance and capabilities. A comprehensive catalog of candidate events, based on the full GBM dataset from 2013—when continuous TTE data first became available—through the present will be presented in a forthcoming paper.

    The burst durations searched, the background estimation windows, and the fixed background interval used in the search are summarized in Table \ref{table: burst durations}. Burst durations were logarithmically spaced, following the methodology described in section \ref{sec: template banks} and Appendix \ref{appendix: match}. 
    \begin{deluxetable}{ccc}
        \tablecaption{Burst durations and background windows used in the search.\label{table: burst durations}}
        \tablehead{
        \colhead{\parbox[c]{2.5cm}{\centering Burst\\Duration (s)}} & 
        \colhead{\parbox[c]{2.5cm}{\centering Background\\Estimation\\Window (s)}} & 
        \colhead{\parbox[c]{3cm}{\centering Constant Background\\Interval (s)}}
        }
        \startdata
        0.003 & 20  & 30  \\
        0.004 & 20  & 30  \\
        0.005 & 20  & 30  \\
        0.007 & 20  & 30  \\
        0.009 & 20  & 30  \\
        0.012 & 20  & 30  \\
        0.016 & 20  & 30  \\
        0.022 & 20  & 30  \\
        0.030 & 80  & 120 \\
        0.040 & 80  & 120 \\
        0.054 & 80  & 120 \\
        0.073 & 80  & 120 \\
        0.098 & 80  & 120 \\
        0.133 & 120 & 180 \\
        0.179 & 120 & 180 \\
        0.242 & 120 & 180 \\
        0.327 & 120 & 180 \\
        0.441 & 120 & 180 \\
        0.596 & 120 & 180 \\
        0.804 & 120 & 180 \\
        1.086 & 120 & 180 \\
        1.466 & 200 & 270 \\
        1.979 & 200 & 270 \\
        2.671 & 200 & 270 \\
        3.606 & 200 & 300 \\
        4.869 & 200 & 300 \\
        6.573 & 200 & 300 \\
        \enddata
        \tablecomments{The background estimation window is the total duration derived from both sides of the gap.}
        \end{deluxetable}

\begin{figure}
        \centering
        \includegraphics[width=1\linewidth]{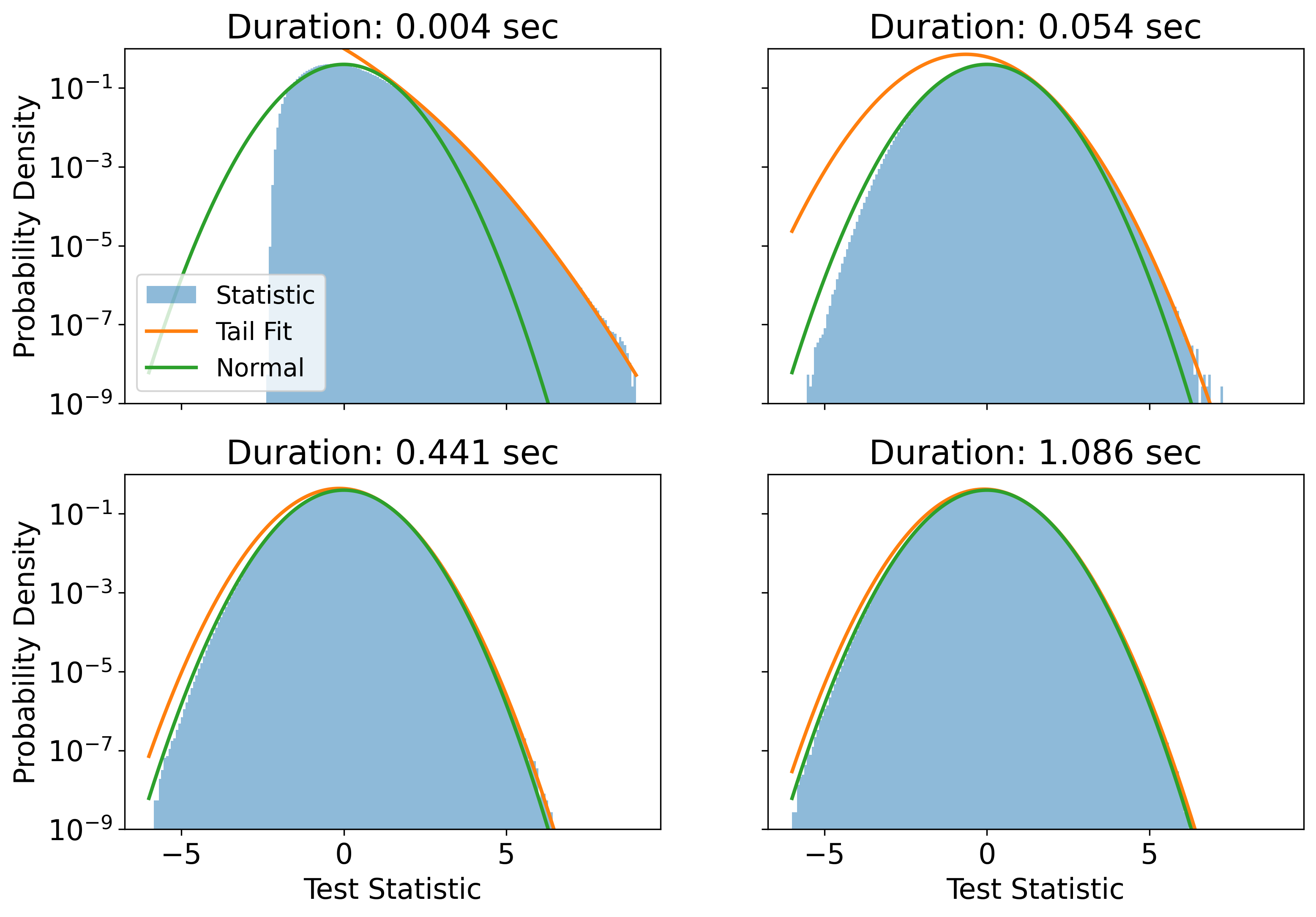}
        \caption{\textbf{Statistic distribution of different burst durations.} The figure shows some of the different distributions obtained when searching for different burst durations using a specific template. The distributions of the test statistic were obtained by Monte Carlo simulations. For the shorter durations, it is apparent that the distribution is deep in the Poisson regime and does not resemble a Normal distribution. As we increase the duration, we accumulate more photons and converge to a Normal distribution. To get a similar interpretation for the different distributions, we fit the tail and calibrate the $p$-value to be the one that would have been obtained from the Normal distribution (see main text and eq. \ref{eq: calibrate dist})}
        \label{fig: calibrate dist}
    \end{figure}

    When running the pipeline, we required triggers to be separated by more than 30 seconds. Thus, when collecting triggers for each duration, if multiple triggers occurred within a 30-second window, only the one with the highest SNR was selected. After applying all the vetoes outlined in Sec. \ref{sec: vetoes}, the burst duration with the maximal SNR was retained. We discard the two longest durations (i.e., larger than 3.606 sec), since our background estimation and drift correction assumptions start to break down, resulting in a long tail in the trigger distribution and therefore resulting in a significant loss in sensitivity (see sec. \ref{sec: drift correction}). Lastly, since we run the pipeline over multiple durations, the statistics for each are different. For example, in the 4 millisecond burst duration search, we sum up only a few photons, resulting in the distribution being in the deep Poisson regime. On the other hand, in the 1.979 seconds burst duration search, we sum up many photons, and we are at the Gaussian statistics regime. This may result in an erroneous systematic selection of shorter burst durations only because the distribution is broader in the Poisson regime than in the Gaussian regime, as can be seen in Fig. \ref{fig: calibrate dist}. Furthermore, since our statistic (Eq. \ref{eq: statistic}) consists of a linear combination of Poisson random variables $d_{t, n}$, each with the different weight $\log(1+T_n(\theta)/b_n)$, the different spectral components of the background, together with the different spectral shapes of the templates, result in a complex statistic structure. Some of the parts of the spectrum are in the Gaussian regime, some deep in the Poisson regime, and some in between. That is to say $T_n/b_n$ is smaller, larger and comparable to $1$, respectively. This means that the issue discussed above should be considered separately for each template.
    To account for this fact, when post-processing the triggers, we calibrate the SNR to give the same p-value as it would have, had it been taken from the normal distribution. This procedure is similar to the Gaussianization transformation described in \cite{ivashtenko2025independent}. Denoting by $f_{\mathcal{N}}$ the probability density function (PDF) of the standard normal distribution and by $f_{\mathcal{S}(\tau, \theta)}$ the PDF of the detection statistic for a burst of duration $\tau$ and template parameters $\theta$. For each trigger, we solve the following equation
    \begin{equation}\label{eq: calibrate dist}
        \int_{\sigma}^{\infty} f_{\mathcal{N}}(x) \dd x = \int_{\text{SNR}}^{\infty} f_{\mathcal{S}(\tau, \theta)}(x) \dd x,
    \end{equation}
    where $\sigma$ is the calibrated SNR. To solve the equation, we used a Monte-Carlo simulation of the distributions for each burst duration and template and fitted an exponential tail. This tail is then used to calculate the integral above.
    
\begin{figure}
        \centering
        \includegraphics[width=1\linewidth]{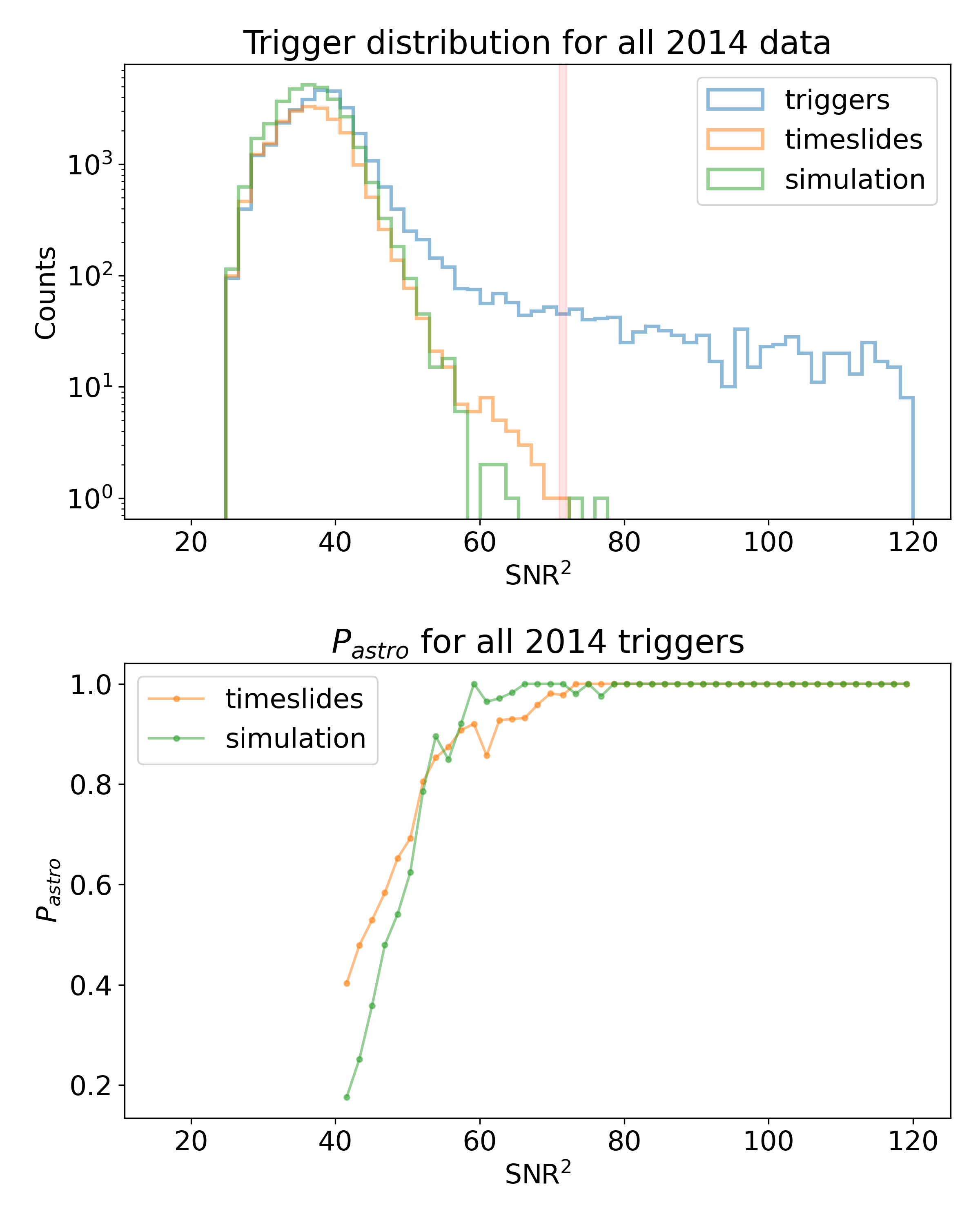}
        \caption{\textbf{Raw trigger distribution for 2014 data.} Upper panel: Histogram of the triggers obtained from running the pipeline over 2014 data, after maximization over the burst duration. The red band indicates the detection threshold for $p_{\text{astro}}=1$. The actual data output is denoted as "triggers", and we also show the timeslides and simulation outputs. More high SNR events were cut from the figures for clarity. Lower panel: $p_{\text{astro}}$ calculated using eq. \ref{eq: pastro} by using either the timeslides or the simulation as the noise distribution. It can be seen that above SNR${}^2\sim 55$ we have $p_{\text{astro}}=0.5$ and above SNR${}^2\sim 70$, marked by the red band, the simulation and timeslides do not trigger on random fluctuations, and we essentially have $p_{\text{astro}}=1$.}
        \label{fig:trig hist}
    \end{figure}

\begin{figure}
        \centering
        \includegraphics[width=1.0\linewidth]{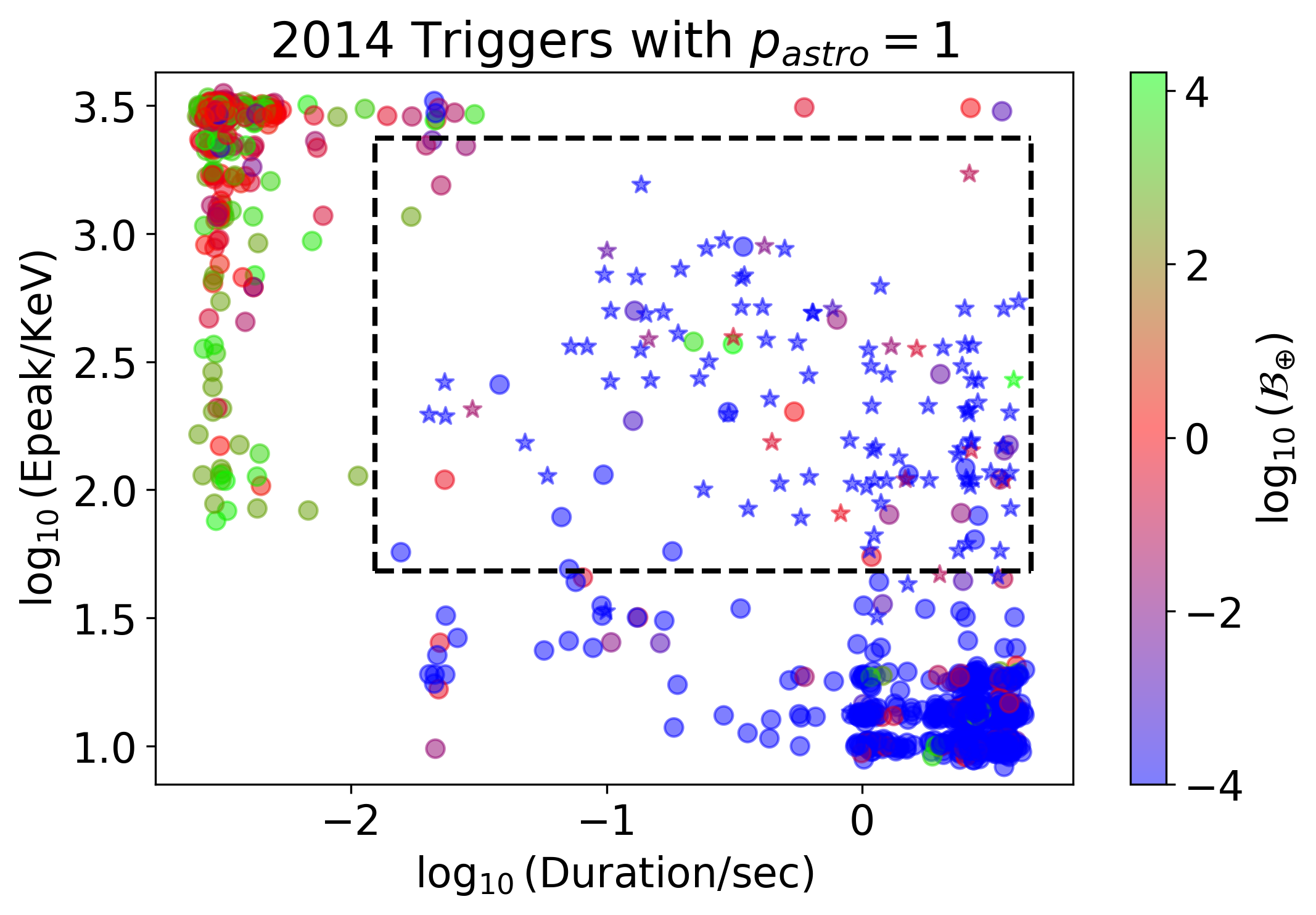}
        \caption{\textbf{Peak energy-duration distribution for 2014 triggers with $p_{\text{astro}}=1$.} This scatter plot displays the peak energy versus duration for triggers with $p_{\text{astro}}=1$ corresponding to SNR${}^2\geq 70$. A small jitter is added to mitigate the overlap caused by the discrete template bank. Color indicates the Earth Bayes factor $\mathcal{B}_{\oplus}$, and star symbols denote triggers classified as GRBs in the GBM catalog. The dashed black rectangle outlines the region used to estimate the number of new GRBs. The duration and high peak energy thresholds help eliminate short, Earth-associated triggers, while a low-energy cut ($E_{\text{peak}} > 40$ KeV) suppresses the excess near the Galactic center, as shown in Fig. \ref{fig:trigger skydist}. It is important to note that genuine GRBs may still occur below this energy threshold, particularly long GRBs with short-duration variability spikes, which can fall into this lower-energy regime.}
        \label{fig: E vs t}
    \end{figure}

\begin{figure}
         \centering
        \includegraphics[width=1\linewidth]{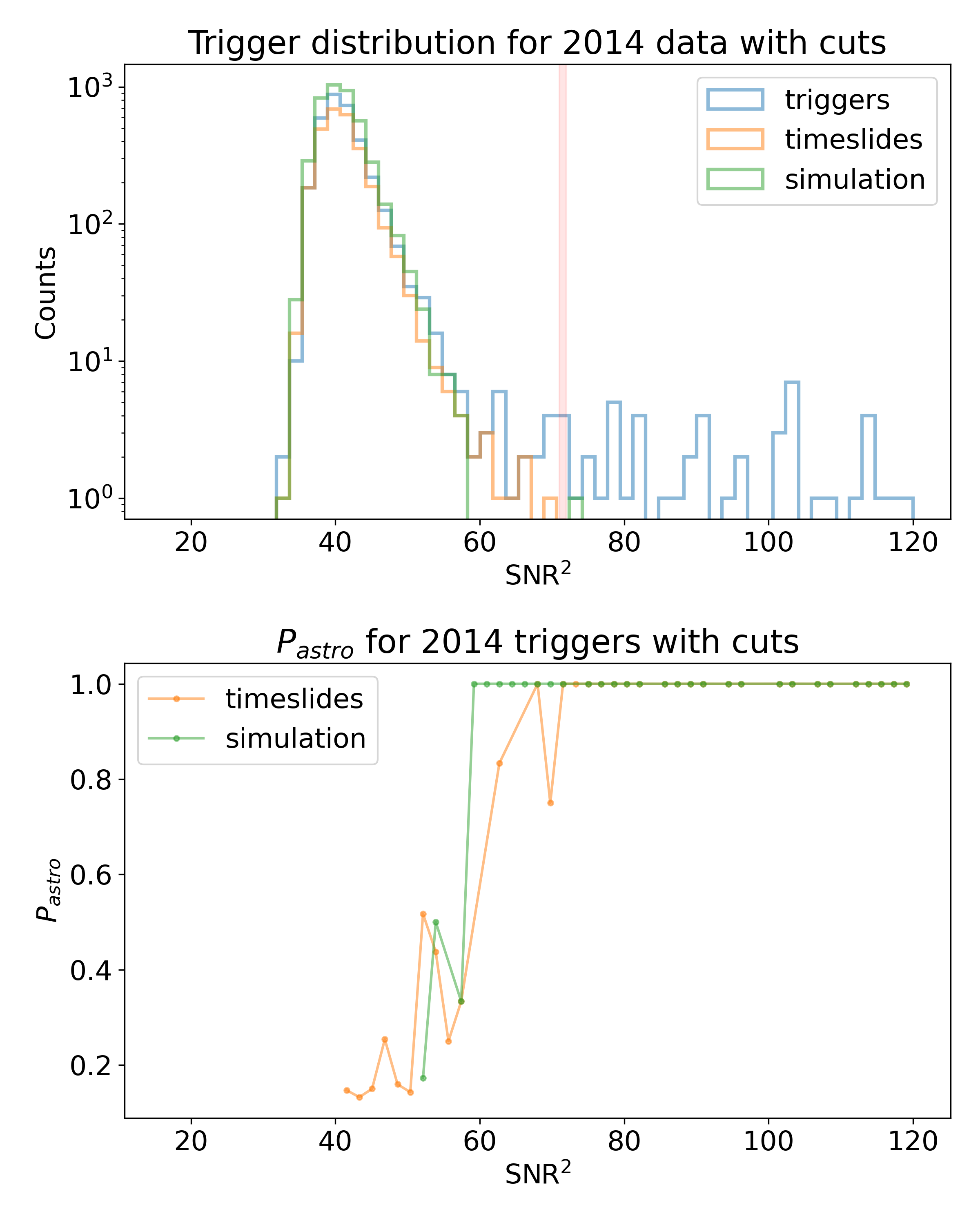}
         \caption{\textbf{Trigger distribution and $p_{\text{astro}}$ after applying energy, time and localization cut.} We took triggers with $40 \leq E_{\text{peak}} \leq 2000$ KeV, $ 0.016 \leq t \leq 3.606$ sec, and $\mathcal{B}_{\oplus}, \mathcal{B}_{\odot}<1$.
         }
         \label{fig:trig hist cut}
     \end{figure}

\begin{figure*}
        \centering
        \includegraphics[width=0.45\linewidth]{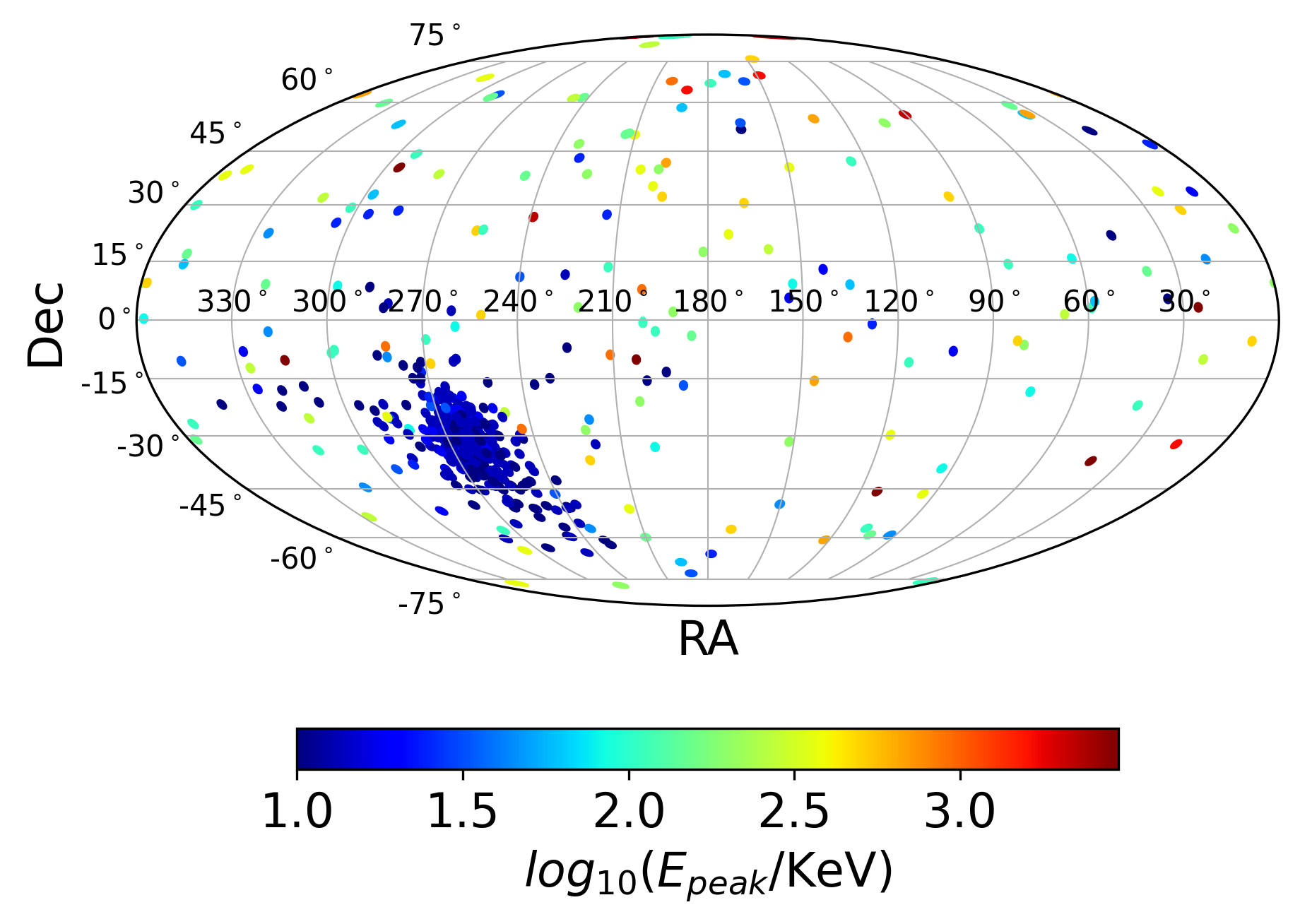}
        \includegraphics[width=0.45\linewidth]{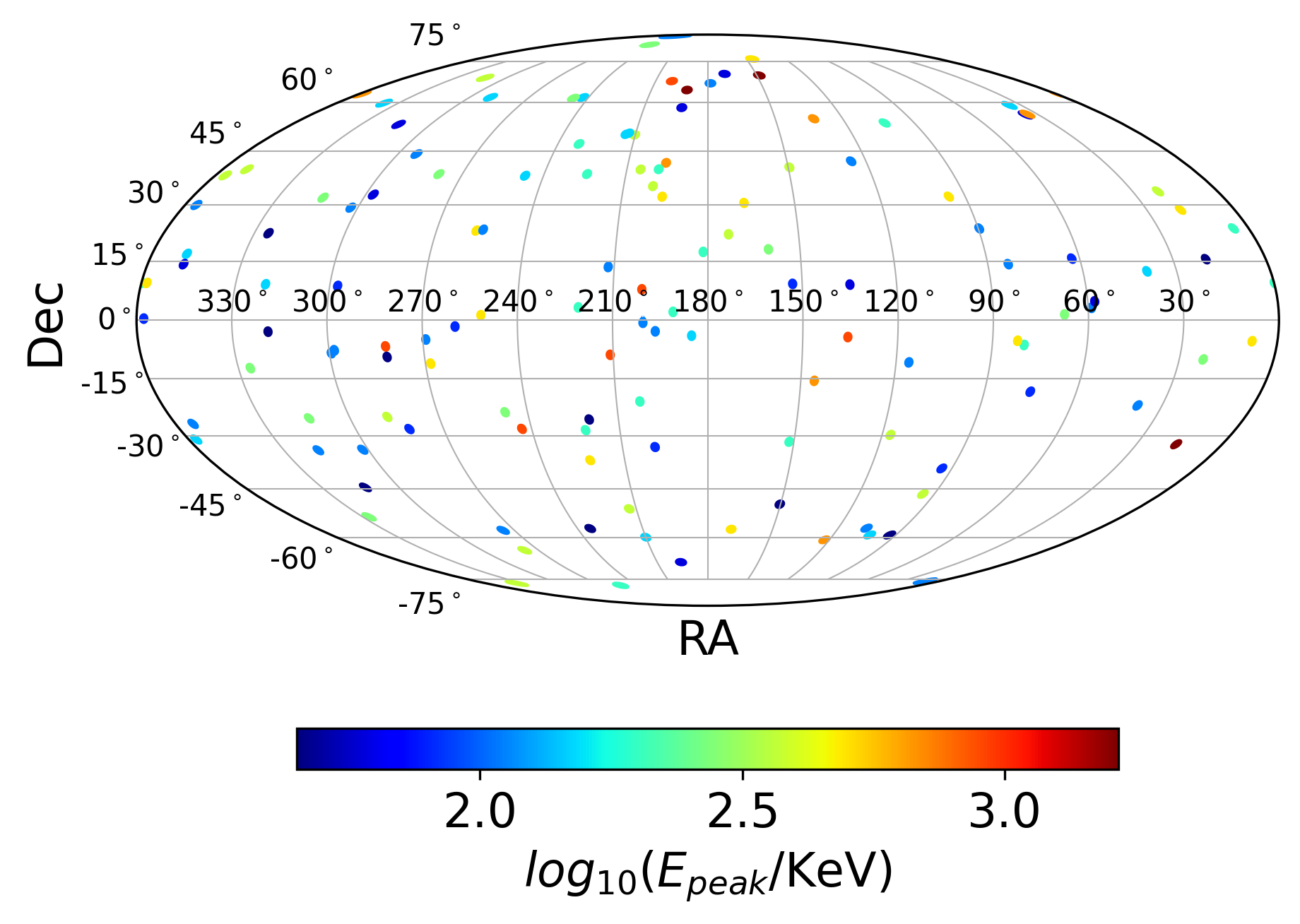}
        \caption{\textbf{Maximum likelihood sky distribution of triggers with $p_{\text{astro}} = 1$.} Left: All triggers with $p_{\text{astro}} = 1$ and $\mathcal{B_{\oplus}},\mathcal{B_{\odot}}<1$. One can see the over-density of triggers close to the galactic center. These triggers are relatively soft, with $E_{\text{peak}}<40$ KeV. Right: Similar to the left panel but now applying a peak energy cuts, keeping only triggers with $40\;\text{KeV}<E_{\text{peak}}<2000\;\text{KeV}$. }
        \label{fig:trigger skydist}
    \end{figure*}

    The resulting multi-scale trigger distribution is shown in figure \ref{fig:trig hist}.
    This distribution includes all detected triggers, encompassing Soft Gamma Repeaters (SGRs), SFs, and TGFs. By applying selection criteria based on trigger properties—such as location statistics (see sec. \ref{sec:locstat}) or peak energy—it is possible to exclude non-GRB events. However, this filtering approach is relatively rudimentary and may remove genuine GRBs. A more robust classification framework is therefore necessary. The development of such a scheme, along with results from the complete GBM dataset, will be presented in a separate paper. Additional discussion of the classification method and its implications can be found in sec. \ref{sec:conc}. Fig. \ref{fig: E vs t} presents a scatter plot of peak energy versus duration, with points color-coded by $\mathcal{B}_{\oplus}$, illustrating the effects of the applied cuts. Distinct clustering is observed in the short-duration, high-energy region (upper left), as well as in the longer-duration, lower-energy region (lower right). However, the plot clearly demonstrates that simple cuts are insufficient for definitive classification. Within the dashed region, where sGRBs are expected, events with high Earth association are still present, and conversely, cataloged GRBs are found outside this region. Furthermore, some events within the sGRB region may instead originate from SGRs, highlighting the limitations of this approach.
    As shown in Fig. \ref{fig:trigger skydist}, applying the selection criteria from Fig. \ref{fig: E vs t} - i.e., excluding the triggers outside the dashed black box, effectively removes the excess of triggers associated with the Galactic center and reduces TGF contamination. The resulting sky distribution is consistent with the expected isotropic distribution of sGRBs. The resulting trigger distribution and $p_{\text{astro}}$ are shown in Fig. \ref{fig:trig hist cut}, and the maximum likelihood sky positions are shown in Fig. \ref{fig:trigger skydist}.

    Compared with the GBM catalog, we recovered 54 out of 57 GRBs with durations up to our maximal timescale of 3.606 seconds, where we associated triggers with temporal separation of 3 seconds symmetrically around the GBM trigger time. The three undetected events are due to the pre-processing filtration of these data segments (two events are at the edge of a file, and for the third event, half of the detectors were inactive). Table \ref{table:catalogs} summarizes the output of our pipeline within the parameter range $E_{\text{peak}}\in [40, 2000]$ KeV and $ 0.016 \leq t \leq 3.606$.    
    Note that one of the 54 catalog GRBs that was recovered has its peak energy estimated to be $33.23$ KeV (bn141230871). Although this value falls below our selection threshold, the event is included in Table \ref{table:catalogs} for completeness. Interestingly, the maximum likelihood estimate of the trigger’s direction lies near the Galactic plane, with the posterior distribution indicating a significant contribution from that region. This again highlights the need for a more robust classification framework to reduce non-GRB contamination in future catalogs and follow-up observations.
    
    The aforementioned parameter range can be used to get a rough estimate of the number of new GRBs our pipeline detected in 2014: we find 27 triggers with $p_{\text{astro}} = 1 $ not present in the GBM catalog.


\section{Conclusions and Discussion} \label{sec:conc}
\begin{figure}
        \centering
        \includegraphics[width=1\linewidth]{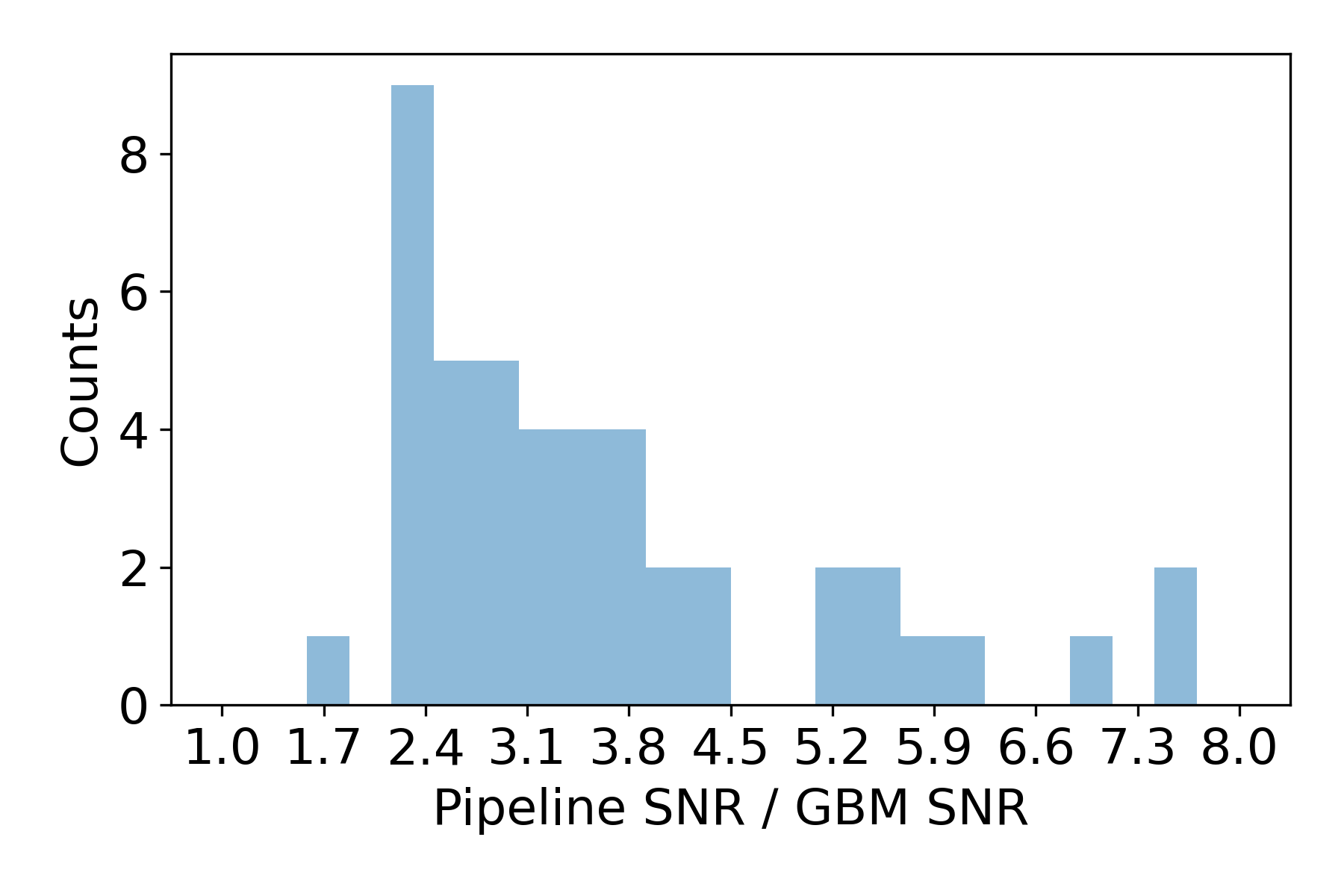}
        \caption{\textbf{SNR ratio between our pipeline and GBM detections for the 54 out of 57 recovered bursts up to 3.606 seconds.} The GBM SNR is the reported significance from the Trigdat file, given in units of $\sigma$ standard deviations, and hence has the same meaning as our SNR.}
        \label{fig:catalogs snr comp}
    \end{figure}

    In this work, we introduced a new pipeline for the detection of sGRBs in Fermi/GBM data and demonstrated its performance using data from 2014. The pipeline successfully recovered the majority of GRBs listed in the GBM catalog and identified 27 additional sGRB candidates with $p_{\text{astro}}=1$. Our pipeline also detected many more transient events, such as TGFs, SFS and SGRs. In addition, the pipeline shows a significant increase in detection SNR compared to the on GBM board triggering algorithm (Fig. \ref{fig:catalogs snr comp}).
    
    The preliminary classification scheme employed in this study—based on selection cuts in peak energy, burst duration, and association probabilities with the Earth and Sun—may lead to the exclusion of some genuine sGRBs and carries a significant risk of misclassification, particularly between GRBs and SGRs. Moreover, the distribution of both signal and background triggers exhibits a strong dependence on burst duration and peak energy, implying that the ratio of true astrophysical event rate to background trigger rate (and thus the inferred value of $p_{\text{astro}}$) is itself a function of these variables. This will be addressed in a forthcoming paper, where we will analyze triggers from the full GBM dataset and present the catalog of events resulting from our pipeline. The expanded dataset will enhance our ability to model the underlying distributions and improve the assignment of astrophysical probabilities across different classes of transients (GRBs, SGRs, solar flares, and TGFs).

    Our enhanced detection capabilities, which yield a $50\%$ increase in the number of sGRB, can play an important role in the search for coincident detections of sGRBs and GWs from LIGO observing runs. Conducting targeted GW searches around the newly identified sGRBs, while incorporating their sky localization posteriors as priors, has the potential to increase the joint detection volume and improve the sensitivity of multi-messenger searches. This will be further enhanced by incorporating the larger sample of sGRB candidates with $0.5<p_{\text{astro}}<1$, which will be available upon the application of our pipeline over the entire GBM dataset.
    
\section*{Acknowledgments}
This research was supported by grant no 2022136 from the United States - Israel Binational Science Foundation
(BSF), Jerusalem, Israel.
BZ is supported by a research grant from the Willner Family Leadership Institute for the Weizmann Institute of Science. 
TV additionally acknowledges support from NSF grants 2012086 and 2309360, the Alfred P. Sloan Foundation through grant number FG-2023-20470, and the Hellman Family Faculty Fellowship during the time this work was done.
\newpage
\appendix

\section{Derivation of the Detection Test Statistic} \label{appendix: test statistic}
    To derive the detection test statistic, we use the likelihood ratio test. For the null hypothesis, we assume that there are only background counts of photons, whereas in the alternative, we also use the fact that there is a signal present. Let us denote the time index by $t$ and the combined detector channel by $n$. We can write the above as a hypothesis-testing problem
    \begin{equation}
        \begin{aligned}
           &\mathcal{H}_0: \; d_{t,n} \sim \text{Poiss}(b_{t,n}); \\
           &\mathcal{H}_1: \; d_{t,n} \sim \text{Poiss}(b_{t,n} + AT_{t,n}(\theta)),
        \end{aligned}
    \end{equation}
    where $d$ is the data, $b$ is the background, $A$ is the amplitude of the signal, and $T(\theta)$ is the signal template that depends on the parameters $\theta$. As shown in \cite{ofek2018optimal}, fixing the amplitude to the detection limit results in an SNR loss of less than $1\%$. In our analysis, we calculate the detection limit for each template by Monte Carlo simulation similar to the process described in \cite{ofek2018optimal}. This allows us to treat the amplitude as known, effectively reducing the number of free parameters. The likelihood ratio, when integrated over the model parameters, is given by
    \begin{align}
        \mathcal{S} = \frac{\int\dd\theta \;p(\theta)p(d|\theta;\mathcal{H}_1)}{p(d|\mathcal{H}_0)},
    \end{align}
    with $p(\theta)$ the prior over the parameters $\theta$ which take to be uniform over the range. Plugging in the Poisson likelihoods, we get
    \begin{align}
        \mathcal{S} &= \int\dd\theta \; p(\theta)
        \left[
        \prod_{t,n} \left(1 + \frac{AT_{t,n}(\theta)}{b_{t,n}}\right)^{d_{t,n}} e^{-AT_{t,n}(\theta)}
        \right] \nonumber \\
        &= \int\dd\theta \; p(\theta)
        e^{\sum_{t,n} \left[d_{t,n}\log\left(1 + \frac{AT_{t,n}(\theta)}{b_{t,n}}\right) - AT_{t,n}(\theta)\right]}.
    \end{align}
    The above integral will be dominated by the template $T_{t,n}(\theta^*)$, where $\theta^*$ are the parameters that describe the signal best. This allows us to approximate the integral as
    \begin{equation}
        \mathcal{S}(\theta) \propto \sum_{t,n} \left[d_{t,n}\log\left(1 + \frac{AT_{t,n}(\theta)}{b_{t,n}}\right) - AT_{t,n}(\theta)\right].
    \end{equation}
    This expression is a function of the data that will ultimately be compared to a threshold to determine whether to reject the null hypothesis. Since any monotonic transformation preserves the ordering of values, such transformations do not affect the decision outcome. We therefore apply two monotonic transformations to simplify the expression and enhance its interpretability. These are the subtraction of the mean of the test given the null and division by the variance of the test given the null
    \begin{align}
         E[\mathcal{S}(\theta)|\mathcal{H}_0] &= \sum_{t,n} \left[b_{t,n}\log\left(1 + \frac{AT_{t,n}(\theta)}{b_{t,n}}\right) - AT_{n,t}(\theta)\right]; \nonumber \\
          V[\mathcal{S}(\theta)|\mathcal{H}_0] &= \sum_{t,n} b_{t,n}\log^2\left(1 + \frac{AT_{t,n}(\theta)}{b_{t,n}}\right).
    \end{align}
    Using the congruence sign for tests with the same meaning, we have
    \begin{equation}
        \mathcal{S}(\theta) \cong \sum_{t,n} (d_{t,n}-b_{t,n})\log\left(1 + \frac{AT_{t,n}(\theta)}{b_{t,n}}\right) \cong \frac{\sum_{t,n} (d_{t,n} - b_{t,n}) \log\left(1 + \frac{AT_{t,n}(\theta)}{b_n}\right)}
        {\sqrt{\sum_{n} b_{t,n} \circledast_t \log^2\left(1 + \frac{AT_{n}(\theta)}{b_n}\right)}}.
    \end{equation}
    The last assumption we make is that the background is constant over the duration of the burst (for the short timescales we consider). This allows us to drop the time index of the terms in the logarithm. The final form of the statistic use in our pipeline is
    \begin{equation}\label{eq: statistic appendix}
        \mathcal{S}_t(\theta) = \frac{\sum_{n} (d_{t,n} - b_{t,n}) \circledast_t \log\left(1 + \frac{AT_{t,n}(\theta)}{b_n}\right)}
        {\sqrt{\sum_{n} b_{t,n} \circledast_t \log^2\left(1 + \frac{AT_{n}(\theta)}{b_n}\right)}},
    \end{equation}
    where we replaced the sum over $t$ by a convolution in order to compute the test statistic efficiently at all times. The convolution provides a computationally efficient way to apply the statistic at each time index of the data. For large enough photon counts, the statistic in eq. \ref{eq: statistic appendix} is approximately a standard normal random variable, although, for low count rates, deviations from normal distribution may be significant. We account for this fact by adjusting the significance obtained from the statistic while computing the false alarm rate of triggers (see main text eq. \ref{eq: calibrate dist} and Fig. \ref{fig: calibrate dist}). For each trigger, we can obtain an optimal estimate of the amplitude of the signal by converting the test statistic in eq. \ref{eq: statistic appendix} units to units of counts. Consider the expectation value given a signal (we absorb the detection limit amplitude into the templates to avoid confusion)
    \begin{equation}
         E[{\mathcal{S}(\theta)}|\mathcal{H}_1] = \frac{A\sum_n T_n \log(1 + T_n/b_n)}{\sqrt{\sum_n b_n\log^2(1+T_n/b_n)}}.
    \end{equation}
    By isolating $A$ we get the conversion factor from probability to counts, so that the statistic in units of amplitude is
    \begin{equation}
        A(\theta) = \frac{\sum_{n,t} (d_{n,t} - b_{n,t}) \log\left(1 + \frac{T_{n}(\theta)}{b_n}\right)}{\sum_n T_n \log(1 + T_n/b_n)}.
    \end{equation}

\section{Match between Templates} \label{appendix: match}
    As discussed in the main text, constructing an efficient template bank for burst searches requires minimizing the number of templates while maximizing the coverage of the relevant parameter space. To quantify the similarity between two templates, we use the ratio of SNRs obtained from the test statistic when each template is applied. This ratio serves as a measure of how well one template can represent the signal modeled by another. Let's assume that we search for a burst with template $T^b$ and that the underlying signal is described by $T^a$ (we ignore the amplitude to simplify the notation). We denote by $\mathcal{S}_a$ the statistic in eq. \ref{eq: statistic appendix} calculated using the template $T^a$. Given that we use the correct template $T^a$, the SNR of $\mathcal{S}_a$ is
    \begin{equation} \label{eq: information content}
        I_a = \frac{E[\mathcal{S}_a|\mathcal{H}_1] - E[\mathcal{S}_a|\mathcal{H}_0]}{\sqrt{V[\mathcal{S}_a|\mathcal{H}_0]}},
    \end{equation}
    which gives in the Poisson case
    \begin{equation}
        I_a = \frac{\sum_n T^a_n \log(1 + T^a_n/b_n) }{\sqrt{\sum_n b_n \log^2(1 + T^a_n/b_n)}}.
    \end{equation}
    If, instead, we had used $T^b$ in the search, while the true signal is $T^a$, the SNR of $\mathcal{S}_b$ would be
    \begin{equation}
        I_{b} = \frac{\sum_n T^a_n \log(1 + T^b_n/b_n) }{\sqrt{\sum_n b_n \log^2(1 + T^b_n/b_n)}}.
    \end{equation}
    The ratio of SNRs gives the match
        \begin{align}
        \mathcal{M}(T^a, T^b) &= \frac{\sum_n T^a_n \log(1+ T^b_n/b_n)}{\sum_n T^a_n \log(1+ T^a_n/b_n)}
        \cdot \sqrt{\frac{\sum_n b_n\log^2(1 + T^a_n/b_n)}{\sum_n b_n\log^2(1 + T^b_n/b_n)}}.
        \end{align}
    For high background counts, the Poisson match reduces to the Gaussian match
    \begin{align}\label{eq: gauss match}
        \mathcal{M}(T^a, T^b) &\approx \frac{\sum_n T^a_nT^b_n/b_n}{\sqrt{\sum_n T^a_nT^a_n/b_n}\sqrt{\sum_n T^b_nT^b_n/b_n}} \nonumber\\
        &= \frac{\langle T^{a}, T^{b} \rangle}{\sqrt{\langle T^{a}, T^{a} \rangle}\sqrt{\langle T^{b}, T^{b} \rangle}},
    \end{align}
    where the sums can now be thought of as inner products. In this regime, the match has the simple interpretation of the cosine of the angle between the two vectors.
    
    In our search, we use the Gaussian form of the match to place our temporal box templates. Suppose that the duration of $T^a$ is $w_a$ and of $T^b$ is $w_b$. Also, without loss of generality, let's assume $w_a < w_b$. Since the templates are normalized boxes, the inner product \ref{eq: gauss match} gives
    \begin{equation}
        \mathcal{M}(T^a, T^b) = \sqrt{\frac{w_a}{w_b}}.
    \end{equation}
    This equation gives the relative SNR between templates with different durations. We define $\alpha = 1 - \mathcal{M}$ to be the mismatch between the templates, and by fixing it to our tolerance, we get $w_1 = w_0/(1-\alpha)^2 = f w_0$. Repeating the argument reveals that the $n$th template will be $w_n = f^n w_0$. To specify the tolerance—i.e., the maximum acceptable mismatch—for our template bank construction, we must account for the fact that the true signal duration present in the data is unknown. Suppose the data $d$ contains a burst of duration $w$, constrained within $w_a\leq w\leq w_b$. The match values between the data and the two nearest templates in duration, $T^a$ and $T^b$, are given by
    \begin{equation}
        \mathcal{M}(T^a, d) = \sqrt{\frac{w_a}{w}}; \quad             \mathcal{M}(T^b, d) = \sqrt{\frac{w}{w_b}}.
    \end{equation}
    The worst-case (minimal) match occurs when these two values are equal, which happens when $w = \sqrt{w_a w_b}$. In this case we have $(1 - \alpha)^4 = w_a/w_b$, which again gives $w_n = f^n w_0$, but this time $f = 1/(1-\alpha)^4$. For logarithmic steps of $f=1.35$ we get a mismatch of $\alpha \approx 0.07$.

\section{Quadratic Correction to the Background Estimation} \label{appendix: bkg}
    As described in Section \ref{sec: bkg est} of the main text, we estimate the background counts using the mean count rate in each detector channel. To ensure that photons from the signal do not contaminate the background estimate, we exclude a central region from the mean calculation, with a width equal to the signal duration (Eq. \ref{eq: bkg estimator}). Since increasing the number of background samples reduces sensitivity due to red noise, we seek to improve the background estimation without using additional samples. We note that Eq. \ref{eq: bkg estimator} is equivalent to fitting a straight line between the estimated background levels on either side of the gap. Instead, if we fit a second-order polynomial, the estimation improves for longer gaps, where the background varies more significantly. Consider the following filter, constructed to calculate the mean around a gap of length $\Delta$
    \begin{equation}
        \vec{w}_1 = \frac{1}{2}[\vec{0}, 1, \vec{0}, 1, \vec{0}],
    \end{equation}
    where the $\vec 0$ are of length $\Delta$, and the vector $\vec b$ is the background estimator $b = 1/N \sum_{\tau=t-N/2}^{t+N/2}d_{\tau}$.
    Using the value $\vec{w}_1\cdot\vec{b}$ as the estimator of the central point, it is similar to our background estimation procedure described in eq. \ref{eq: bkg estimator}. Adding another filter
    \begin{equation}
        \vec{w}_2 = \frac{1}{2}[1, \vec0, \vec0, \vec0, 1],
    \end{equation}
    we can achieve a better estimation by constructing a linear combination of the two filters
    \begin{equation}
        \vec{F} = a_1\vec{w}_1 + a_2\vec{w}_2.
    \end{equation}
    To find the coefficients, we require the removal of any linear or quadratic contribution to the estimator
    \begin{align}
        \vec{F}\cdot \vec{1} = 1; \quad
        \vec{F}\cdot \vec{b} = 0; \quad
        \vec{F}\cdot \vec{b}^2 = 0.
    \end{align}
    Solving these equations, we find that, $a_1 = 4/3$ and $a_2 = -1/3$ so that
    \begin{equation}
        \vec{F} = \frac{4\vec{w}_1-\vec{w}_2}{3},
    \end{equation}
    hence eq. \ref{eq: bkg quad} follows.
    
\section{Drift Correction} \label{appendix: drift correction}
    Here, we give a brief review of the drift correction presented in \cite{zackay2021detecting}, and show that in the Poisson regime, contrary to the Gaussian regime, the loss in SNR due to background misestimation scales linearly with the error. We start by calculating the SNR loss due to incorrect background; for this, let us define the "wrong" statistic
    \begin{equation}
        \mathcal{S}_w = \frac{\sum_n(d_n - b_n^w)\log(1+T_n/b_n^w)}{\sqrt{\sum_n b_n^w \log^2(1+T_n/b_n^w)}},
    \end{equation}
    where we are considering a single time bin, $n$ is the usual detector/channel index, and for brevity, we absorbed the amplitude into the template. The "wrong" nature of this statistic is due to the normalization by the wrong, or rather approximate, background. The true SNR of this statistic is
    \begin{equation}
        I_w = \frac{\sum_n T_n \log(1 + T_n / b^w_n)}{\sqrt{\sum_n b^c_n \log^2(1 + T_n/b_n^w)}},
    \end{equation}
    and if we were to naively take the wrong background as the correct one (similar to what we have in eq. \ref{eq: statistic}), we would get
    \begin{equation}
        I_{w, \text{computed}} = \frac{\sum_n T_n \log(1 + T_n / b^w_n)}{\sqrt{\sum_n b^w_n \log^2(1 + T_n/b_n^w)}}.
    \end{equation}
    On the other hand, had we known the exact background, the correct statistic would be 
    \begin{equation}
        \mathcal{S}_c = \frac{\sum_n(d_n - b_n^c)\log(1+T_n/b_n^c)}{\sqrt{\sum_n b_n^c \log^2(1+T_n/b_n^c)}},
    \end{equation}
    with SNR
    \begin{equation}
        I_c = \frac{\sum_n T_n \log(1 + T_n/b_n^c)}{\sqrt{\sum_n b_n^c \log^2(1+T_n/b_n^c)}}.
    \end{equation}
    Writing the background as $b_n^w = (1+\epsilon_n)b_n^c$, the ratio gives
    \begin{align}
        \left(\frac{I_c}{I_w}\right)^2 = \left(\frac{\sum_n T_n \log(1 + T_n/b_n^c)}{\sum_n T_n \log(1 + T_n / (1+\epsilon_n)b_n^c)}\right)^2 \cdot
        \frac{\sum_n b^c_n \log^2(1 + T_n/(1+\epsilon_n)b_n^c)}{\sum_n b_n^c \log^2(1+T_n/b_n^c)}.
    \end{align}
    In the limit $T_n/b_n \ll 1$, the ratio reduces to
    \begin{equation}
        \left(\frac{I_c}{I_w}\right)^2 = \frac{\sum_n T_n^2/(1+\epsilon_n)^2 b_n^c}{(\sum_n T_n^2/(1+\epsilon_n) b_n^c)^2}\cdot \sum_n T_n^2/b_n,
    \end{equation}
    which is, up to normalization, the expression in \cite{zackay2021detecting} eq. 33. In the other limit, $T_n/b_n \gg 1$, which is relevant in the low count Poisson regime, we get
    \begin{align}
        \left(\frac{I_c}{I_w}\right)^2 &\approx \left(\frac{\sum_n T_n \log(T_n/b_n^c)}{\sum_n T_n \log(T_n / b_n^c) - \sum_n T_n\log(1 + \epsilon_n)} \right)^2 \cdot
        \frac{\sum_n b^c_n \left(\log(T_n/b_n^c)- \log(1 + \epsilon_n)\right)^2}{\sum_n b_n^c \log^2(T_n/b_n^c)} \nonumber \\
        &\approx 
        \frac{1}{(1 - \sum_n \tilde{T}_n\epsilon_n)^2} \cdot \left( 1 -2\sum_n \tilde{b}_n \log(T_n/b_n^c)\epsilon_n + \mathcal{O}(\epsilon_n^2)\right),
    \end{align}
    where we defined $\tilde{T}_n = T_n / \sum_n T_n\log(T_n/b_n^c)$ and $\tilde{b}_n = b_n^c / \sum_n b_n^c\log^2(T_n/b_n^c)$ for ease of notation. Continuing the expansion
    \begin{align}
        \left(\frac{I_c}{I_w}\right)^2 &\approx 
        (1 + 2\sum_n\tilde{T}_n\epsilon_n + \mathcal{O}(\epsilon_n^2))\left( 1 -2\sum_n \tilde{b}_n \log(T_n/b_n^c)\epsilon_n + \mathcal{O}(\epsilon_n^2)\right) \nonumber \\
        &\approx
        1 + 2\sum_n (\tilde{T}_n-\tilde{b}_n\log(T_n/b_n^c))\epsilon_n + \mathcal{O}(\epsilon_n^2).
    \end{align}
    The coefficient of the linear term is not symmetric and, hence, does not vanish identically.

    \subsection{Generalization of the drift correction for the non-Gaussian case}
    While the first order error in the S/N of the optimal search is not algebraically vanishing, the overall argument why the loss in the power of the test is quadratic in $\epsilon_n$ is. The general argument is that if $\epsilon_n$ is zero, the test's power is maximal. Hence, the change is $O(\epsilon_n^2)$. The correction, scaling the test statistic, could equivalently be thought of as a change in the threshold (to adapt it such that we will have a constant false alarm rate). Hence, the correct form of the correction in the general case is some monotone transformation that would fix the false-alarm rate of the test at all times.
    Although we do not have a rigorous proof that this correction is valid in all regimes, we did verify that in our case \ref{Eq.Drift_correction} fixes the empirical false alarm rate to a good enough approximation.
    
\section{Sky posterior} \label{appendix: sky posterior}
    We can get an estimate of the sky position of a trigger by calculating the posterior probability of the sky position. Recall the hypotheses are
    \begin{align*}
        \mathcal{H}_0&: d_n \sim \text{Pois}(b_n); \\
        \mathcal{H}_1&: d_n \sim \text{Pois}(b_n + AT_n(\Omega, \xi)),
    \end{align*}
    where we denote by $\Omega$ the sky position and by $\xi$ the spectral parameters. The posterior probability of the sky position is
    \begin{align*}
        P(\Omega | d; \mathcal{H}_1) &\propto \int \text{d}\xi \; P(d | \Omega, \xi;\mathcal{H}_1) P(\Omega) p(\xi) \\
        &\propto \int \text{d}\xi \; P(d | \Omega, \xi;\mathcal{H}_1) \\
        &\propto \int \text{d}\xi \; \frac{P(d | \Omega, \xi;\mathcal{H}_1)}{P(d | \mathcal{H}_0)},
    \end{align*}
    where we have assumed a uniform prior on both the spectral parameters and the sky position, and in the last step, we divided by the likelihood of the null hypothesis (as it is constant with respect to the signal parameters). Now, plugging in the Poisson likelihood, we get
    \begin{align}\label{eq: appendix posterior}
        &P(\Omega | d; \mathcal{H}_1) \propto \\
        &\propto \int \text{d}\xi \; \prod_n \frac{e^{-(b_n + AT_n(\Omega, \xi))} (b_n + AT_n(\Omega, \xi))^{d_n}/d_n!}{e^{-b_n} b_n^{d_n}/d_n!} \\
        &\propto \int \text{d}\xi \; \prod_n e^{-AT_n(\Omega, \xi)} \left(1 + \frac{AT_n(\Omega, \xi)}{b_n}\right)^{d_n} \\
        &= \int \text{d}\xi \; \exp{\sum_n \left[d_n\log\left(1 + \frac{AT_n(\Omega, \xi)}{b_n}\right) - AT_n(\Omega, \xi)\right]}.
    \end{align}

\section{Bayesian Model Selection} \label{appendix: bayesian model selection}
    Bayesian model selection is a criterion to distinguish between two competing models based on the ratio of model probabilities given the observed data. In our setting, we try to distinguish between models with different parameters. The parameters are the sky position of the triggers, which we will denote as $\Omega$. The hypotheses of our models are
    \begin{align}
        \mathcal{H}_0&: d_n \sim \text{Pois}(b_n); \nonumber\\
        \mathcal{H}_{{\odot}}&: d_n \sim \text{Pois}(b_n + AT_n(\Omega, \xi)), \; \Omega \in \Omega_{\odot}; \nonumber\\
        \mathcal{H}_{\oplus}&: d_n \sim \text{Pois}(b_n + AT_n(\Omega, \xi)), \; \Omega \in \Omega_{\oplus}; \nonumber\\
        \mathcal{H}_1&: d_n \sim \text{Pois}(b_n + AT_n(\Omega, \xi)), \; \Omega \notin \Omega_{\odot, \oplus}, \nonumber\\
    \end{align}
    where $\Omega_{\odot}$ is the set of angles covering the Sun (and similarly for the Earth for $\Omega_{\oplus}$) and $\xi$ refers to the set of spectral parameters (the Band function parameters described in section \ref{sec: template banks}).
    We also included, for completeness, the null hypothesis, which has no free parameters. Let us take, without loss of generality, the Earth in our derivation (a similar result holds for the Sun). The probability ratio of the signal coming from the Sun to the probability that it does not come from the Sun or Earth is
    \begin{equation}
        \mathcal{B}_{\oplus} = \frac{P(\mathcal{H}_{\oplus}|d)}{P(\mathcal{H}_1|d)} = \frac{\pi(\mathcal{H}_{\oplus})}{\pi(\mathcal{H}_1)} \frac{\int d\Omega d\xi\; p(\Omega) p(\xi) L(d|\mathcal{H}_{\oplus};\Omega, \xi) } {\int d\Omega d\xi\; p(\Omega) p(\xi) L(d|\mathcal{H}_1;\Omega, \xi)},
    \end{equation}
    where $\pi(\mathcal{H}_{\oplus})$ ($\pi(\mathcal{H}_1)$) is the prior probability of the Earth (not Earth or Sun) model. This quantity is also known as the Bayes factor. Notice that these are precisely the steps we took in the sky posterior computation. The Bayes factor can be written simply as the ratio of the integrated posterior on the sky regions of interest
   \begin{equation}
        \mathcal{B}_{\oplus} = \frac{P(\mathcal{H}_{\oplus}|d)}{P(\mathcal{H}_1|d)} \frac {\int d\Omega \; p(\Omega | d; \mathcal{H}_{1}) \; I(\Omega \in \Omega_{\oplus}) }{\int d\Omega \; p(\Omega | d; \mathcal{H}_{1}) \; I(\Omega \notin \Omega_{\odot, \oplus}) }.
    \end{equation}

\section{SVD for Fast Matched Filtering}\label{appendix: drift SVD}
\begin{figure}
        \centering
        \includegraphics[width=0.4\linewidth]{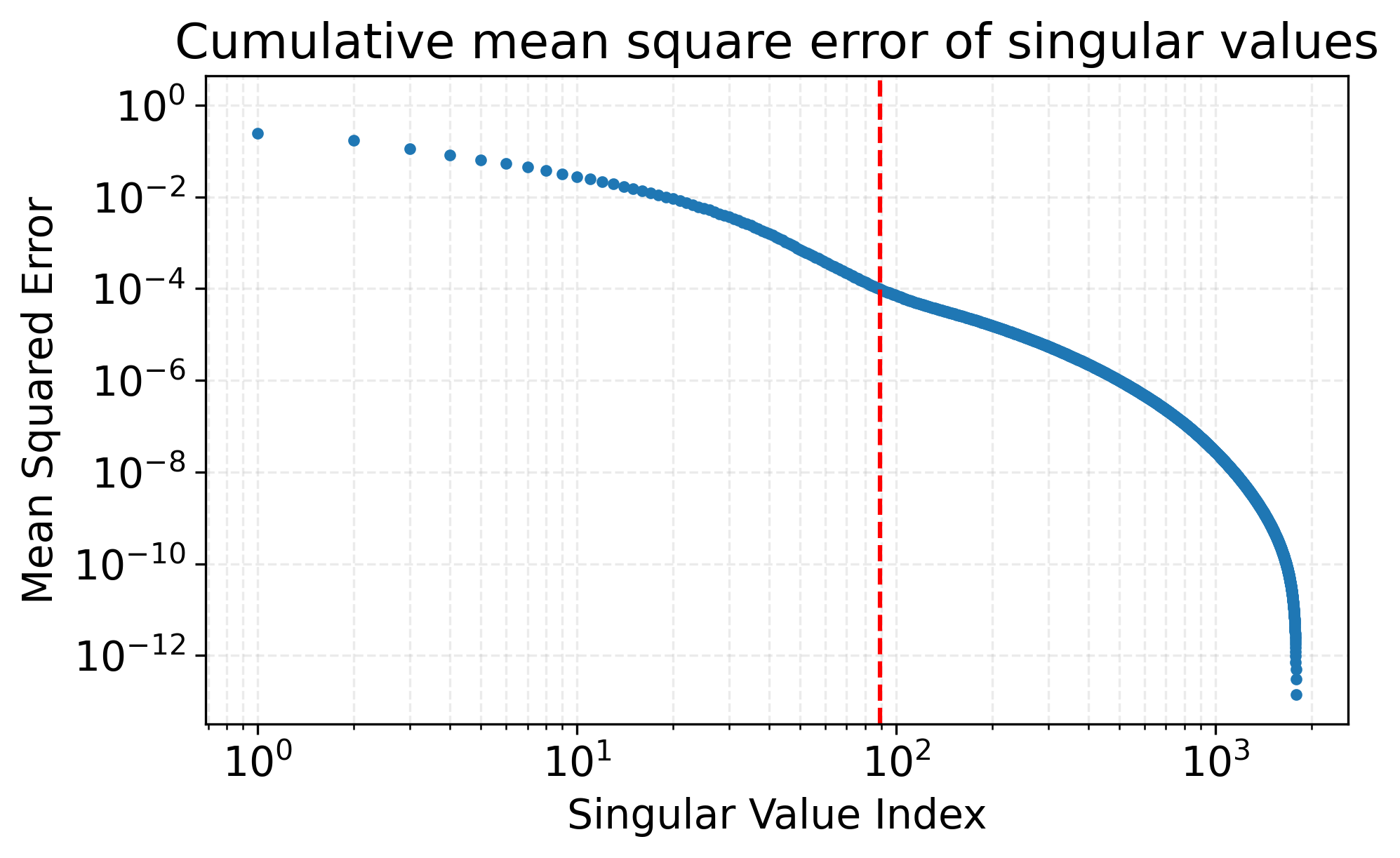}
        \includegraphics[width=0.4\linewidth]{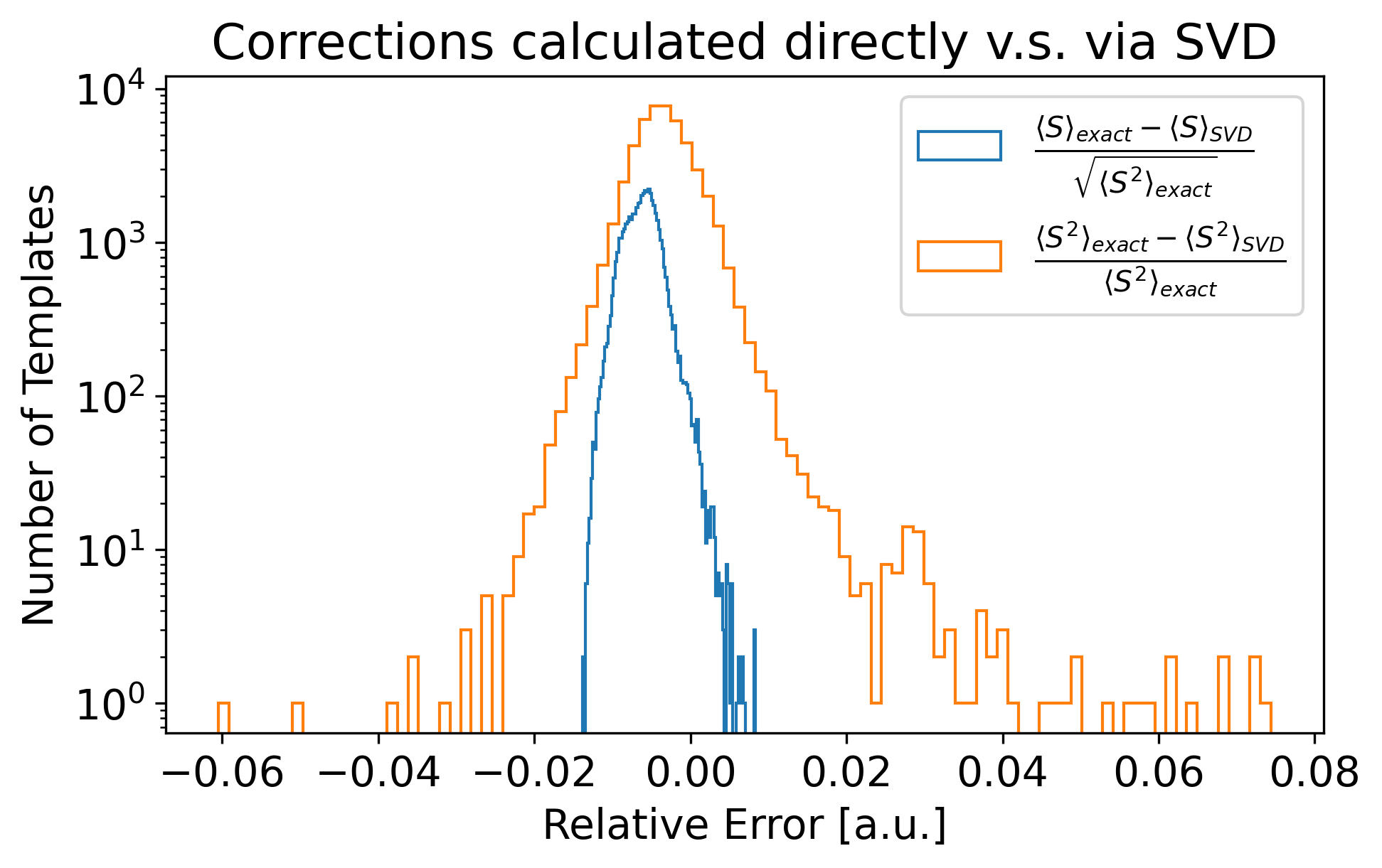}
        \caption{Comparison of the dense bank SVD for GRB170817. Left: Mean square error (normalized) for the reconstruction of the full test statistic in the dense bank ($5\times 10^4$ templates). One can see that $\sim 100$ singular vectors are sufficient for a $10^{-4}$ error. Right: Background corrections for GRB170817 using a direct calculation v.s. using 100 singular values.}
        \label{fig: appendix svd localization}
    \end{figure}

\begin{figure}
         \centering
         \includegraphics[width=0.4\linewidth]{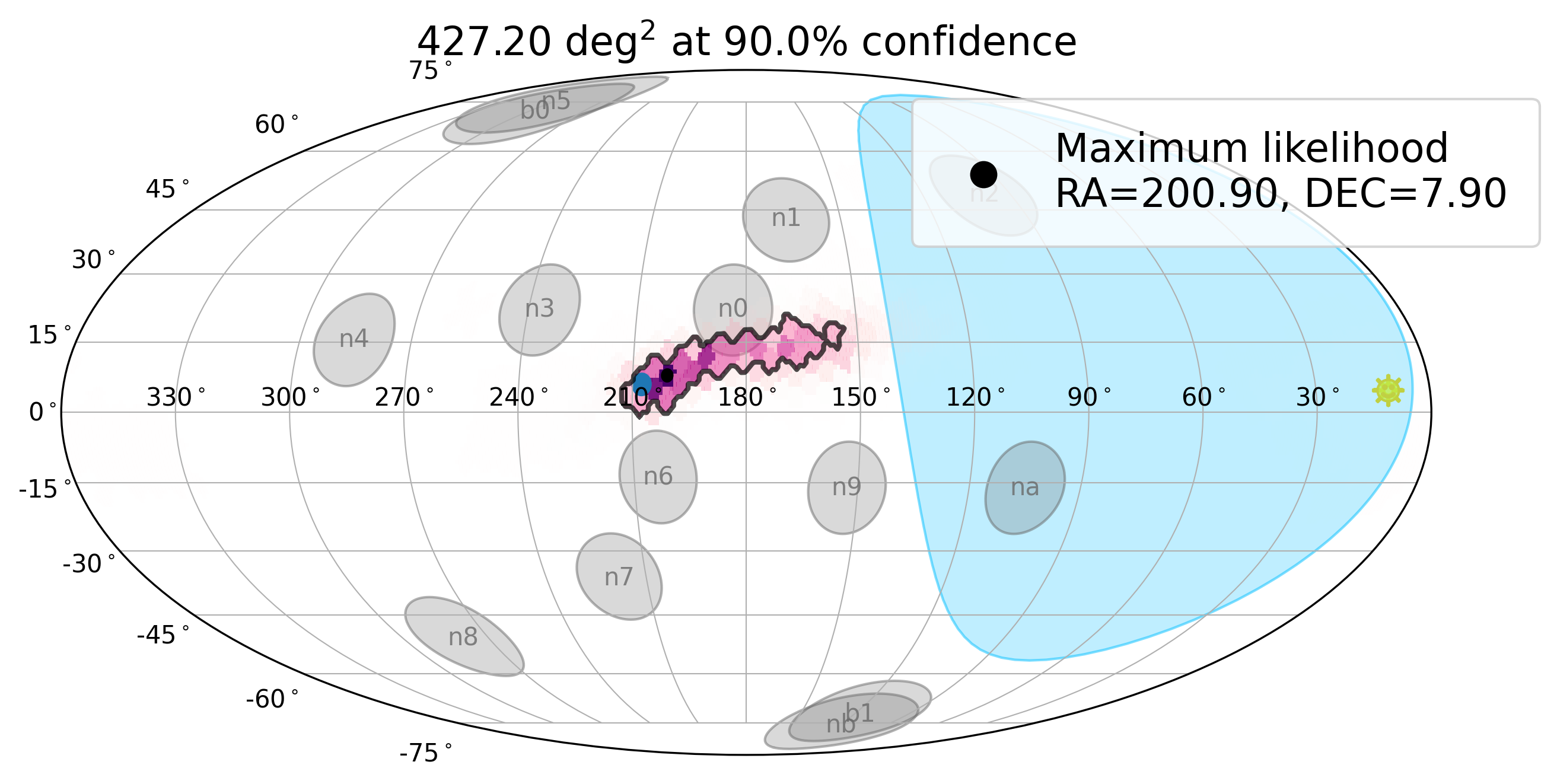}
         \includegraphics[width=0.4\linewidth]{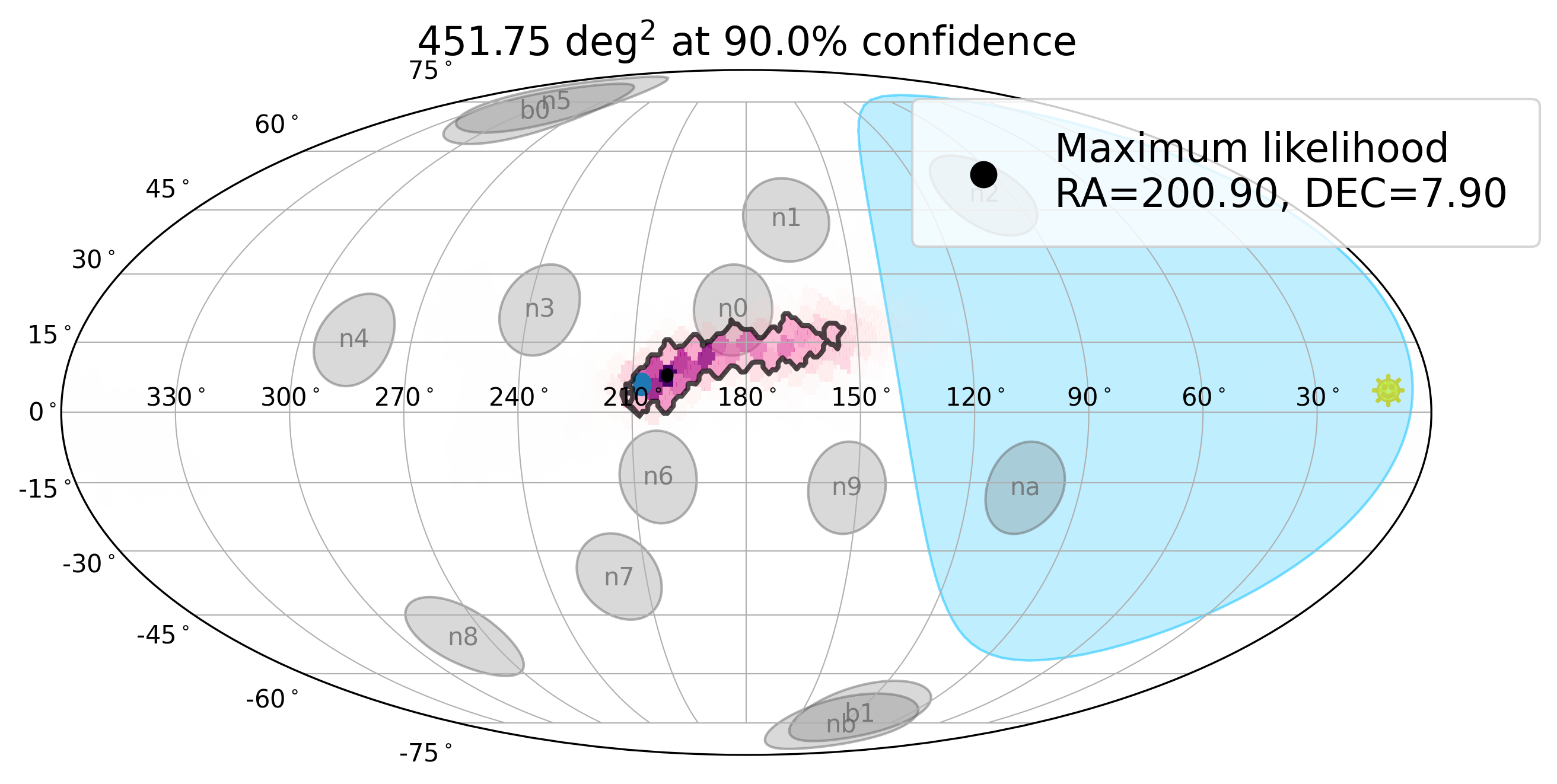}
        \includegraphics[width=0.18\linewidth]{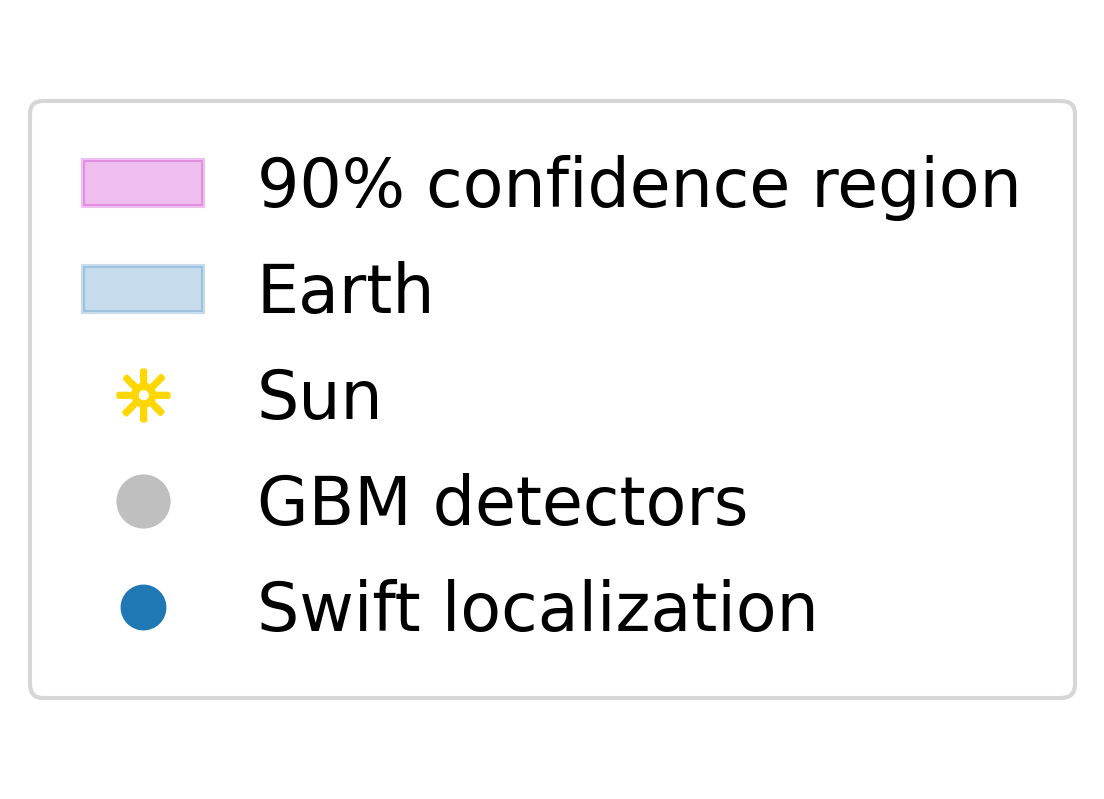}
         \caption{GRB140402007 sky localization with and without drift correction. Left: No correction, right: amplitude is drift corrected, resulting in the confidence region slightly increasing.}
         \label{fig: 140402 localization}
     \end{figure}

    When using large template banks, calculating the statistics at many different time bins and averaging to get the corrections 
    \begin{equation}
        \tilde{\mathcal{S}}_t(\theta) = \frac{\mathcal{S}_t(\theta) - \left\langle \mathcal{S}_t(\theta) \right\rangle_t}{\sqrt{\left\langle \mathcal{S}_t^2(\theta) \right\rangle_t-\left\langle \mathcal{S}_t(\theta) \right\rangle_t^2}},
    \end{equation}
    may become too computationally expensive to allow a full search. To overcome this we can use SVD to find a small number of basis vectors that we can use instead of the full template bank and use them to construct the full statistic. Let us write the statistic \ref{eq: statistic appendix} at the time of a trigger in vector notation
    \begin{equation}
         \vec{\tilde{d}}_t = \vec{d}_t - \vec{b}_t; \quad \vec{f}_t = \frac{\log(1 + \vec{T}(\theta)/\vec{b})}{\sqrt{\vec{b}_t\cdot\log^2(1 + \vec{T}(\theta)/\vec{b})}}, 
    \end{equation}
    so that
    \begin{equation}
         \mathcal{S}_t(\theta) = \vec{\tilde d}_t \cdot \vec{f}_t(\theta). 
    \end{equation}
    Let us now take the SVD of the filter
    \begin{equation}
        \left\langle \mathcal{S}_t(\theta) \right\rangle_t = \langle \vec{\tilde d} \cdot \vec{f} \rangle_t = \sum_k \vec{\alpha}_k \langle \vec{\tilde d}_t \cdot \vec{v}_{tk}^T \rangle_t.
    \end{equation}
    \begin{equation}
        \left\langle \mathcal{S}^2_t(\theta) \right\rangle_t = \langle (\vec{\tilde d} \cdot \vec{f})^2 \rangle_t = \sum_{k,j} \vec{\alpha}_k \langle (\vec{\tilde d}_t \cdot \vec{v}_{tk}^T) (\vec{\tilde d}_t \cdot \vec{v}_{tj}^T) \rangle_t \vec{\alpha}_j .
    \end{equation}
    We may keep a small number, $k$, of basis vectors for which we know how to construct the full $ \mathcal{S}_t(\theta)$. In practice, we choose the number of basis vectors corresponding to a reconstruction error of $10^{-4}$ which is of the order of $100$. Figure \ref{fig: appendix svd localization} shows a comparison of the directly calculated dense bank corrections and the approximated SVD corrections.
    Note that this approach is also useful when calculating the posteriors. Since the amplitude in the posterior \ref{eq: appendix posterior} depends on the statistic value, we would like to correct background errors. The corrected value is
    \begin{equation}
         \tilde{A}^*(\theta) = \frac{{\sqrt{\langle \mathcal{S}^2_{t^*}(\theta) \rangle_t-\left\langle \mathcal{S}_{t^*}(\theta) \right\rangle_t^2}}\left[\sum_n (d_{n,t^*}-b_{n,t^*})\log(1 + T_n(\theta)/b_{n,t^*})\right] + \langle \mathcal{S}_{t^*}(\theta) \rangle_t \sqrt{\sum_n b_{n,t^*}\log^2(1 + T_n(\theta)/b_{n,t^*})}}{\sum_n T_{n}(\theta)\log(1 + T_n(\theta)/b_{n,t^*})}
     \end{equation}
     As an illustrative example, Fig. \ref{fig: 140402 localization} shows the effect of applying this correction to GRB 140402A, which was also localized by Swift BAT. In this case, the correction leads to a small increase in the size of the confidence region, while the maximum likelihood value remains unchanged.   

\section{Contaminant Signal Bank}\label{appendix: glitch bank}
\begin{figure}
        \centering
        \includegraphics[scale=0.65]{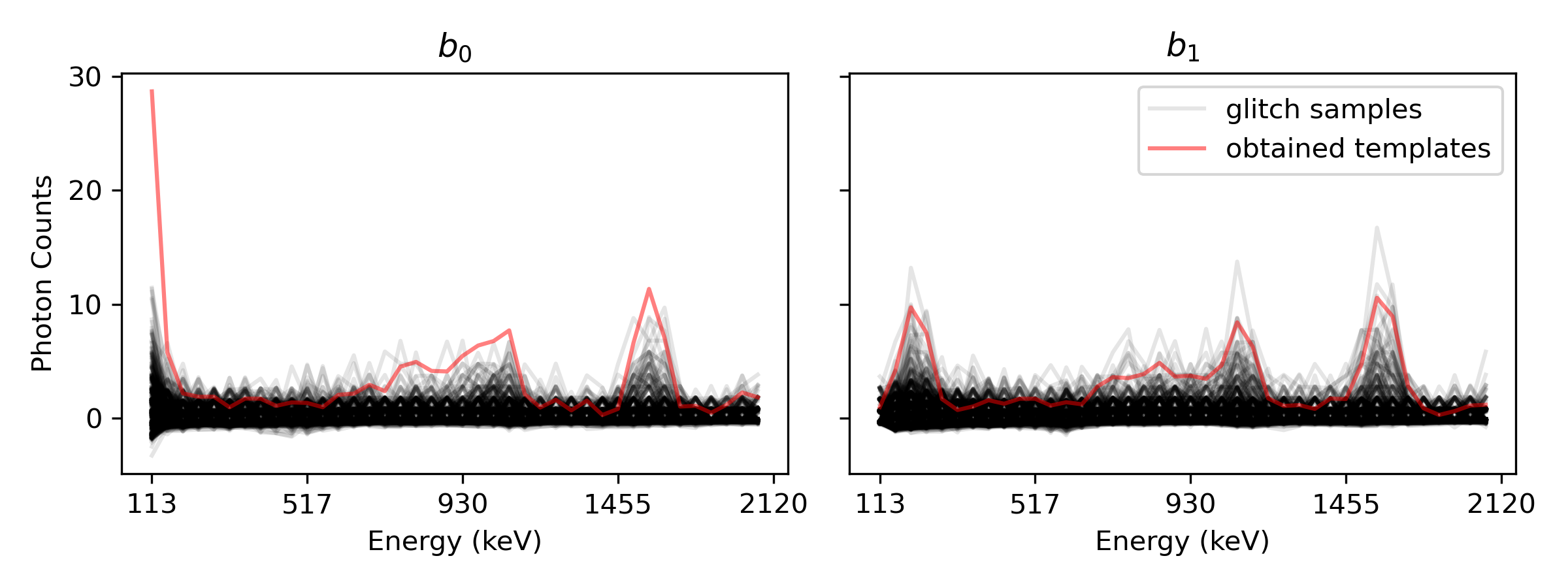}
        \caption{BGO contaminant templates overlaid on the sample used to create it using SVD. The x-axis in the figure was plotted only in the lower energy channels as the higher channels do not contain a signal.}
        \label{fig: bgo glitches}
    \end{figure}

    As mentioned in the main text, we identified spurious signals appearing in different detectors. For the NAI detectors, these signals present themselves in the lowest energy channels and are modeled simply as a Gaussian in the first three channels of the detector, and are absent any signal in the rest of the detectors. For the BGO detectors, a common repeating signal has a more complicated shape. Due to its high occurrence rate and the fact that it presents itself only in one detector, we filter it out. To model its shape, we collected a large sample of triggers having this shape for both the BGO detectors and used SVD on the sample. We then use the first singular component as our BGO contaminant template. The sample and the resulting shapes are presented for both detectors in Fig. \ref{fig: bgo glitches}

\section{Contaminant Removal Tests Histograms}\label{app: vetoes}
    Here we provide illustrations of the performance of the specialized tests developed in section \ref{sec: vetoes}. These include the single detector test, the particle precipitation test and the occultation step test. The histogram of the timing glitch test was provided in the main text, Fig. \ref{fig:timing glitches}.
    \begin{figure}
        \centering
        \includegraphics[scale=0.45]{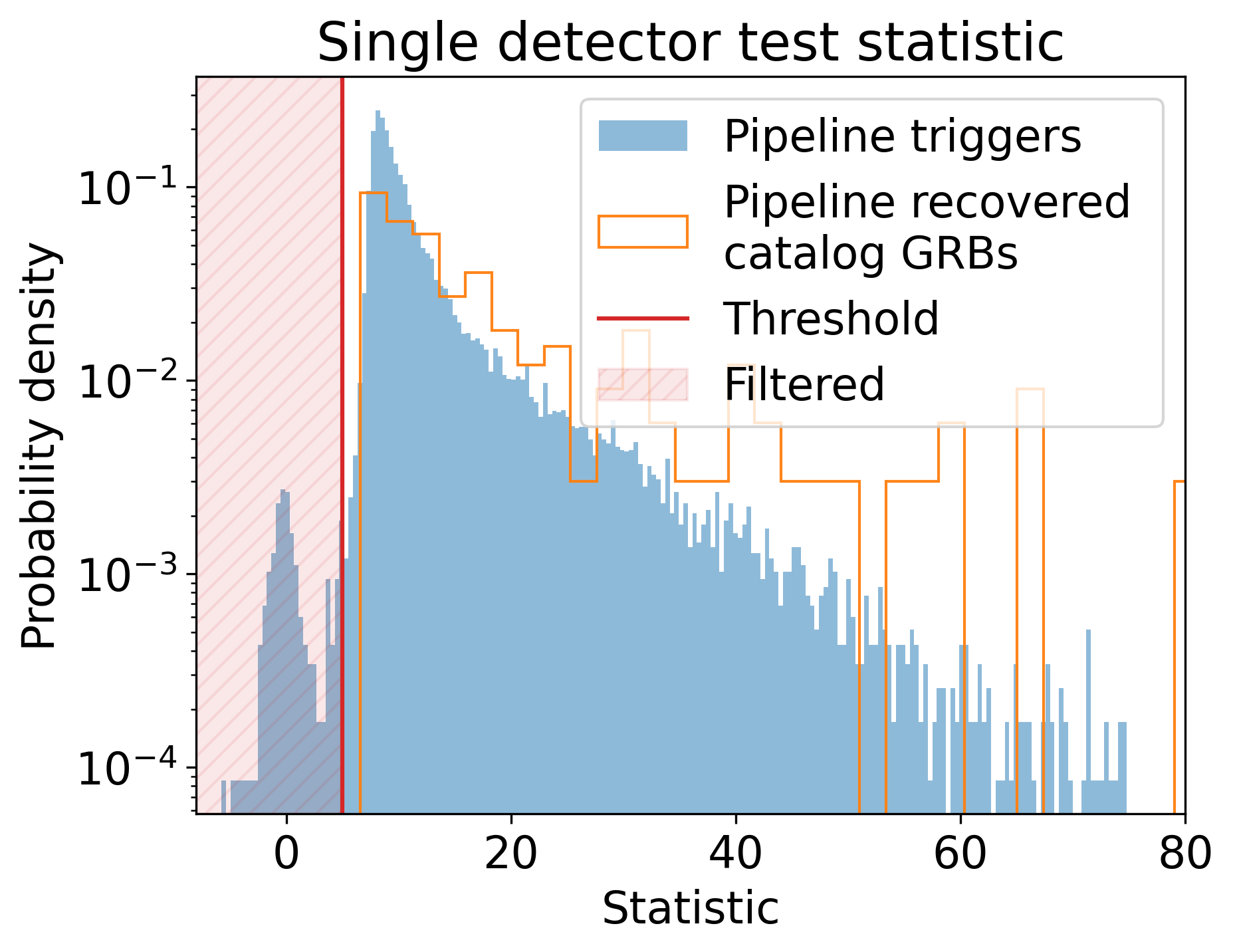}
        \includegraphics[scale=0.45]{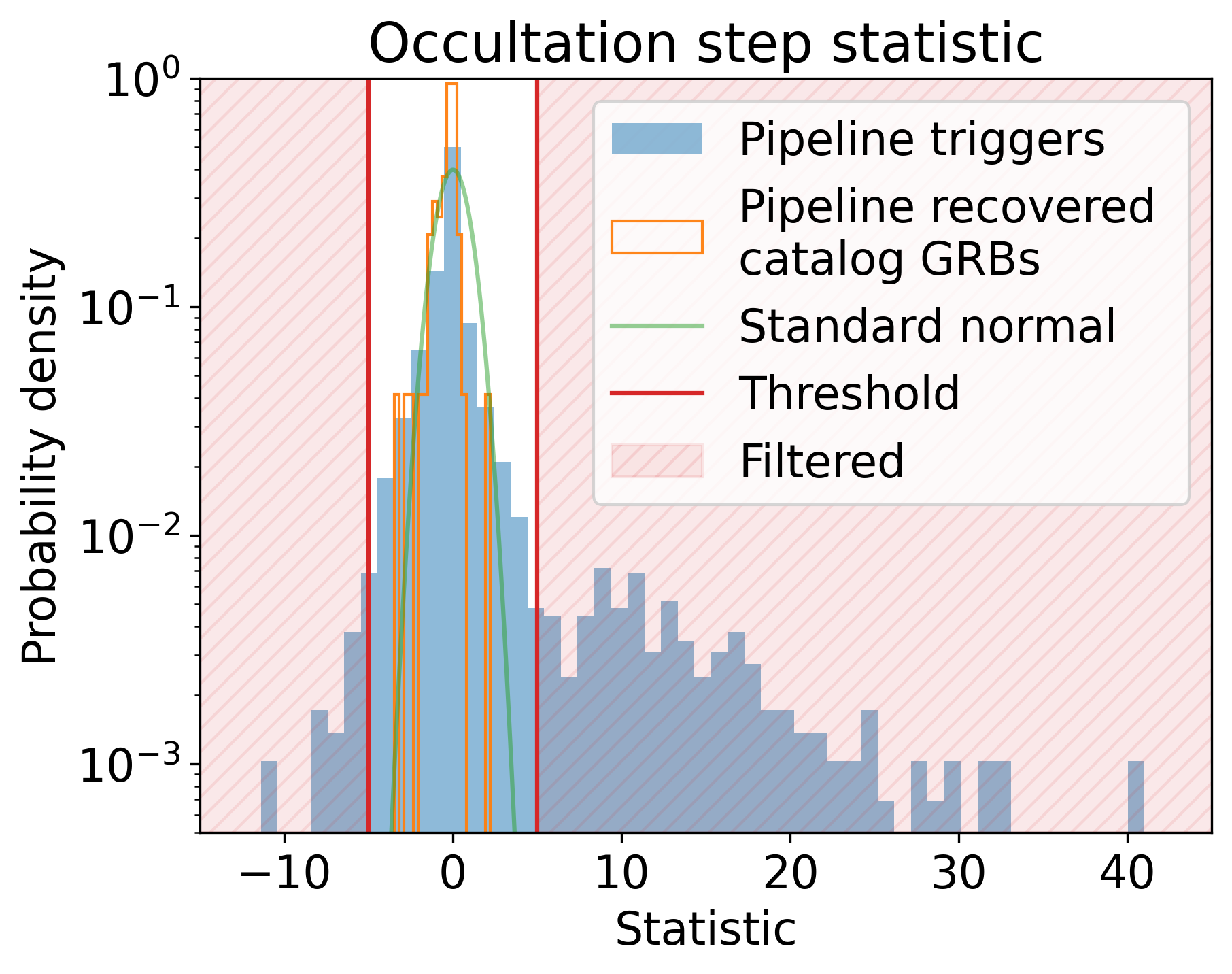}
        \includegraphics[scale=0.45]{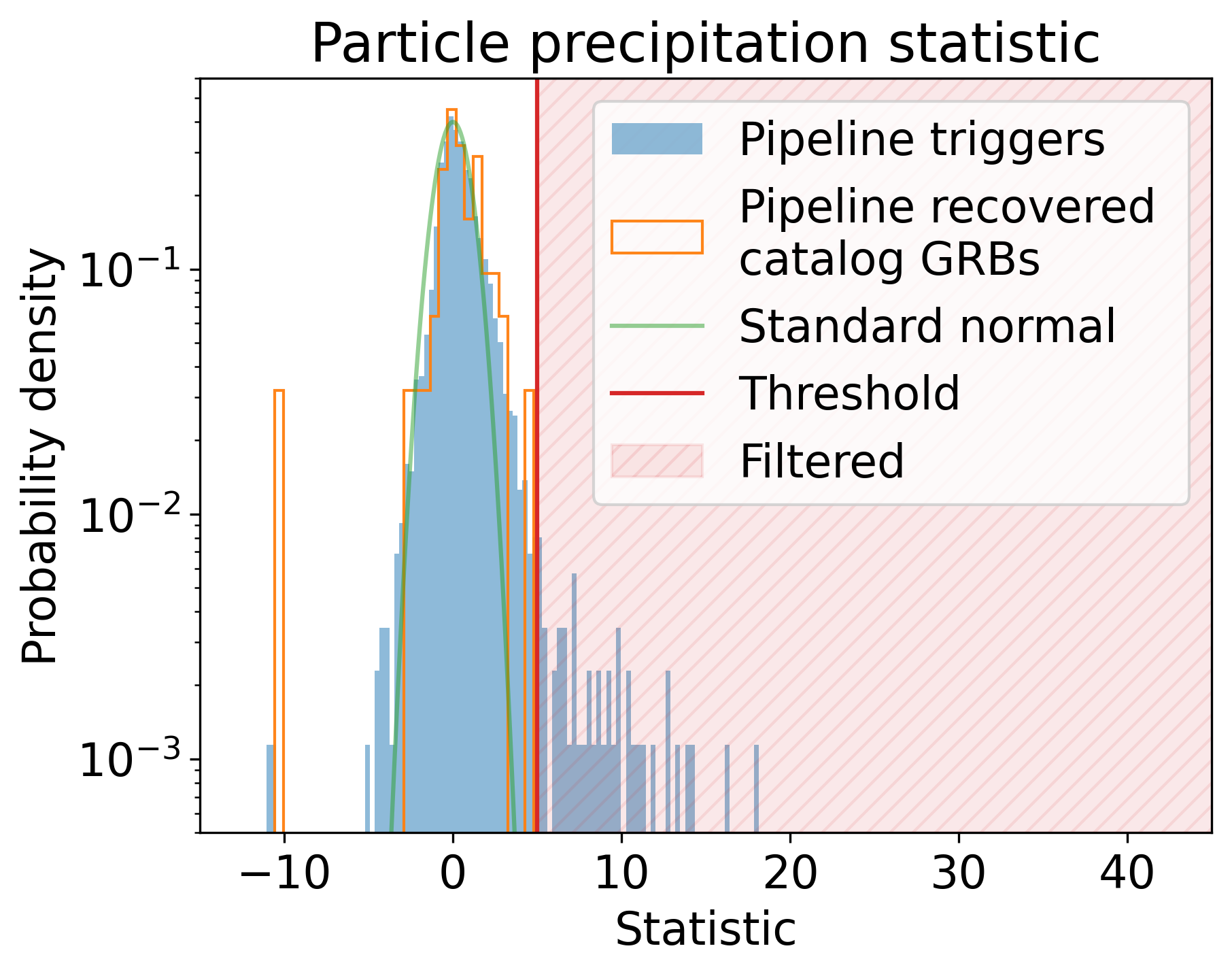}
        \caption{\textbf{Contaminant Removal Tests}. The histograms illustrate the performance of the specialized tests developed to filter out contaminants from the trigger distribution. In all figures, the blue curves represent the distribution of all triggers, while the orange curves correspond to the distribution of the test statistic for triggers that were also identified as GRBs in the GBM catalog. Where relevant, the green curve shows the corresponding normal distribution probability density function. The selected thresholds are indicated by red lines, and the triggers filtered out by the tests fall within the red-shaded bands. In all tests, we chose the $4\sigma$ thresholds. The histograms show that the catalog GRBs roughly follow the distribution of triggers we keep.}
        \label{fig:enter-label}
    \end{figure}

\newpage
\section{Tables} \label{appendix: tables}

    \begin{table*}[t]
    \small
    \setlength{\tabcolsep}{3pt}
    \renewcommand{\arraystretch}{1.5}
    \caption{Fermi observation context for the dates in Fig.~2. \\Taken from https://fermi.gsfc.nasa.gov/ssc/observations/timeline/posting/}
    \label{tab:fermi_context}
    \begin{tabularx}{\textwidth}{l p{2.5cm} p{2.5cm} l p{7.5cm}}
    \hline
    \textbf{Obs ID} & \textbf{Start (UTC)} & \textbf{Stop (UTC)} & \textbf{Mode} & \textbf{Observation Details} \\
    \hline
    070100-56-1 & 2014-10-15 01:06:00 & 2014-10-16 00:00:00 & Pointed & Modified survey for galactic center, Pole position = 68.70, 64.40. Target position (J2000): 268.44, 24.28  \\
    080400-45-1 & 2016-03-31 00:10:00 & 2016-04-07 00:00:00 & Survey & 50deg 6am rock; orbit period 5718 sec; -/+ 50deg rock angle over 2 orbits \\
    090400-66-1 & 2017-05-18 00:10:00 & 2017-05-24 01:57:00 & Survey & 50deg 6am rock; orbit period 5715 sec; -/+ 50deg rock angle over 2 orbits \\
    090400-71-1 & 2017-06-15 00:10:00 & 2017-06-22 00:00:00 & Survey & 50deg 6am rock; orbit period 5715 sec; -/+ 50deg rock angle over 2 orbits \\
    090400-82-1	& 2017-07-20 00:55:00 & 2017-07-27 00:00:00 & Survey & 50deg 6am rock; orbit period 5715 sec; -/+ 50deg rock angle over 2 orbits \\
    100601-1-1 & 2018-01-16 17:57:00 & 2018-01-17 23:12:00 & ToO Pointed & Offset pointing position. Target position (J2000): 171.6, -55.5 \\
    100403-6-1 & 2018-04-26 00:10:00 & 2018-04-30 14:35:00 & Survey & +50 deg only rock \\
    110403-28-1 & 2019-01-03 17:20:00 & 2019-01-07 05:21:00 & Survey & -50 deg only rock \\
    150401-16-1 & 2022-12-15 01:27:00 & 2022-12-22 00:00:00 & Survey & +50, -50 two orbit rock, 5710s \\
    160416-10-1 & 2023-10-26 01:25:00 & 2023-10-30 00:15:00 & Survey & +50, -50 two orbit rock, 5710s \\ \\
    140400-3-1 & 2021-09-23 01:03:00 & 2021-09-27 23:55:00 & Survey & 50deg; orbit period 5713 sec; -/+ 50deg rock angle over 2 orbits \\
    \hline
    \end{tabularx}
    \tablenotetext{}{The observation times in the table are ordered in the same order as the times used in Fig \ref{fig: gbm vs balrog}. The last entry shows the observation details of the time used to generate our template bank.}
    \end{table*}

    \centerwidetable
    \begin{longrotatetable}
    \begin{deluxetable*}{lcccc|ccccccccc}
        \rowcolors{2}{gray!10}{white}
        \tabletypesize{\small}
        \tablecaption{
            New GRB candidates and a comparison of GBM GRBs and our pipeline detected catalog GRBs of the 2014 search. \label{table:catalogs}}
        \tablewidth{0pt}
        \tablehead{
            \multicolumn{5}{c}{\textbf{GBM}} & 
            \multicolumn{7}{c}{\textbf{Pipeline}} &
            \multicolumn{1}{c}{\textbf{Subthreshold/Other}\tablenotemark{\scriptsize f}}
            \\
            \colhead{Name\tablenotemark{\scriptsize a}} & 
            \colhead{T90 (s)} & 
            \colhead{SNR ($\sigma$)\tablenotemark{\scriptsize b}} & 
            \colhead{RA (deg)} & 
            \colhead{Dec (deg)} & 
            \colhead{Duration (s)\tablenotemark{\scriptsize c}} & 
            \colhead{SNR ($\sigma$)} & 
            \colhead{RA (deg)\tablenotemark{\scriptsize d}} & 
            \colhead{Dec (deg)\tablenotemark{\scriptsize e}} & 
            \colhead{$E_{\text{peak}}$ (KeV)\tablenotemark{\scriptsize f}} & 
            \colhead{$\mathcal{B}_{\oplus}$} & 
            \colhead{$\mathcal{B}_{\odot}$} &
            \colhead{}
        }
        \startdata
            2014-01-01 07:50:54.12 & - & - & - & - & 2.671 & 8.42 & 325.31 & 22.44 & 44.86 & 0.00 & 0.00 & - \\
            bn140105065 & 1.09 & 6.40 & 208.22 & 50.17 & 0.242 & 21.44 & 210.17 & 49.79 & 366.86 & 0.00 & 0.00 & - \\
            bn140105748 & 0.58 & 4.70 & 252.88 & 19.03 & 3.606 & 19.15 & 256.83 & 23.13 & 495.30 & 0.00 & 0.00 & - \\
            bn140109771 & 0.70 & 4.80 & 102.74 & 29.76 & 0.327 & 16.21 & 195.73 & 41.64 & 668.72 & 0.00 & 0.00 & - \\
            bn140110411 & 0.77 & 4.60 & 50.64 & -69.29 & 0.327 & 11.06 & 334.96 & -76.88 & 366.86 & 0.30 & 0.20 & - \\
            bn140122597 & 3.58 & 4.80 & 56.08 & 15.09 & 2.671 & 17.58 & 58.98 & 3.16 & 110.41 & 0.00 & 0.00 & - \\
            2014-01-24 18:00:09.62 & - & - & - & - & 0.133 & 11.90 & 222.52 & -36.77 & 495.30 & 0.00 & 0.00 & GBM subthreshold \\
            2014-01-29 07:06:50.56 & - & - & - & - & 0.133 & 10.79 & 79.91 & -6.35 & 201.26 & 0.00 & 0.00 & - \\
            bn140129499 & 0.13 & 8.20 & 183.40 & -10.32 & 0.054 & 23.71 & 185.10 & -3.97 & 149.06 & 0.00 & 0.00 & - \\
            2014-02-02 21:48:46.53 & - & - & - & - & 0.022 & 8.39 & 272.66 & 88.03 & 110.41 & 0.98 & 0.00 & - \\
            bn140209313 & 1.41 & 5.40 & 81.33 & 32.49 & 0.596 & 235.61 & 89.66 & 23.69 & 110.41 & 0.00 & 0.00 & - \\
            2014-02-09 22:03:12.19 & - & - & - & - & 2.671 & 9.78 & 281.68 & -28.26 & 81.78 & 0.00 & 0.07 & - \\
            bn140211091 & 3.46 & 5.00 & 115.84 & -13.59 & 1.979 & 18.46 & 115.91 & -10.83 & 110.41 & 0.00 & 0.00 & - \\
            bn140216331 & 2.43 & 5.10 & 194.04 & 31.46 & 0.596 & 12.25 & 204.93 & 39.76 & 366.86 & 0.00 & 0.00 & - \\
            bn140224382 & 2.30 & 4.70 & 2.78 & 20.37 & 2.671 & 25.91 & 330.99 & 27.19 & 271.72 & 0.00 & 0.00 & - \\
            2014-02-25 04:03:20.64 & - & - & - & - & 3.606 & 10.03 & 348.73 & 17.04 & 149.06 & 0.01 & 0.00 & - \\
            2014-03-03 06:44:35.49 & - & - & - & - & 0.073 & 12.02 & 297.48 & 8.78 & 81.78 & 0.00 & 0.00 & - \\
            2014-03-07 00:43:25.41 & - & - & - & - & 0.022 & 17.68 & 148.46 & 69.51 & 1645.78 & 0.09 & 0.00 & GBM subthreshold \\
            bn140320092 & 2.30 & 5.60 & 281.86 & -11.17 & 0.242 & 17.94 & 281.97 & -6.71 & 902.86 & 0.00 & 0.00 & - \\
            2014-03-23 20:48:42.70 & - & - & - & - & 2.671 & 10.08 & 313.85 & 53.01 & 60.57 & 0.00 & 0.00 & - \\
            bn140329272 & 0.06 & 10.70 & 92.35 & -41.08 & 0.073 & 32.42 & 94.57 & -46.41 & 366.86 & 0.00 & 0.00 & - \\
            bn140402007 & 0.32 & 5.70 & 207.59 & 5.97 & 0.098 & 20.52 & 200.90 & 7.90 & 902.86 & 0.00 & 0.00 & - \\
            2014-04-05 14:44:14.78 & - & - & - & - & 1.979 & 9.81 & 326.22 & -12.30 & 271.72 & 0.00 & 0.03 & - \\
            2014-04-12 17:31:24.51 & - & - & - & - & 0.327 & 10.93 & 243.31 & -28.21 & 902.86 & 0.00 & 0.00 & GBM subthreshold \\
            bn140428906 & 0.32 & 7.40 & 2.01 & 68.17 & 0.133 & 37.84 & 358.58 & 62.82 & 668.72 & 0.00 & 0.00 & - \\
            bn140501139 & 0.26 & - & 171.88 & 24.64 & 0.133 & 15.49 & 167.47 & 30.51 & 495.30 & 0.00 & 0.00 & - \\
            bn140511095 & 1.41 & 10.70 & 329.76 & -30.06 & 0.098 & 23.57 & 313.76 & -25.40 & 271.72 & 0.00 & 0.00 & - \\
            2014-05-13 03:18:50.18 & - & - & - & - & 3.606 & 9.06 & 178.49 & 66.61 & 110.41 & 0.06 & 0.00 & - \\
            bn140518709 & 0.70 & 4.70 & 244.04 & -77.86 & 0.133 & 11.21 & 277.42 & -64.21 & 366.86 & 0.00 & 0.00 & - \\
            bn140526571 & 0.06 & 6.00 & 142.99 & -10.95 & 0.098 & 17.60 & 145.65 & -15.57 & 668.72 & 0.00 & 0.00 & - \\
            2014-05-27 14:56:06.45 & - & - & - & - & 1.466 & 8.45 & 357.72 & 0.40 & 81.78 & 0.02 & 0.00 & - \\
            bn140605377 & 0.51 & 8.30 & 121.79 & -53.86 & 0.179 & 47.37 & 122.45 & -60.64 & 668.72 & 0.00 & 0.00 & - \\
            2014-06-06 10:58:13.56 & - & - & - & - & 0.327 & 12.98 & 224.26 & 38.46 & 201.26 & 0.00 & 0.00 & GBM subthreshold, Swift, ACS \\
            bn140610487 & 0.96 & 6.20 & 199.05 & 35.91 & 0.179 & 16.51 & 199.57 & 35.10 & 366.86 & 0.00 & 0.00 & - \\
            2014-06-10 16:18:39.11 & - & - & - & - & 1.086 & 13.27 & 219.84 & -25.70 & 44.86 & 0.00 & 0.00 & GBM subthreshold \\
            bn140619475 & 2.82 & 5.60 & 132.70 & -9.70 & 0.596 & 42.96 & 135.77 & -4.31 & 902.86 & 0.00 & 0.00 & - \\
            bn140619490 & 0.45 & 5.80 & 304.96 & -39.18 & 0.133 & 17.83 & 247.45 & -23.79 & 271.72 & 0.00 & 0.00 & - \\
            2014-06-22 11:11:55.76 & - & - & - & - & 0.073 & 9.54 & 281.88 & -9.43 & 44.86 & 0.54 & 0.00 & - \\
            bn140624423 & 0.10 & 7.70 & 23.16 & -0.56 & 0.022 & 58.16 & 22.39 & -10.07 & 271.72 & 0.00 & 0.00 & - \\
            bn140626843 & 1.79 & 6.10 & 120.81 & 38.78 & 1.466 & 31.69 & 125.61 & 42.09 & 110.41 & 0.00 & 0.00 & - \\
            2014-07-01 18:38:26.69 & - & - & - & - & 0.098 & 11.41 & 38.08 & -21.99 & 110.41 & 0.00 & 0.00 & GBM subthreshold \\
            2014-07-04 19:24:42.31 & - & - & - & - & 1.466 & 8.63 & 301.99 & -33.86 & 110.41 & 0.00 & 0.00 & - \\
            2014-07-07 05:59:44.82 & - & - & - & - & 3.606 & 9.09 & 320.46 & 9.21 & 149.06 & 0.00 & 0.00 & - \\
            bn140710537 & 0.38 & 5.60 & 2.80 & -38.88 & 0.441 & 17.12 & 357.43 & -31.08 & 149.06 & 0.34 & 0.00 & - \\
            bn140716306 & 3.33 & 5.20 & 215.23 & 57.01 & 2.671 & 19.76 & 231.55 & 47.12 & 201.26 & 0.00 & 0.00 & - \\
            2014-07-19 05:28:18.47 & - & - & - & - & 1.086 & 9.36 & 35.92 & 55.97 & 60.57 & 0.82 & 0.00 & - \\
            bn140720158 & 0.32 & 7.80 & 175.03 & -32.31 & 0.022 & 17.81 & 202.31 & -20.95 & 201.26 & 0.00 & 0.00 & - \\
            bn140724533 & 0.90 & 5.20 & 314.73 & -1.85 & 0.441 & 11.30 & 299.43 & -8.44 & 110.41 & 0.00 & 0.00 & - \\
            bn140807500 & 0.51 & 5.20 & 200.16 & 26.49 & 0.327 & 66.51 & 195.93 & 32.20 & 495.30 & 0.00 & 0.00 & - \\
            bn140831215 & 0.70 & 4.70 & 280.37 & 25.64 & 0.242 & 10.59 & 278.58 & 38.44 & 271.72 & 0.00 & 0.00 & - \\
            bn140831374 & 3.58 & 4.80 & 4.29 & 44.01 & 1.086 & 28.84 & 351.03 & 39.80 & 366.86 & 0.00 & 0.00 & - \\
            bn140901821 & 0.18 & 13.40 & 15.82 & -32.76 & 0.133 & 206.81 & 16.11 & -32.36 & 1645.78 & 0.00 & 0.00 & - \\
            bn140930134 & 3.26 & 4.70 & 41.61 & 57.58 & 1.466 & 20.49 & 21.03 & 33.66 & 366.86 & 0.03 & 0.00 & - \\
            bn141004973 & 2.56 & 4.80 & 76.75 & 12.82 & 0.441 & 33.85 & 62.52 & 15.80 & 81.78 & 0.00 & 0.00 & - \\
            bn141005217 & 3.39 & 5.00 & 291.09 & 36.10 & 2.671 & 53.42 & 292.81 & 44.17 & 110.41 & 0.00 & 0.00 & - \\
            bn141011282 & 0.08 & 35.00 & 257.94 & -9.68 & 0.179 & 64.23 & 251.51 & 1.28 & 495.30 & 0.00 & 0.00 & - \\
            bn141020439 & 1.60 & 4.60 & 265.49 & -1.68 & 0.804 & 10.98 & 268.31 & -11.14 & 495.30 & 0.00 & 0.00 & - \\
            2014-10-21 14:54:13.07 & - & - & - & - & 0.016 & 8.46 & 348.43 & 14.37 & 60.57 & 0.00 & 0.00 & - \\
            2014-10-23 04:24:57.77 & - & - & - & - & 0.179 & 11.28 & 195.86 & -68.46 & 60.57 & 0.00 & 0.00 & - \\
            bn141026742 & 2.56 & 5.70 & 132.82 & 62.48 & 1.979 & 22.03 & 102.92 & 53.45 & 201.26 & 0.00 & 0.00 & - \\
            2014-10-27 03:24:33.67 & - & - & - & - & 0.073 & 9.58 & 93.03 & -59.05 & 44.86 & 0.00 & 0.00 & - \\
            bn141031998 & 0.16 & 6.20 & 133.08 & -33.68 & 0.133 & 13.79 & 117.16 & -29.82 & 366.86 & 0.02 & 0.00 & - \\
            bn141102112 & 0.02 & 5.10 & 223.23 & -17.42 & 0.030 & 15.25 & 220.87 & 3.27 & 201.26 & 0.04 & 0.53 & - \\
            bn141102536 & 2.62 & 7.80 & 208.61 & -47.10 & 0.441 & 43.31 & 212.95 & -50.83 & 366.86 & 0.00 & 0.00 & - \\
            bn141105406 & 1.28 & 6.40 & 16.91 & 29.17 & 0.596 & 57.53 & 18.44 & 28.67 & 495.30 & 0.00 & 0.00 & - \\
            bn141111435 & 1.73 & 4.80 & 51.22 & 43.02 & 0.327 & 10.15 & 34.53 & 56.21 & 668.72 & 0.00 & 0.00 & - \\
            bn141113346 & 0.45 & 5.20 & 171.02 & 80.26 & 0.441 & 16.63 & 200.65 & 67.39 & 902.86 & 0.01 & 0.00 & - \\
            2014-11-20 11:29:29.44 & - & - & - & - & 3.606 & 10.64 & 150.31 & -49.43 & 44.86 & 0.15 & 0.00 & GBM subthreshold \\
            bn141122087 & 1.28 & 5.70 & 9.71 & -20.02 & 0.098 & 17.08 & 8.01 & -5.38 & 495.30 & 0.00 & 0.00 & - \\
            bn141124277 & 0.51 & 4.90 & 135.07 & 78.18 & 0.441 & 17.49 & 145.34 & 76.03 & 495.30 & 0.00 & 0.00 & - \\
            bn141126233 & 0.90 & 4.80 & 243.87 & 59.99 & 0.596 & 11.66 & 247.34 & 61.45 & 271.72 & 0.00 & 0.00 & - \\
            bn141128962 & 0.27 & 12.70 & 321.80 & -35.76 & 0.054 & 30.90 & 317.84 & -34.01 & 110.41 & 0.00 & 0.00 & - \\
            bn141202470 & 1.41 & 5.60 & 143.07 & 54.16 & 1.086 & 79.11 & 132.93 & 54.76 & 668.72 & 0.00 & 0.00 & - \\
            bn141205337 & 1.28 & 5.10 & 92.86 & 37.88 & 0.596 & 20.64 & 82.10 & -5.26 & 495.30 & 0.00 & 0.00 & - \\
            bn141208632 & 0.96 & 5.90 & 359.26 & 26.44 & 0.073 & 13.17 & 356.40 & 38.08 & 366.86 & 0.00 & 0.00 & - \\
            bn141213300 & 0.77 & 5.90 & 248.19 & 18.06 & 0.242 & 24.91 & 254.61 & 23.37 & 110.41 & 0.00 & 0.00 & - \\
            2014-12-17 04:22:13.27 & - & - & - & - & 0.804 & 9.45 & 95.72 & 32.23 & 495.30 & 0.03 & 0.00 & - \\
            2014-12-19 06:47:56.80 & - & - & - & - & 2.671 & 8.88 & 196.55 & -2.85 & 110.41 & 0.00 & 0.00 & - \\
            bn141222298 & 2.75 & 9.90 & 178.04 & -57.35 & 3.606 & 158.69 & 169.17 & -57.20 & 495.30 & 0.00 & 0.00 & - \\
            2014-12-26 01:00:08.93 & - & - & - & - & 2.671 & 12.50 & 75.02 & -18.39 & 81.78 & 0.02 & 0.00 & GBM subthreshold \\
            bn140109877 & 3.33 & 5.10 & 24.09 & -25.05 & - & - & - & - & - & - & -  & -\\
            bn140616165 & 0.51 & 5.60 & 104.94 & -70.51 & - & - & - & - & - & - & -  & -\\
            bn140912664 & 2.30 & - & 303.50 & 59.56 & - & - & - & - & - & - & -  & -\\
            bn141230871${}^*$ & 0.22 & 6.70 & 246.93 & -40.18 & - & - & - & - & - & - & - & -\\
\enddata
        \tablenotetext{a}{\textit{Name:} The trigger name as appears in the GBM trigger catalog. If it is a new trigger, the trigger time is displayed instead.}
        \tablenotetext{b}{\textit{SNR ($\sigma$):} The significance of the trigger at triggering, as appears in the GBM data product: TRIGDAT file. It is reported in units of standard deviations of standard normal distribution, hence it has the same meaning as our SNR.}
        \tablenotetext{c}{\textit{Duration:} The burst duration that had the maximal SNR in our search.}
        \tablenotetext{d}{\textit{RA (deg):} The maximum likelihood estimation of the right ascension, obtained by the sky posterior.}
        \tablenotetext{e}{\textit{DEC (deg):} The maximum likelihood estimation of the declination, obtained by the sky posterior.}
        \tablenotetext{e}{$E_{\text{peak}}$ (KeV): The maximum likelihood estimation of the peak energy, obtained by the densest template bank.}
        \tablenotetext{f}{\textit{Subthreshold/Other:} Indicated if another subthreshold search or other instrument has detected this candidate.}
        \tablenotetext{*}{The peak energy of this GRB was estimated in our pipeline to be $33.23$ KeV and was included in this table for completeness.}
        \end{deluxetable*}
        \end{longrotatetable}

\bibliography{bibliography}{}

\begin{thebibliography}{}
\expandafter\ifx\csname natexlab\endcsname\relax\def\natexlab#1{#1}\fi
\providecommand{\url}[1]{\href{#1}{#1}}
\providecommand{\dodoi}[1]{doi:~\href{http://doi.org/#1}{\nolinkurl{#1}}}
\providecommand{\doeprint}[1]{\href{http://ascl.net/#1}{\nolinkurl{http://ascl.net/#1}}}
\providecommand{\doarXiv}[1]{\href{https://arxiv.org/abs/#1}{\nolinkurl{https://arxiv.org/abs/#1}}}

\bibitem[{Abbott {et~al.}(2017{\natexlab{a}})Abbott, Abbott, Abbott, Acernese,
  Ackley, Adams, Adams, Addesso, Adhikari, Adya, {et~al.}}]{abbott2017gw170817}
Abbott, B.~P., Abbott, R., Abbott, T., {et~al.} 2017{\natexlab{a}}, Physical
  review letters, 119, 161101

\bibitem[{Abbott {et~al.}(2017{\natexlab{b}})Abbott, Abbott, Abbott, Acernese,
  Ackley, Adams, Adams, Addesso, Adhikari, Adya,
  {et~al.}}]{abbott2017gravitational}
---. 2017{\natexlab{b}}, The Astrophysical Journal Letters, 848, L13

\bibitem[{{Band} {et~al.}(1993){Band}, {Matteson}, {Ford}, {Schaefer},
  {Palmer}, {Teegarden}, {Cline}, {Briggs}, {Paciesas}, {Pendleton}, {Fishman},
  {Kouveliotou}, {Meegan}, {Wilson}, \& {Lestrade}}]{1993ApJ...413..281B}
{Band}, D., {Matteson}, J., {Ford}, L., {et~al.} 1993, \apj, 413, 281,
  \dodoi{10.1086/172995}

\bibitem[{Berger(2014)}]{berger2014short}
Berger, E. 2014, Annual review of Astronomy and Astrophysics, 52, 43

\bibitem[{Blackburn {et~al.}(2015)Blackburn, Briggs, Camp, Christensen,
  Connaughton, Jenke, Remillard, \& Veitch}]{blackburn2015high}
Blackburn, L., Briggs, M.~S., Camp, J., {et~al.} 2015, The Astrophysical
  Journal Supplement Series, 217, 8

\bibitem[{Blanchard {et~al.}(2016)Blanchard, Berger, \&
  Fong}]{blanchard2016offset}
Blanchard, P.~K., Berger, E., \& Fong, W.-f. 2016, The Astrophysical Journal,
  817, 144

\bibitem[{Briggs {et~al.}(2013)Briggs, Xiong, Connaughton, Tierney,
  Fitzpatrick, Foley, Grove, Chekhtman, Gibby, Fishman,
  {et~al.}}]{briggs2013terrestrial}
Briggs, M.~S., Xiong, S., Connaughton, V., {et~al.} 2013, Journal of
  Geophysical Research: Space Physics, 118, 3805

\bibitem[{Burgess {et~al.}(2018)Burgess, Yu, Greiner, \&
  Mortlock}]{burgess2018awakening}
Burgess, J.~M., Yu, H.-F., Greiner, J., \& Mortlock, D.~J. 2018, Monthly
  Notices of the Royal Astronomical Society, 476, 1427

\bibitem[{Burns(2019)}]{burns2019fermi}
Burns, E. 2019, Default journal, 871

\bibitem[{Cano {et~al.}(2017)Cano, Wang, Dai, \& Wu}]{cano2017observer}
Cano, Z., Wang, S.-Q., Dai, Z.-G., \& Wu, X.-F. 2017, Advances in Astronomy,
  2017, 8929054

\bibitem[{Connaughton {et~al.}(2015)Connaughton, Briggs, Goldstein, Meegan,
  Paciesas, Preece, Wilson-Hodge, Gibby, Greiner, Gruber,
  {et~al.}}]{connaughton2015localization}
Connaughton, V., Briggs, M., Goldstein, A., {et~al.} 2015, The Astrophysical
  Journal Supplement Series, 216, 32

\bibitem[{Dwyer {et~al.}(2012)Dwyer, Smith, \& Cummer}]{dwyer2012high}
Dwyer, J.~R., Smith, D.~M., \& Cummer, S.~A. 2012, Space Science Reviews, 173,
  133

\bibitem[{Eichler {et~al.}(1989)Eichler, Livio, Piran, \&
  Schramm}]{eichler1989nucleosynthesis}
Eichler, D., Livio, M., Piran, T., \& Schramm, D.~N. 1989, Nature, 340, 126

\bibitem[{Fletcher {et~al.}(2011)Fletcher, Dennis, Hudson, Krucker, Phillips,
  Veronig, Battaglia, Bone, Caspi, Chen, {et~al.}}]{fletcher2011observational}
Fletcher, L., Dennis, B.~R., Hudson, H.~S., {et~al.} 2011, Space science
  reviews, 159, 19

\bibitem[{Fong \& Berger(2013)}]{fong2013locations}
Fong, W.-f., \& Berger, E. 2013, The Astrophysical Journal, 776, 18

\bibitem[{Fong {et~al.}(2009)Fong, Berger, \& Fox}]{fong2009hubble}
Fong, W.-f., Berger, E., \& Fox, D. 2009, The Astrophysical Journal, 708, 9

\bibitem[{Galama {et~al.}(1998)Galama, Vreeswijk, Van~Paradijs, Kouveliotou,
  Augusteijn, B{\"o}hnhardt, Brewer, Doublier, Gonzalez, Leibundgut,
  {et~al.}}]{galama1998unusual}
Galama, T.~J., Vreeswijk, P., Van~Paradijs, J., {et~al.} 1998, nature, 395, 670

\bibitem[{Goldstein {et~al.}(2016)Goldstein, Burns, Hamburg, Connaughton,
  Veres, Briggs, Hui, \& Collaboration}]{goldstein2016updates}
Goldstein, A., Burns, E., Hamburg, R., {et~al.} 2016, arXiv preprint
  arXiv:1612.02395

\bibitem[{Goldstein {et~al.}(2022)Goldstein, Cleveland, \&
  Kocevski}]{GbmDataTools}
Goldstein, A., Cleveland, W.~H., \& Kocevski, D. 2022, Fermi GBM Data Tools:
  v1.1.1.
\newblock \url{https://fermi.gsfc.nasa.gov/ssc/data/analysis/gbm}

\bibitem[{Goldstein {et~al.}(2017)Goldstein, Veres, Burns, Briggs, Hamburg,
  Kocevski, Wilson-Hodge, Preece, Poolakkil, Roberts,
  {et~al.}}]{goldstein2017ordinary}
Goldstein, A., Veres, P., Burns, E., {et~al.} 2017, The Astrophysical Journal
  Letters, 848, L14

\bibitem[{Goldstein {et~al.}(2019)Goldstein, Hamburg, Wood, Hui, Cleveland,
  Kocevski, Littenberg, Burns, Canton, Veres, {et~al.}}]{goldstein2019updates}
Goldstein, A., Hamburg, R., Wood, J., {et~al.} 2019, arXiv preprint
  arXiv:1903.12597

\bibitem[{Gonz{\'a}lez(2010)}]{gonzalez2010measurement}
Gonz{\'a}lez, {\'A}. 2010, Mathematical geosciences, 42, 49

\bibitem[{Goodman(1986)}]{goodman1986gamma}
Goodman, J. 1986, Astrophysical Journal, Part 2-Letters to the Editor (ISSN
  0004-637X), vol. 308, Sept. 15, 1986, p. L47-L50., 308, L47

\bibitem[{Gorski {et~al.}(2005)Gorski, Hivon, Banday, Wandelt, Hansen,
  Reinecke, \& Bartelmann}]{gorski2005healpix}
Gorski, K.~M., Hivon, E., Banday, A.~J., {et~al.} 2005, The Astrophysical
  Journal, 622, 759

\bibitem[{Gottlieb {et~al.}(2018)Gottlieb, Nakar, Piran, \&
  Hotokezaka}]{gottlieb2018cocoon}
Gottlieb, O., Nakar, E., Piran, T., \& Hotokezaka, K. 2018, Monthly Notices of
  the Royal Astronomical Society, 479, 588

\bibitem[{Harmon {et~al.}(2002)Harmon, Fishman, Wilson, Paciesas, Zhang,
  Finger, Koshut, McCollough, Robinson, \& Rubin}]{harmon2002burst}
Harmon, B., Fishman, G., Wilson, C., {et~al.} 2002, The Astrophysical Journal
  Supplement Series, 138, 149

\bibitem[{Harry {et~al.}(2009)Harry, Allen, \&
  Sathyaprakash}]{harry2009stochastic}
Harry, I.~W., Allen, B., \& Sathyaprakash, B. 2009, Physical Review
  D—Particles, Fields, Gravitation, and Cosmology, 80, 104014

\bibitem[{Horack {et~al.}(1991)Horack, Fishman, Meegan, Wilson, \&
  Paciesas}]{horack1991batse}
Horack, J., Fishman, G., Meegan, C., Wilson, R., \& Paciesas, W. 1991, in AIP
  Conference Proceedings, Vol. 265, 373--377

\bibitem[{Ivashtenko \& Zackay(2025)}]{ivashtenko2025independent}
Ivashtenko, O., \& Zackay, B. 2025, arXiv preprint arXiv:2501.03019

\bibitem[{Kapadia {et~al.}(2020)Kapadia, Caudill, Creighton, Farr, Mendell,
  Weinstein, Cannon, Fong, Godwin, Lo, {et~al.}}]{kapadia2020self}
Kapadia, S.~J., Caudill, S., Creighton, J.~D., {et~al.} 2020, Classical and
  Quantum Gravity, 37, 045007

\bibitem[{{Klebesadel} {et~al.}(1973){Klebesadel}, {Strong}, \&
  {Olson}}]{1973ApJ...182L..85K}
{Klebesadel}, R.~W., {Strong}, I.~B., \& {Olson}, R.~A. 1973, \apjl, 182, L85,
  \dodoi{10.1086/181225}

\bibitem[{Kocevski {et~al.}(2018)Kocevski, Burns, Goldstein, Dal~Canton,
  Briggs, Blackburn, Veres, Hui, Hamburg, Roberts,
  {et~al.}}]{kocevski2018analysis}
Kocevski, D., Burns, E., Goldstein, A., {et~al.} 2018, The Astrophysical
  Journal, 862, 152

\bibitem[{Kouveliotou {et~al.}(1993)Kouveliotou, Meegan, Fishman, Bhat, Briggs,
  Koshut, Paciesas, \& Pendleton}]{kouveliotou1993identification}
Kouveliotou, C., Meegan, C.~A., Fishman, G.~J., {et~al.} 1993, Astrophysical
  Journal, Part 2-Letters (ISSN 0004-637X), vol. 413, no. 2, p. L101-L104.,
  413, L101

\bibitem[{Kulkarni {et~al.}(1998)Kulkarni, Frail, Wieringa, Ekers, Sadler,
  Wark, Higdon, Phinney, \& Bloom}]{kulkarni1998radio}
Kulkarni, S., Frail, D., Wieringa, M., {et~al.} 1998, Nature, 395, 663

\bibitem[{Kulkarni(2005)}]{kulkarni2005modeling}
Kulkarni, S.~R. 2005, arXiv preprint astro-ph/0510256

\bibitem[{Lesage {et~al.}(2023)Lesage, Veres, Briggs, Goldstein, Kocevski,
  Burns, Wilson-Hodge, Bhat, Huppenkothen, Fryer, {et~al.}}]{lesage2023fermi}
Lesage, S., Veres, P., Briggs, M., {et~al.} 2023, The Astrophysical journal
  letters, 952, L42

\bibitem[{Levan(2018)}]{10.1088/2514-3433/aae164}
Levan, A. 2018, Gamma-Ray Bursts, 2514-3433 (IOP Publishing),
  \dodoi{10.1088/2514-3433/aae164}

\bibitem[{Li \& Paczy{\'n}ski(1998)}]{li1998transient}
Li, L.-X., \& Paczy{\'n}ski, B. 1998, The Astrophysical Journal, 507, L59

\bibitem[{Lien {et~al.}(2016)Lien, Sakamoto, Barthelmy, Baumgartner, Cannizzo,
  Chen, Collins, Cummings, Gehrels, Krimm, {et~al.}}]{lien2016third}
Lien, A., Sakamoto, T., Barthelmy, S.~D., {et~al.} 2016, The Astrophysical
  Journal, 829, 7

\bibitem[{Lindanger {et~al.}(2021)Lindanger, Marisaldi, Sarria, {\O}stgaard,
  Lehtinen, Skeie, Mezentzev, Kochkin, Ullaland, Yang,
  {et~al.}}]{lindanger2021spectral}
Lindanger, A., Marisaldi, M., Sarria, D., {et~al.} 2021, Journal of Geophysical
  Research: Atmospheres, 126, e2021JD035347

\bibitem[{Lunnan {et~al.}(2014)Lunnan, Chornock, Berger, Laskar, Fong, Rest,
  Sanders, Challis, Drout, Foley, {et~al.}}]{lunnan2014hydrogen}
Lunnan, R., Chornock, R., Berger, E., {et~al.} 2014, The Astrophysical Journal,
  787, 138

\bibitem[{Meegan {et~al.}(2009)Meegan, Lichti, Bhat, Bissaldi, Briggs,
  Connaughton, Diehl, Fishman, Greiner, Hoover, {et~al.}}]{meegan2009fermi}
Meegan, C., Lichti, G., Bhat, P., {et~al.} 2009, The Astrophysical Journal,
  702, 791

\bibitem[{Metzger(2020)}]{metzger2020kilonovae}
Metzger, B.~D. 2020, Living Reviews in Relativity, 23, 1

\bibitem[{Neyman \& Pearson(1933)}]{neyman1933ix}
Neyman, J., \& Pearson, E.~S. 1933, Philosophical Transactions of the Royal
  Society of London. Series A, Containing Papers of a Mathematical or Physical
  Character, 231, 289

\bibitem[{Ofek \& Zackay(2018)}]{ofek2018optimal}
Ofek, E.~O., \& Zackay, B. 2018, The Astronomical Journal, 155, 169

\bibitem[{Paciesas {et~al.}(1999)Paciesas, Meegan, Pendleton, Briggs,
  Kouveliotou, Koshut, Lestrade, McCollough, Brainerd, Hakkila,
  {et~al.}}]{paciesas1999fourth}
Paciesas, W.~S., Meegan, C.~A., Pendleton, G.~N., {et~al.} 1999, The
  Astrophysical Journal Supplement Series, 122, 465

\bibitem[{Paczynski(1986)}]{paczynski1986gamma}
Paczynski, B. 1986, Astrophysical Journal, Part 2-Letters to the Editor (ISSN
  0004-637X), vol. 308, Sept. 15, 1986, p. L43-L46., 308, L43

\bibitem[{Preece {et~al.}(2000)Preece, Briggs, Mallozzi, Pendleton, Paciesas,
  \& Band}]{preece2000batse}
Preece, R.~D., Briggs, M.~S., Mallozzi, R.~S., {et~al.} 2000, The Astrophysical
  Journal Supplement Series, 126, 19

\bibitem[{Rodi {et~al.}(2014)Rodi, Cherry, Case, Camero-Arranz, Chaplin,
  Finger, Jenke, \& Wilson-Hodge}]{rodi2014earth}
Rodi, J., Cherry, M., Case, G., {et~al.} 2014, Astronomy \& Astrophysics, 562,
  A7

\bibitem[{Salafia {et~al.}(2018)Salafia, Ghisellini, Ghirlanda, \&
  Colpi}]{salafia2018interpreting}
Salafia, O.~S., Ghisellini, G., Ghirlanda, G., \& Colpi, M. 2018, Astronomy \&
  Astrophysics, 619, A18

\bibitem[{Savchenko {et~al.}(2017)Savchenko, Ferrigno, Kuulkers, Bazzano,
  Bozzo, Brandt, Chenevez, Courvoisier, Diehl, Domingo,
  {et~al.}}]{savchenko2017integral}
Savchenko, V., Ferrigno, C., Kuulkers, E., {et~al.} 2017, The Astrophysical
  Journal Letters, 848, L15

\bibitem[{Smartt {et~al.}(2017)Smartt, Chen, Jerkstrand, Coughlin, Kankare,
  Sim, Fraser, Inserra, Maguire, Chambers, {et~al.}}]{smartt2017kilonova}
Smartt, S., Chen, T.-W., Jerkstrand, A., {et~al.} 2017, Nature, 551, 75

\bibitem[{Tanvir {et~al.}(2017)Tanvir, Levan, Gonz{\'a}lez-Fern{\'a}ndez,
  Korobkin, Mandel, Rosswog, Hjorth, D’Avanzo, Fruchter, Fryer,
  {et~al.}}]{tanvir2017emergence}
Tanvir, N.~R., Levan, A.~J., Gonz{\'a}lez-Fern{\'a}ndez, C., {et~al.} 2017, The
  Astrophysical Journal Letters, 848, L27

\bibitem[{Venumadhav {et~al.}(2019)Venumadhav, Zackay, Roulet, Dai, \&
  Zaldarriaga}]{venumadhav2019new}
Venumadhav, T., Zackay, B., Roulet, J., Dai, L., \& Zaldarriaga, M. 2019,
  Physical Review D, 100, 023011

\bibitem[{Vianello {et~al.}(2015)Vianello, Lauer, Younk, Tibaldo, Burgess,
  Ayala, Harding, Hui, Omodei, \& Zhou}]{vianello2015multi}
Vianello, G., Lauer, R.~J., Younk, P., {et~al.} 2015, arXiv preprint
  arXiv:1507.08343

\bibitem[{Von~Kienlin {et~al.}(2014)Von~Kienlin, Meegan, Paciesas, Bhat,
  Bissaldi, Briggs, Burgess, Byrne, Chaplin, Cleveland,
  {et~al.}}]{von2014second}
Von~Kienlin, A., Meegan, C.~A., Paciesas, W.~S., {et~al.} 2014, The
  Astrophysical Journal Supplement Series, 211, 13

\bibitem[{Von~Kienlin {et~al.}(2020)Von~Kienlin, Meegan, Paciesas, Bhat,
  Bissaldi, Briggs, Burns, Cleveland, Gibby, Giles, {et~al.}}]{von2020fourth}
Von~Kienlin, A., Meegan, C., Paciesas, W., {et~al.} 2020, The Astrophysical
  Journal, 893, 46

\bibitem[{{W{\k{a}}s} {et~al.}(2010){W{\k{a}}s}, {Bizouard}, {Brisson},
  {Cavalier}, {Davier}, {Hello}, {Leroy}, {Robinet}, \&
  {Vavoulidis}}]{2010CQGra..27a5005W}
{W{\k{a}}s}, M., {Bizouard}, M.-A., {Brisson}, V., {et~al.} 2010, Classical and
  Quantum Gravity, 27, 015005, \dodoi{10.1088/0264-9381/27/1/015005}

\bibitem[{Woosley(1993)}]{woosley1993gamma}
Woosley, S.~E. 1993, Astrophysical Journal, Part 1 (ISSN 0004-637X), vol. 405,
  no. 1, p. 273-277., 405, 273

\bibitem[{Zackay {et~al.}(2021)Zackay, Venumadhav, Roulet, Dai, \&
  Zaldarriaga}]{zackay2021detecting}
Zackay, B., Venumadhav, T., Roulet, J., Dai, L., \& Zaldarriaga, M. 2021,
  Physical Review D, 104, 063034

\bibitem[{Zhang(2019)}]{zhang2019delay}
Zhang, B. 2019, Frontiers of Physics, 14, 1

\end{thebibliography}
\bibliographystyle{aasjournal}

\end{document}